\documentclass[sigplan,screen,10pt,nonacm]{acmart}
\newtoggle{comments}
\newtoggle{people}

\usepackage{graphicx}
\graphicspath{./figs}
\usepackage{pifont}
\usepackage{xspace}
\usepackage{pgfplotstable}
\usepackage[normalem]{ulem}

\usepackage{subcaption}
\usepackage{pgfplots}
\usepackage{mdframed} %

\usepackage{amsthm}
\usepackage{amsmath,amsfonts}
\usepackage{algorithmic}
\usepackage{tikzsymbols} \usetikzlibrary{positioning}
\usepackage{textcomp}
\usepackage{bbm}

\usepackage{arydshln}
\usepackage{paralist}
\usepackage{float}
\usepackage{enumitem} \setlist{nosep}
\usepackage{multirow}
\usepackage{multicol}
\usepackage{diagbox}

\usepackage{booktabs}
\usepackage{breakcites}

\usepackage{tabularx}
\usepackage{array}

\usepackage{dashbox}

\pgfplotsset{compat=1.7}
\usepackage{tikz}

\usepackage{dsfont}

\usepackage[
    lambda,
    operators,
    advantage,
    sets,
    adversary,
    landau,
    probability,
    notions,
    logic,
    ff,
    mm,
    primitives,
    events,
    oracles,
    complexity,
    asymptotics,
    keys]{support/cryptocode}

\usepackage[capitalize]{cleveref}

\definecolor{greenish}{rgb}{0.0, 0.5, 0.0}
\definecolor{Orange}{rgb}{1.0, 0.5, 0.0}

\iftoggle{people}{

}{

}

\definecolor{blue-violet}{rgb}{0.54, 0.17, 0.89}
\definecolor{forest-green}{rgb}{0.13, 0.54, 0.13}
\definecolor{forest-green-2}{rgb}{0,0.47,0.43}

\iftoggle{comments}{

\newcommand{\todo}[1]{\textcolor{red}{TODO:#1}}
\newcommand{\secnotes}[1]{
\textcolor{purple}{
\noindent
------------------------- Section Notes ------------------------}
#1
\noindent \textcolor{purple}{\hrule}\vspace{0.2cm}}

\newcommand{\highlightchange}[1]{\textcolor{blue-violet}{#1}}
\newcommand{\verifytext}[1]{\textcolor{brown}{#1}}

\newcommand{\revision}[1]{\textcolor{forest-green}{#1}}
\newcommand{\revisiontwo}[1]{{\color{forest-green-2}#1}}

}{

\newcommand{\todo}[1]{}
\newcommand{\secnotes}[1]{}

\newcommand{\highlightchange}[1]{#1}
\newcommand{\verifytext}[1]{#1}

\newcommand{\revision}[1]{#1}
\newcommand{\revisiontwo}[1]{#1}
}

\newcommand{\eg}{e.g.,\xspace}
\newcommand{\ie}{i.e.,\xspace}
\newcommand{\parhead}[1]{\noindent \textbf{#1}}

\newcommand{\iseq}{\overset{?}{=}}
\newcommand{\isin}{\overset{?}{\in}}
\newcommand{\isnotin}{\overset{?}{\notin}}

\newcommand{\append}{\Leftarrow}

\newcommand{\name}{TRIP\xspace}
\newcommand{\sysname}{Votegral\xspace}

\renewcommand{\L}{\mathds{L}} %
\newcommand{\A}{\mathds{A}} %
\newcommand{\R}{\mathds{R}} %
    \renewcommand{\O}{\mathds{O}} %
    \newcommand{\OSD}{OSD\xspace} %
    \newcommand{\K}{\mathds{K}} %
    \renewcommand{\P}{\mathds{P}} %
\newcommand{\V}{\mathds{V}} %
\newcommand{\VSD}{VSD\xspace} %

\providecommand{\E}{\ensuremath{\mathbf{E}}} %
\providecommand{\pc}{\pckeystyle{pc}} %
\newcommand{\fc}{\tilde{c}} %
\newcommand{\Pfc}{\tilde{P}} %

\newcolumntype{b}{>{\hsize=.75\hsize}X}
\newcolumntype{s}{>{\hsize=.25\hsize}X}

\newtheorem{theorem}{Theorem}
\newtheorem{hybrid}{Hybrid}

\newtheorem{lemma}[theorem]{Lemma}

\newtheorem{definition}[theorem]{Definition}
\providecommand{\mix}{\pcalgostyle{Mix}}
\providecommand{\mixk}{\pcalgostyle{MixK}}
\providecommand{\tagg}{\pcalgostyle{Tag}}
\providecommand{\sign}{\pcalgostyle{Sign}}

\providecommand{\pubkey}{\pcalgostyle{PubKey}}
\providecommand{\dkg}{\pcalgostyle{DKG}}
\providecommand{\elgamal}{\pcalgostyle{EG}}

\createprocedureblock{gameblock}{center,boxed}{}{}{linenumbering}
\renewcommand{\pcadvantagesuperstyle}[1]{\mathrm{#1}}

\newcommand{\cResist}{\ensuremath{\pcalgostyle{C\textsf{-}Resist}}\xspace}
\newcommand{\cResistIdeal}{\ensuremath{\pcalgostyle{C\textsf{-}Resist\textsf{-}Ideal}}\xspace}

\newcommand{\IndVer}{\ensuremath{\pcalgostyle{IV}\xspace}}
\newcommand{\IndVerTarget}{\ensuremath{\pcalgostyle{IV^*}\xspace}}
\newcommand{\IndVerTargetStrong}{%
\ensuremath{\pcalgostyle{IV^{\dagger}}}\xspace}
\newcommand{\IterIndVer}{\ensuremath{\pcalgostyle{I\textsf{-}IV}\xspace}}
\newcommand{\IterIndVerStrong}{\ensuremath{\pcalgostyle{S\textsf{-}I\textsf{-}IV}\xspace}}

\newcommand{\Notify}{\ensuremath{\mathsf{Notify}}\xspace}
\newcommand{\Setup}{\ensuremath{\mathsf{Setup}}\xspace}
\newcommand{\CheckIn}{\ensuremath{\mathsf{CheckIn}}\xspace}
\newcommand{\RealCred}{\ensuremath{\mathsf{RealCred}}\xspace}
\newcommand{\CheckOut}{\ensuremath{\mathsf{CheckOut}}\xspace}
\newcommand{\FakeCred}{\ensuremath{\mathsf{FakeCred}}\xspace}
\newcommand{\FakeCreds}{\ensuremath{\mathsf{FakeCreds}}\xspace}
\newcommand{\Activate}{\ensuremath{\mathsf{Activate}}\xspace}
\newcommand{\Vote}{\ensuremath{\mathsf{Vote}}\xspace}
\newcommand{\Tally}{\ensuremath{\mathsf{Tally}}\xspace}
\newcommand{\VoterVerify}{\ensuremath{\mathsf{VoterVerify}}\xspace}
\newcommand{\Extract}{\ensuremath{\mathsf{Extract}}\xspace}

\newcommand{\VoterInteraction}{\ensuremath{\mathsf{VoterInteraction}}\xspace}
\newcommand{\IdealTally}{\ensuremath{\mathsf{Ideal\textsf{-}Tally}}\xspace}

\newcommand{\IA}{\mathcal{I}} %
\newcommand{\PA}{\pdv} %
\newcommand{\CA}{\cdv} %

\newcommand{\Vc}{\ensuremath{\mathbf{V}_\CA}\xspace}
\newcommand{\goal}{\ensuremath{\mathsf{goal}}\xspace}
\newcommand{\intent}{\ensuremath{\mathsf{intent}}\xspace}
\newcommand{\params}{\ensuremath{\mathsf{params}}\xspace}
\renewcommand{\pp}{\ensuremath{\mathsf{pp}}\xspace}
\newcommand{\infra}{\ensuremath{\textit{infra}}\xspace}
\newcommand{\happy}{\ensuremath{\mathsf{happy}}\xspace}
\newcommand{\advoutcome}{\ensuremath{\mathsf{adv\_outcome}}\xspace}
\newcommand{\Win}{\ensuremath{\mathsf{Win}}\xspace}
\newcommand{\EventE}{\ensuremath{\mathsf{E}}\xspace}
\newcommand{\RegSub}{\ensuremath{\mathsf{RegSub}}\xspace}
\newcommand{\VoteSub}{\ensuremath{\mathsf{VoteSub}}\xspace}
\newcommand{\EnvGuess}{\ensuremath{\mathsf{EnvGuess}}\xspace}
\newcommand{\DupCatch}{\ensuremath{\mathsf{DupCatch}}\xspace}

\newcommand{\local}{\ensuremath{\mathsf{local}}\xspace}
\newcommand{\posted}{\ensuremath{\mathsf{posted}}\xspace}
\newcommand{\wins}{\ensuremath{\mathsf{successes}}\xspace}

\newtoggle{longpaper} %
\usepackage{xr} %
\toggletrue{longpaper}

\newcommand{\LongRefPrefix}{L-}
\iftoggle{longpaper}{}{%
  \externaldocument[\LongRefPrefix]{extended-version} %
}

\newcommand{\apxref}[1]{%
  \iftoggle{longpaper}{%
    \cref{#1}%
  }{%
  \href{https://arxiv.org/abs/2202.06692}{\cref*{\LongRefPrefix#1}, extended version} \cite{merino2025TRIPPreprint}%
  }%
}

\crefformat{section}{§#2#1#3}

\begin{document}

\title{\name: Coercion-resistant Registration for E-Voting \\ with Verifiability and Usability in \sysname}

\clearpage{}%
\def\epfl{%
  \institution{EPFL}
  \city{Lausanne}
  \country{Switzerland}
}
\def\kcl{%
  \institution{King's College London}
  \city{London}
  \country{United Kingdom}
}
\def\bu{%
  \institution{Boston University}
  \city{Boston}
  \country{USA}
}
\def\harvard{%
  \institution{Harvard University}
  \city{Cambridge}
  \country{USA}
}
\def\armasuisse{%
  \institution{Armasuisse S+T}
  \city{Bern}
  \country{Switzerland}
}

\def\yale{%
  \institution{Yale University}
  \city{New Haven}
  \country{USA}
}

\author{Louis-Henri Merino}
\affiliation{\epfl}

\author{Simone Colombo}
\affiliation{\kcl}

\author{Rene Reyes}
\affiliation{\bu}

\author{Alaleh Azhir}
\affiliation{\harvard}

\author{Shailesh Mishra}
\affiliation{\epfl}

\author{Pasindu Tennage}
\affiliation{\epfl}

\author{Mohammad Amin Raeisi}
\affiliation{\yale}

\author{Haoqian Zhang}
\affiliation{\epfl}

\author{Jeff R. Allen}
\affiliation{\epfl}

\author{Bernhard Tellenbach}
\affiliation{\armasuisse}

\author{Vero Est\-ra\-da-Gali\-ña\-nes}
\affiliation{\epfl}

\author{Bryan Ford}
\affiliation{\epfl}

\author{~}
\affiliation{\country{~}}
\clearpage{}%

\renewcommand{\shortauthors}{Merino et al.}
\renewcommand{\shorttitle}{\name: Coercion-resistant Registration for E-Voting}






\begin{abstract}
Online voting is convenient and flexible,
but amplifies the risks of voter coercion and vote buying.
One promising mitigation strategy
enables voters to give a coercer
fake voting credentials,
which silently cast votes that do not count.
Current systems along these lines
make problematic assumptions
about credential issuance,
however,
such as
strong trust in a registrar and/or in voter-controlled hardware,
or expecting voters to interact with multiple registrars.
\sysname is
the first coercion-resistant voting architecture
that leverages the physical security of in-person registration
to address these credential-issuance challenges,
amortizing the convenience costs of in-person registration
by reusing credentials across successive elections.
\sysname's registration component, \name,
gives voters a kiosk in a privacy booth
with which to print real and fake credentials on paper,
eliminating dependence on trusted hardware in credential issuance.
The voter learns and can verify in the privacy booth which credential is real,
but real and fake credentials thereafter appear indistinguishable to others.
Only voters actually under coercion,
a hopefully-rare case,
need to trust the kiosk.
To achieve verifiability,
each paper credential encodes an interactive zero-knowledge proof,
which is sound in real credentials but unsound in fake credentials.
Voters observe the difference in
the order of printing steps,
but need not understand the technical details.
Experimental results with our prototype suggest that \sysname
is practical and sufficiently scalable for real-world elections.
User-visible latency of credential issuance in \name
is at most 19.7 seconds even on resource-constrained kiosk hardware,
making it suitable for registration at remote locations or on battery power.
A companion usability study indicates that \name's usability
is competitive with other e-voting systems including some lacking coercion resistance,
and formal proofs support \name's combination of
coercion-resistance and verifiability.
 \end{abstract}

\maketitle

\section{Introduction}\label{sec:introduction}

Democracy is the operating system of 
a free self-governing civilization:
it holds governments accountable to their citizens 
and enables correction and renewal through the peaceful transfer of power,
periodically ``rebooting'' the government, so to speak.
Tech-savvy idealists often dream
that large-scale distributed computing systems
could improve or even positively transform democracy,
enabling more convenient large-scale direct 
participation via mechanisms like 
liquid democracy~\cite{
ford2020LiquidPerspective,
blum16liquid}
or online citizens' assemblies~\cite{
Antonin2021CitizensAssembly}.
However, today's computing systems
are not even \emph{safe to use} in basic democratic processes --
such as popular voting --
according to the overwhelming consensus of today's experts~\cite{
committee2018SecuringtheVote,
springall2014EstoniaEVotingAnalysis,appel2020BallotMarkingDevices}.
Further,
large-scale social platforms
not only failed to live up to
their one-time promise to help ``democratize'' the world~\cite{khondker2011ArabSpringMedia},
but have come to appear increasingly
\emph{anti-democratic} in effect if not intent~\cite{epstein2015SearchElectionManipulation,ribeiro2020YoutubeRadicalization,flamino2023TwitterPoliticalPolarization},
feeding an increasingly-prevalent technology backlash~\cite{katsomitros2025TechAntiTrustBreakup}.

One central challenge for national voting systems
is resolving the conflicting goals of
transparency and ballot secrecy~\cite{
Fitzgerald2016SecretBallotAtRisk,
park2021BadToWorse,
benaloh1994ReceiptfreeSecretBallot,
benaloh1987VerifiableSecretBallot,
juels2010CoercionResistantElections}.
Transparency requires convincing voters
that their ballots were cast and counted properly,
while ballot secrecy ensures that voters can express their true preferences,
free from improper influence such as coercion or vote buying.
In the best practice of in-person voting,
voters get ballot secrecy by marking paper ballots alone in a privacy booth
at an official polling site.
Voters obtain ``end-to-end'' transparency
by first depositing their marked ballots in a ballot box,
then relying on election observers to monitor the subsequent handling
and tallying of all the ballots.

While in-person voting is widely 
accepted by experts~\cite{
committee2018SecuringtheVote},
it has posed significant 
inconveniences to voters,
including long wait times~\cite{
famighetti2016ElectionWaitTimes,
wilder2021VoterSuppression},
temporary polling place closures~\cite{
2024ElectionIssues},
and even voter intimidation near
polling locations~\cite{
wilder2021VoterSuppression}.
In-person voting is additionally difficult 
for those traveling and living abroad, 
such as expatriates or deployed military~\cite{
serdult2015SwissOnlineVotingHistory}
and during crises like the recent pandemic~\cite{
weiser2020PandemicVotingDifficulty}.

\emph{Remote} voting---%
at any location of the voter's choosing---%
promises to improve convenience and 
increase voter turnout \cite{
qvortrup2005WhyMailInVoting,
serdult2015SwissOnlineVotingHistory}.
Because the voting location is uncontrolled, however,
remote voting normally compromises ballot secrecy
and hence resistance to coercion or vote-buying~\cite{
wilder2021VoterSuppression,
castro2022GangInfluenceVoting,
Gonzalez2020GuatemalaVoteBuying,
frye2019RussiaVoteBuying,
filipovich2021MexicoVoterCoercion,
robertson2022NorthCarolinaVoterFraud,
2024MoldovanVoteBuying}.
An abusive spouse might insist on 
their partner voting their way under supervision,
a party activist might offer to purchase and 
``help'' fill mail-in ballots, 
as seen in the United States~\cite{
graff2021NorthCarolinaAbsenteeBallots,
2022NCAbsenteeFraud},
or a foreign power might attempt the same
at scale, as seen recently in Moldova~\cite{
2024MoldovanVoteBuying}.
Postal voting also compromises transparency
by subjecting ballots to unpredictable delays
and potential loss~\cite{johnson2021MailBallotsBostonDesertion,
2021BallotMailDesertion,killer2019SwissVotingAnalysis}.

Remote electronic voting or \emph{e-voting},
using a voter's preferred device,
promises further convenience benefits and
avoids postal issues~\cite{ehin2022InternetVotingEstonia,
pammett2013CanadaOnlineVotingPromises,
serdult2015SwissOnlineVotingHistory}.
E-voting can even offer voters
greater transparency via ``end-to-end'' cryptographic proofs,
verifiable on their devices,
that their vote was cast and tallied correctly~\cite{cortier2024BeleniosCaI,benaloh1987VerifiableSecretBallot,adida2008Helios}.
However, such proofs, or \emph{receipts},
can also enable voters to prove
how they voted to a coercer or vote
buyer~\cite{benaloh1994ReceiptfreeSecretBallot,
juels2010CoercionResistantElections,
park2021BadToWorse}.
Blockchain and cryptocurrency-based techniques such as ``dark DAOs'' might even fund
vote-buying anonymously,
at scale and across borders,
with no realistic prospect for accountability or deterrence~\cite{austgen2023DarkDAOPreprint,
Ronne2025ToolboxCoercion}.

One seminal proposal to counter
coercion in e-voting is to provide voters with
both real and
\emph{fake credentials}~\cite{juels2010CoercionResistantElections}.
Only the voter knows which credential is real,
enabling them to
give or sell fake credentials to a coercer or vote buyer
while secretly casting their true vote using the real credential.
A key limitation of this appealing idea, however,
is that it not so much \emph{solves} but rather \emph{shifts}
the most security-critical, delicate, and challenging stage
from voting time to credential-issuance or \emph{registration} time.
Efforts to make fake credentials practical often make strong
and arguably-unrealistic assumptions that all voters have special trusted hardware,
such as expensive smart cards,
at registration~\cite{clarkson2008Civitas,
feier2014CRSmartCards,
estaji2020UsableCRSmartCards,
neumann2013SmartCards2}.
A common implicit assumption is that a coercer cannot
simply confiscate the voter's trusted hardware
and allow its use only under their supervision.
Related approaches either compromise transparency
by assuming a fully-trusted
registrar~\cite{juels2010CoercionResistantElections},
or achieve transparency at the cost of
requiring voters to interact with multiple registration authorities~\cite{clarkson2008Civitas} --
a usability challenge that
has yet to be rigorously studied.

We present \sysname,
the first coercion-resistant online
voting system with end-to-end verifiability
and systematic evidence of potential usability.
This paper focuses primarily on \name,
\sysname's novel registration component.
\name enables voters to obtain verifiable real
and \emph{indistinguishable} fake voting 
credentials on paper.
\name addresses three key challenges:
1) issuing verifiably real voting credentials
without requiring voters to have a trusted device
during registration,
2) issuing fake credentials that are
distinguishable only to the voter,
and
3) materializing these real and fake credentials usably
with only inexpensive paper materials.

\name leverages an in-person process 
to give voters a co\-er\-cion-free environment
in which to create voting credentials.
Unlike in-person voting,
voters may register
at any time convenient to them, and
may use their credentials to cast votes
in multiple successive elections.
To address the first challenge of
issuing verifiable real credentials,
voters interact with a kiosk in a privacy booth.
The kiosk proves to the voter that this credential is in fact real
using an \emph{interactive} zero-knowledge proof (IZKP),
although the voter need not understand the details.
To address the second challenge of creating
fake credentials distinguishable only to the voter,
the voter and the kiosk follow a visibly distinct process
in which the kiosk forges a \emph{false} IZKP
that is subsequently indistinguishable from the real one.
The voter thus knows which credential is real
but is unable to prove that fact to anyone,
and can safely give or sell fake credentials to a coercer.

To address the third challenge of
materializing real and fake credentials,
kiosks print all credentials on paper.
Paper-based credentials are inexpensive,
making it more easy and cost-effective
for most voters to create a few fake credentials
including for reasons other than coercion risks.
Voters can activate credentials
on any device they choose,
including a trusted friend's device
if their own is under a coercer's control.
Printed envelopes supplied to voters in the booth
form a part of each credential,
simplifying the voter's task of choosing a random challenge
for each IZKP,
and serving to conceal sensitive secret keys during credential transport.

While this paper focuses on e-voting,
we view this work as a stepping-stone toward secure 
large-scale \emph{democratic computing systems}
usable by self-governing groups, such as
user communities, organizations, and nations.
Many decentralized projects,
for example,
wish to incorporate democratic 
self-governance into their designs for
decentralized autonomous organizations (DAOs),
but lack the means to ensure that voting
participants are \emph{real people} acting in
their own interests~\cite{Tanusree2024DAOsInPractice,ohlhaver2025IdenaHistoryCrisis}.
This is one of the most important and challenging instances of the well-known Sybil attack problem~\cite{douceur2002TheSybilAttack},
which has motivated considerable systems research in the past~\cite{yu06sybilguard,
yu08sybillimit,yu09dsybil,
nakamoto08bitcoin,
tran2009SybilResilientContentVoting,
gilad2017Algorand},
though none of this prior work has addressed coercion.
Integrating \name's coercion-resistance mechanism
into in-person \emph{pseudonym parties}~\cite{
ford2008OnlineAccountablePseudonyms}
as a proof-of-personhood protocol \cite{BorgeFord2017PoP,ford2020PersonhoodDigitalDemocracy,siddarth20who},
in particular,
could help address this challenge
and enable truly democratic DAOs
and other democratic computing platforms
in the future.
\cref{apx:broader-app} further discusses
the broader potential systems applications of this work.

We implemented a prototype of \name consisting 
of 2,633 lines of Go~\cite{go-lang}.
Our \sysname and \name prototypes focus
on the cryptographic path,
which represents the dominant
computation and latency cost
in e-voting systems.
We evaluated \sysname
against three state-of-the-art e-voting systems:
(1) Civitas, an end-to-end verifiable and
coercion-resistant system based on fake credentials~\cite{clarkson2008Civitas},
(2) the Swiss Post's verifiable but non-coercion-resistant
system~\cite{2021SwissPostProofs}, and
(3) VoteAgain, a coercion-resistant system
based on deniable re-voting~\cite{lueks2020VoteAgain}.
We find that \sysname's end-to-end latency
is comparable to that of Swiss Post and VoteAgain,
and significantly improves upon Civitas.
For \name, we also implement the use of
peripherals to determine \name's
voter-observable latency across several setups:
(1) a Point of Sale Kiosk, %
(2) a Raspberry Pi 4,
(3) a Macbook Pro, and
(4) a Mini PC. %
We find that \name's voter-observable
latency is slowest on the Kiosk
at 19.7 seconds and fastest on the
Macbook Pro at 15.8 seconds,
both suitable inside a booth where
voters typically spend a few minutes.

To refine and validate our design,
we incorporate feedback from
two preliminary user studies
involving 77 participants.
We also summarize key insights from
our main user study involving 150 participants,
which we conducted on this 
design~\cite{evoteconscience}.
We find that
83\% of participants successfully created and
used their real credential to cast a mock vote.
Additionally,
47\% of participants exposed to a malicious kiosk,
with security education,
could identify and report it.%

We summarize important limitations 
of \sysname in \cref{sec:protocol:limitations}.

This paper makes the following primary contributions:
\begin{itemize}[leftmargin=*]
    \item
	\name,
	the first coercion-resistant, user-studied
        voter registration scheme to offer a concrete
        realization of real and fake credentials while
        maintaining verifiability.
    \item
    The first use of paper transcripts of
    interactive zero-knowl\-edge proofs to achieve
    verifiability during registration,
    avoiding reliance on trusted hardware.
    \item
    Security proofs demonstrating that \name satisfies
    verifiability and coercion-resistance.
    \item
    An implementation and evaluation of
    \name against state-of-the-art baselines on
    multiple hardware platforms.
\end{itemize}

\section{Background and motivation}\label{sec:background}

Voting systems represent
one of the most-critical components
in the functioning of a modern democracy.
The high stakes of national elections
inevitably attract all manner of self-serving tactics from those wishing to acquire or maintain power.
These stakes
attract hundreds of millions of dollars in investment
from those wishing to influence national power~\cite{2024MoldovanVoteBuying,
Press2024MuskOneMillionADay},
and even create multi-billion-dollar markets
merely around \emph{predicting} election outcomes~\cite{Poteriaieva2024ElectionPolymarket}.

Central requirements for a
national voting system include
transparency, ballot secrecy, and usability~\cite{
Fitzgerald2016SecretBallotAtRisk,
park2021BadToWorse,
benaloh1994ReceiptfreeSecretBallot,
benaloh1987VerifiableSecretBallot,
juels2010CoercionResistantElections}.
The importance of transparency in elections is evident
from the contrastingly non-transparent \emph{sham elections}
routinely held by dictatorships to support predetermined outcomes~\cite{
wienerbronner2014NorthKoreaElections,zavadskaya2024RussianElection}.
Ballot secrecy was first recognized gradually 
across several countries,
but quickly became standard internationally after Australia's introduction
of the modern secret ballot in 1856~\cite{crook2011ModernSecretBallot}.
Moreover, election systems must be comprehensible and reliably usable 
by all voters -- including those with special needs or disabilities -- 
to ensure broad, equitable participation~\cite{
legislation2002HelpAmericaVote}.
Despite efforts towards digitalizing voting,
in-person voting with paper ballots remains the accepted best practice,
as further discussed in \apxref{apx:baseline}.

\subsection{Remote voting with end-to-end verifiability}

E-voting systems
aim to enhance transparency
by providing
cryptographic end-to-end verifiability of
the voting and tallying process~\cite{
adida2008Helios,kusters2020Ordinos,
2021SwissPostProofs,
cortier2019CastAsIntended}.
In a simplistic but illustrative sketch of such a process,
the voter's personal device encrypts the voter's choices
and posts the encrypted ballot to a 
\emph{public bulletin board} or PBB that anyone can read,
such as a tamper-evident log~\cite{crosby2009TamperEvidentLog}
or blockchain~\cite{nist2018BlockchainOverview}.
Each encrypted ballot is posted along with voter identity information sufficient
to ensure eligibility and prevent multiple votes.
After the voting deadline,
each of several \emph{talliers} reads the ballot ciphertexts from the PBB,
shuffles them to anonymize the ballots 
(scrubbing their linkage to voter identities),
then strips one layer off each ballot's encryption,
and posts the shuffled ballots back onto the PBB along with
a zero-knowledge proof that the shuffle was
performed correctly~\cite{neff2001VerifiableShuffling,
groth2010VerifiableSecretShuffle}.
After these shuffles and decryptions,
anyone may tally the cleartext ballots left on the PBB
and verify the series of proofs that they correspond 
to the originally-cast ballots,
providing \emph{universal verifiability} of the process.
Voters may additionally use their 
private voting materials
to check that their own choices were correctly encrypted and recorded
on the PBB, and included in the tally,
for \emph{individual verifiability}.

While this e-voting sketch
offers individual and universal verifiability,
it unfortunately compromises ballot secrecy
by effectively giving voters a \emph{receipt}~\cite{
benaloh1994ReceiptfreeSecretBallot,
juels2010CoercionResistantElections},
allowing them to prove how (and whether) they voted.
A coercer need merely check 
the PBB to see whether 
the voter cast a ballot, and if so, 
demand the voter's receipt
to confirm compliance.

\subsection{Coercion resistance in e-voting systems}

Most proposals to address coercion 
and vote buying
rely on either re-voting or fake credentials.
Re-voting lets voters override coerced votes
by casting their intended vote later
in secret~\cite{post2010DeniableRevoting,
achenbach2015JCJDeniableRevoting,
lueks2020VoteAgain,azurmendi2020NetVote}.
Estonia is currently the only nation
to deploy an e-voting system with 
coercion-resistance
using the re-voting approach~\cite{ehin2022InternetVotingEstonia},
but this approach comes at the cost of end-to-end 
verifiability~\cite{springall2014EstoniaEVotingAnalysis}.
Moreover, re-voting can be 
circumvented by coercing a last-minute vote,
or by seizing key voting materials --
such as the voter's ID card 
used in Estonia's system~\cite{estonia2024EVotingSteps,estonia2024IDcard} --
until after the election.

The \emph{fake credentials} approach~\cite{juels2010CoercionResistantElections}
instead relies on %
enabling voters to obtain both real and fake credentials.
A voter may use a fake credential under a coercer's supervision,
or give or sell fake credentials to a coercer or vote buyer.
A voter may cast their intended vote in secret at any time 
with their real credential,
avoiding the weakness in re-voting of requiring the real vote
to follow all coerced votes.
A limitation of fake credentials, however,
is that it time-shifts key unsolved problems earlier
from voting to registration time.

Making fake credentials practical requires physically embodying
or \emph{materializing} credentials and their issuance,
in a way that ordinary people can understand
and use~\cite{
kulyk2020HumanFactorsCoercion,
neumann2012CivitasRealWorld,
neumann2013SmartCards2}.
One challenge is that
the voter must know which credential is real,
and must personally be able to \emph{verify} it as real for transparency,
but then must be unable to \emph{prove} this fact to anyone else.
A coercer must also not know how many fake credentials the voter has or can create;
otherwise a coercer can simply demand \emph{all} of the voter's credentials.

Usability-focused approaches to materializing fake credentials
have generally followed Estonia's lead by assuming trusted hardware,
such as smart cards that store both real and fake credentials
under different PINs~\cite{clarkson2008Civitas,neumann2012CivitasRealWorld,
neumann2013SmartCards2,feier2014CRSmartCards,
Krivoruchko2007CRVoterRegistration}.
Even under the as-yet-unproven assumption that
sufficiently-powerful and secure smart cards can be developed,
they will be expensive,
making even one per voter difficult to justify economically --
let alone \emph{several} per voter,
as would be required to achieve real coercion resistance.
If governments issue only one smart card per voter,
then, as with Estonia's e-ID cards,
a coercer can simply ``offer'' to ``keep safe'' (\ie confiscate)
the voter's smart card and allow its use only under supervision,
negating its effective coercion resistance.

Registration-time transparency is another key challenge.
Why can't a compromised election authority
just issue fake %
smart cards, for example,
which silently issue \emph{only} fake credentials,
while the authority retains the secret keys required
to cast ``real'' votes on behalf of all voters?

\subsection{E-voting as a computer systems challenge}

To design e-voting systems
that are usable, practical,
and above all safe to deploy,
we feel it is necessary to approach the problem
holistically as a \emph{computer systems} research challenge.
In that sense,
we must build and maintain a clear picture
of the entire ``end-to-end'' architecture and system design,
considering together 
many important constraints
both technical
(\eg security, privacy, performance, scalability, code correctness)
and non-technical
(\eg usability,
political and legal constraints,
and other human factors).

E-voting needs and critically builds on cryptography,
for example,
but the vast cryptographic literature on e-voting
habitually makes key assumptions
that work well in cryptographic proofs
but are unrealistic in practice --
such as that ordinary voters
can perform complex cryptographic calculations
``in their heads'' without electronic devices
(see \apxref{apx:lifetime}).
Usability and usable security are thus crucial,
but the problem is not just user-interface (UI) design either.
While a typical UI designer
or human-factors researcher might readily
identify many usability issues
and make stylistic and process improvements,
a UI designer
without a broad, end-to-end systems-architecture
and security perspective
would never have made the decision we made in \sysname
not only to avoid depending on,
but even to depend on the \emph{absence of},
personal devices at registration time
(see \apxref{apx:devices}).
These complex interdependencies
make end-to-end systems research on e-voting
a challenging prospect,
but we see no other way forward
towards solving the problem.
\section{\sysname design overview}

This section provides a high-level 
overview of \sysname's architecture 
to provide context needed to understand \name.
We defer a detailed technical description of \name,
the focus and central contribution of this paper,
to \cref{sec:scheme}.

\subsection{High-level election process}

\begin{figure*}[t]
  \centering
  \includegraphics[width=0.7\linewidth]{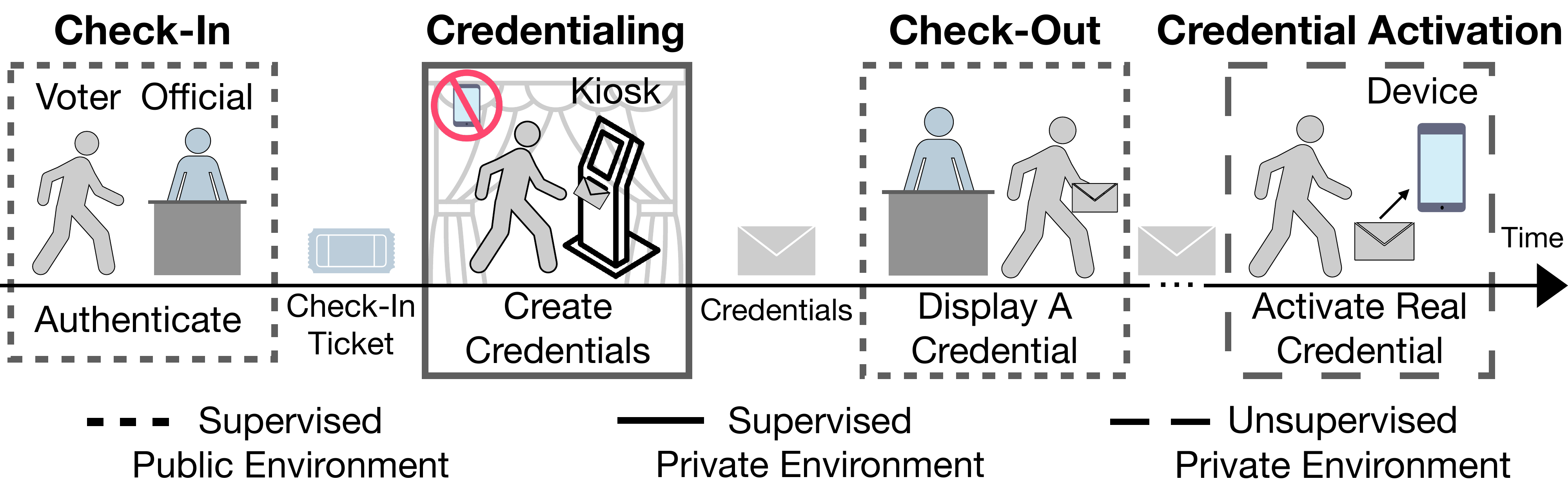}
  \captionsetup{justification=centerfirst}
  \caption{\textbf{Voter-registration workflow in the \name architecture}\\
  In this workflow,
  the voter 
  (1)~authenticates to an official at a check-in desk
  to obtain a check-in ticket, %
  (2)~enters a supervised private environment 
    to create a real and any potential fake credentials,
  (3)~presents one of their credentials to an official
  	at the check-out desk, and
  (4)~activates their real credential on a device they trust.
  }
  \Description{A four-panel diagram shows the TRIP voter workflow from left to right along a timeline. Panel 1, ``Check‑In'' (dashed box = supervised public), depicts a voter authenticating with an official at a desk and receiving a check‑in ticket. Panel 2, ``Credentialing'' (solid box = supervised private), shows the voter alone at a kiosk inside a booth marked with a no‑recording symbol, where paper credentials are created and placed in envelopes. Panel 3, ``Check‑Out'' (dashed box = supervised public), shows the voter presenting one envelope to an official for scanning. Panel 4, ``Credential Activation'' (thick dashed box = unsupervised private), shows the voter later activating the real credential on a personal device. A legend explains that border styles indicate environment types.}
  \label{fig:design:voter-workflow}
\end{figure*}
 
Any voting system generally includes
\emph{setup}, \emph{voting}, and \emph{tallying} phases.
In the setup phase,
the election authority establishes the \emph{roster} or list of eligible voters,
and the set of choices that each voter is entitled to make,
then prepares the system to accept ballots.
In the voting phase,
voters cast ballots via one or more voting \emph{channels}
or ballot-submission methods.
For clarity we will assume here that e-voting via \sysname is the only channel,
but \cref{apx:registration} outlines some considerations applicable
if \sysname were to be integrated with other channels 
such as in-person and/or postal voting.
Once the voting deadline passes,
the system accepts no more ballots and tallying begins,
where all properly-submitted ballots are counted and the results published.
In systems like \sysname supporting fake credentials,
it is crucial that the tallying phase count 
only votes cast using real credentials.

\parhead{Registration.}
Like any coercion-resistant design with fake credentials,
\sysname assumes that voters \emph{register} for e-voting before they can cast ballots.
\sysname more uniquely requires that this registration for e-voting be done in person.
\sysname's \emph{registration for e-voting} may, but need not necessarily,
coincide with \emph{registration to vote},
as further detailed in \cref{apx:registration}.
In US-style elections where voters must register to vote anyway,
voters might in principle register for e-voting via \sysname at the same time.
In Europe-style elections where registration \emph{to vote} is normally automatic,
registration \emph{for e-voting} might be a separate step
required only of those wishing to opt-in to e-voting.
Regardless, registration for e-voting
should normally be required at most once every several elections,
thereby amortizing the costs of in-person registration
across multiple successive uses of the same credentials.

In \sysname's design,
we consider in-person registration %
to be not just a step technically needed to create real 
and fake credentials, but also an educational opportunity
for voters to learn and ask questions about the e-voting system,
and an opportunity for voters to report and discuss
actual attempts at coercion or other voting irregularities
in a protected environment.
We will focus here on the actual process of credential creation, however --
first from the voter's perspective, ignoring technical details,
in the next subsection.

\parhead{Renewal.}
While we expect \name credentials to be reusable
across successive elections,
they will normally expire at some point,
and need to be renewed by again registering in-person.
While credential lifetime is a policy choice outside this paper's scope,
\apxref{apx:lifetime} discusses some credential lifetime considerations,
including the tension between voter convenience
and recovery from ``surprise coercion'' situations.

\subsection{Voter-facing design of e-voting registration}\label{sec:design:voter-facing}

Registering to use \sysname involves the following main steps,
which \cref{fig:design:voter-workflow} illustrates at a high level.

\parhead{Instructional Video.}
In the registrar's office,
voters first watch an instructional video,
covering the credential creation processes and 
the purpose of fake credentials.

\parhead{Check-In.}
A registration official verifies each voter's eligibility,
then gives the voter a \emph{check-in ticket},
which permits the voter to enter a privacy booth and use the kiosk inside.
Depending on applicable anti-recording policy,
voters might be asked to turn off or check in personal devices
before entering the privacy booth,
as further discussed in \apxref{apx:devices}.

\parhead{Privacy Booth.}
\begin{figure}[t]
    \centering
    \begin{subfigure}{.24\linewidth}
        \centering
        \includegraphics[width=\linewidth]{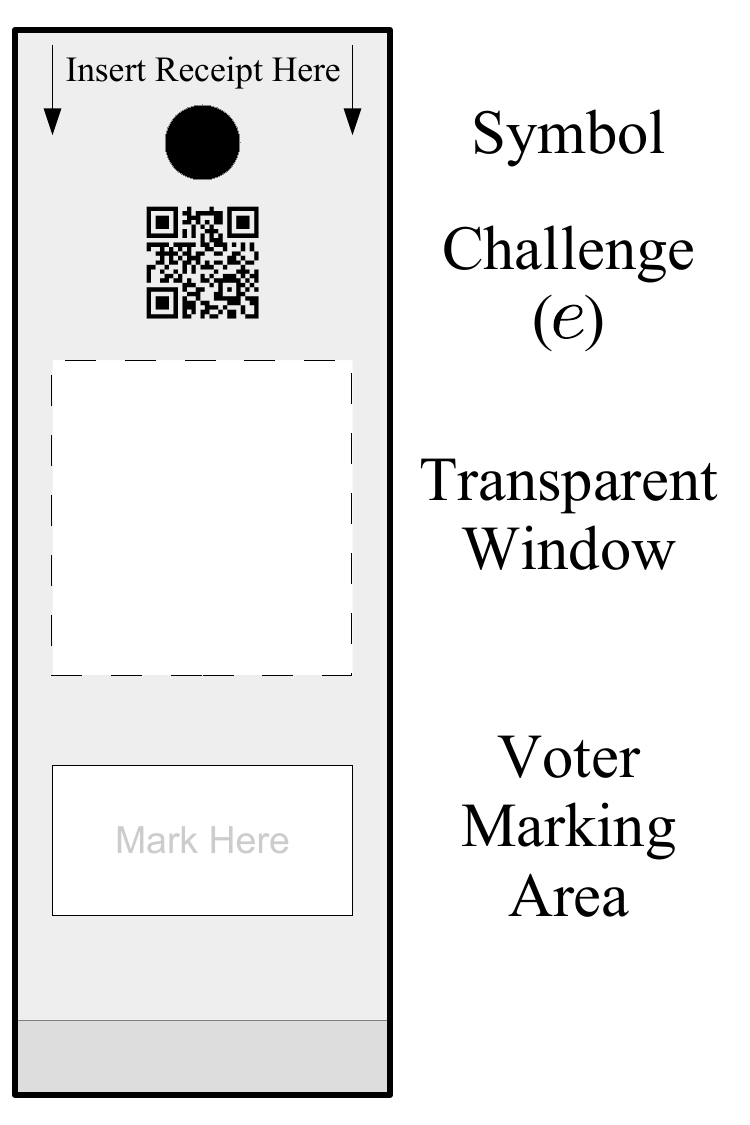}
        \caption{Envelope}
        \label{fig:envelope}
    \end{subfigure}%
    \begin{subfigure}{.24\linewidth}
        \centering
        \includegraphics[width=\linewidth]{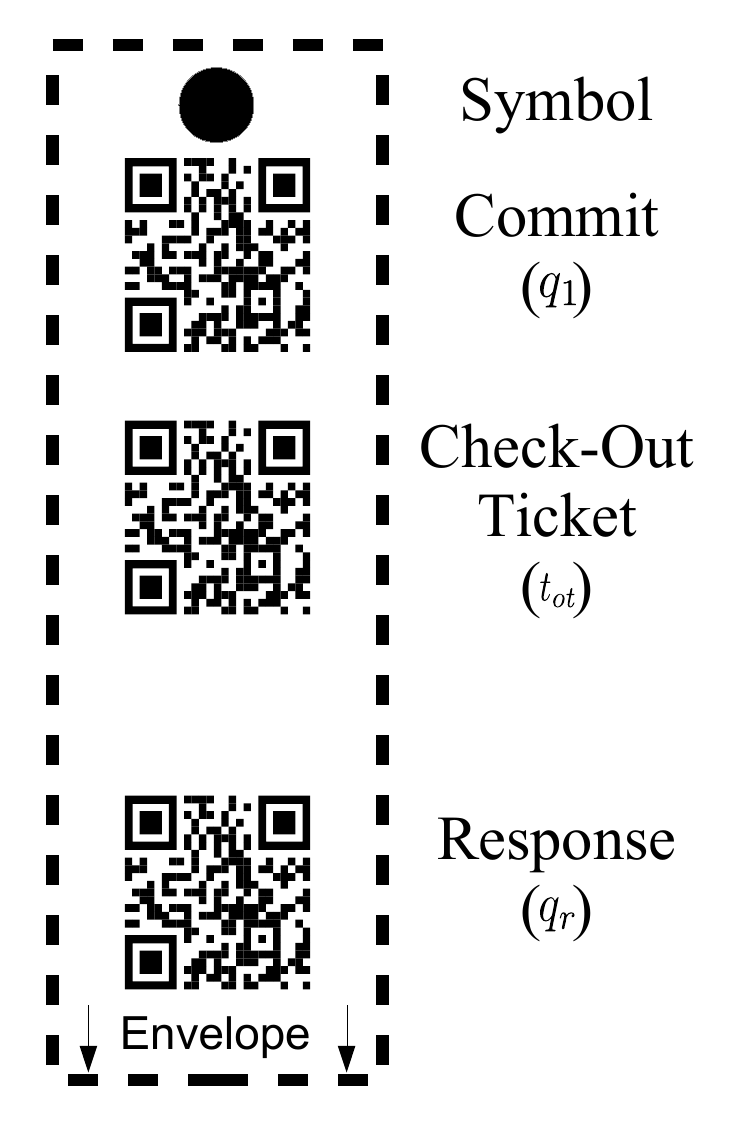}
        \caption{Receipt}
        \label{fig:receipt}
    \end{subfigure}%
    \begin{subfigure}{.24\linewidth}
        \centering
        \includegraphics[width=\linewidth]{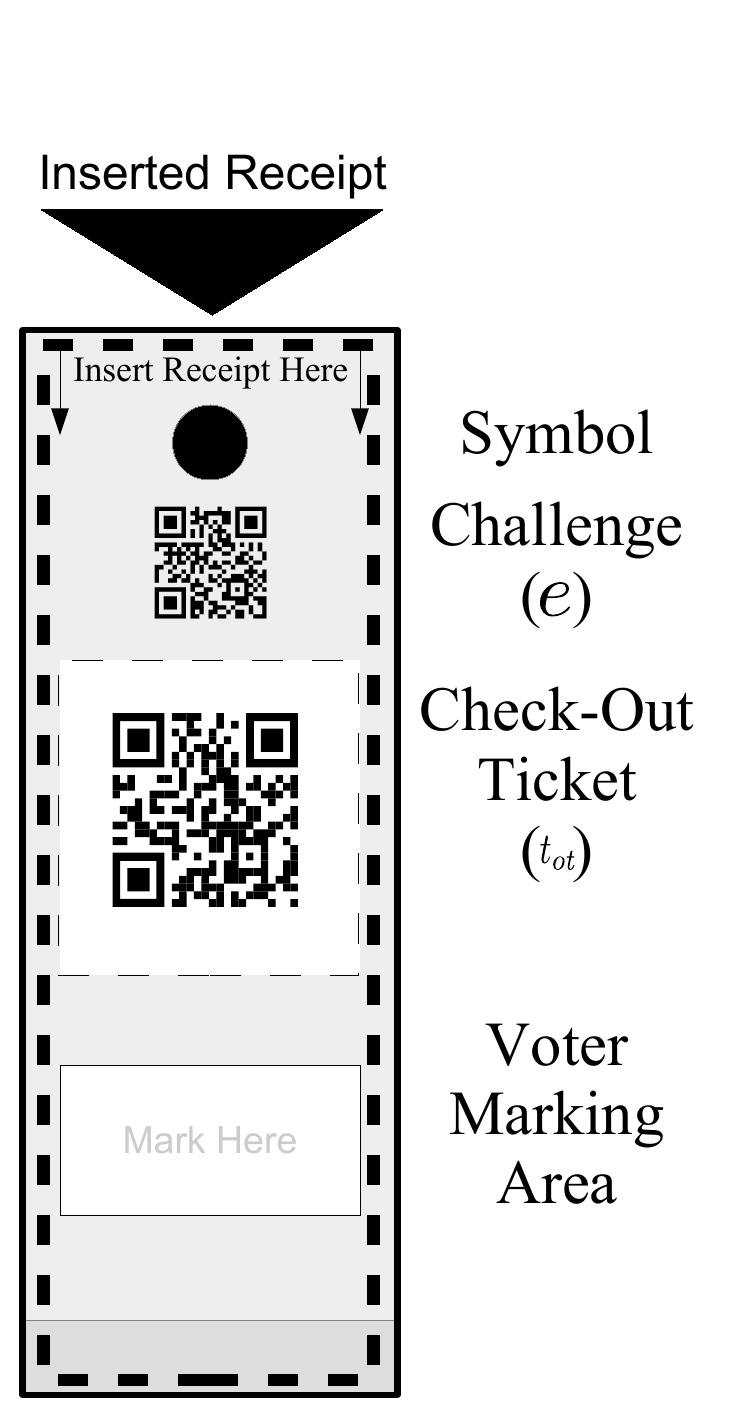}
        \caption{Transport}
        \label{fig:transport}
    \end{subfigure}%
    \begin{subfigure}{.24\linewidth}
        \centering
        \includegraphics[width=\linewidth]{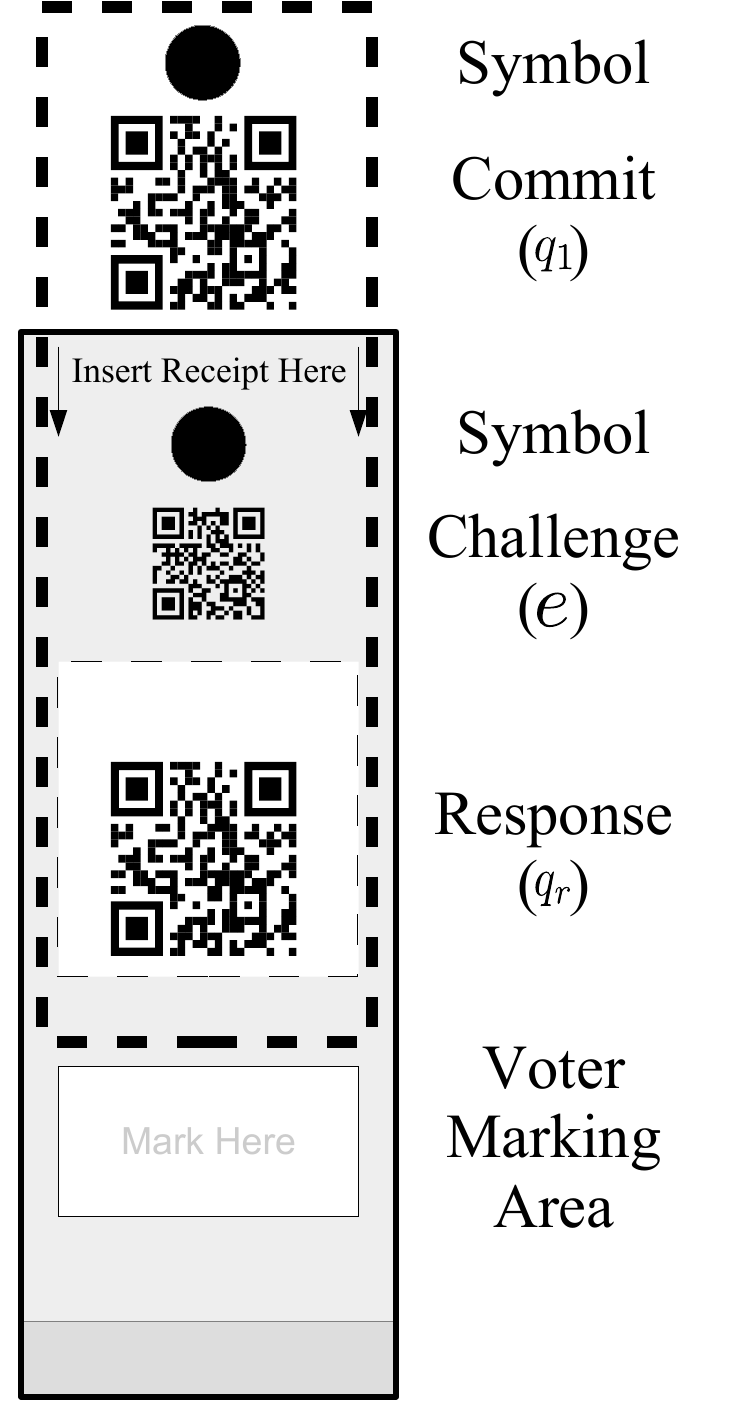}
        \caption{Activate}
        \label{fig:activate}
    \end{subfigure}%
    \vspace{3mm}
    \captionsetup{justification=centerfirst}
    \caption{\textbf{Registration receipt and envelope design} \\
    A paper credential consists of an envelope (a) and printed receipt (b). 
    Voters mark envelopes to discern their 
    real credential from fake ones.
    They transport each credential by placing 
    the receipt in the envelope (c), and activate it on their devices
    by lifting the receipt one-third of its length (d).
    }
    \Description{A four‑part schematic explains the paper credential's components and states. (a) Envelope: a tall envelope labeled ``Insert Receipt Here,'' with a symbol, and a QR code, a transparent window, and a ``Voter Marking Area'' where the voter can write a mark. (b) Receipt: a long strip of paper with three vertically stacked QR codes—top ``Commit (q_c)'' next to the symbol, middle ``Check‑Out Ticket (t_ot),'' and bottom ``Response (q_r).'' (c) Transport state: the receipt is fully inserted into the envelope so only the middle QR code (check‑out ticket) is visible through the window; the commit and response codes are hidden. (d) Activate state: the receipt is lifted about one‑third of its length so the top commit and bottom response QR codes become visible, together with the envelope’s challenge QR; these are scanned to activate the credential.}
    \label{fig:receipt_envelope}
\end{figure}
Inside the booth is
a touchscreen kiosk
that can scan and print machine-readable codes, 
along with a pen and an ample supply of envelopes.
Each envelope has a see-through area and
printed markings as depicted in~\cref{fig:envelope}.

\parhead{Real Credential Creation.}
Voters follow a 4-step process to create their real credential.
The voter first scans the barcode 
on their check-in ticket (Step 1).
The kiosk prints a symbol and a 
QR code on receipt paper (Step 2).
The voter selects an envelope with a 
matching symbol and scans its QR code (Step 3).
The kiosk then prints two additional QR codes (Step 4),
completing the receipt shown in \cref{fig:receipt}.

The voter inserts the receipt
inside the scanned envelope 
for transport (\cref{fig:transport}),
forming their real paper credential.
The voter marks the credential,
for instance by labeling it as `$\mathcal{R}$',
to distinguish it from 
any fake credentials.
If a coercer learns or guesses that the voter follows this practice,
then at the next registration the voter might
mark a fake credential `$\mathcal{R}$'
and mark their real credential `$\mathcal{RR}$'
or in any other memorable way.
Because each voter privately marks both real and fake credentials,
only voters themselves know their own convention.

\parhead{Fake Credential Creation.}
A voter wishing to create a fake 
credential does so in 2 steps:
the voter picks and scans an envelope 
(Step 1, \cref{fig:envelope}),
then the kiosk prints
the entire receipt
(Step 2, \cref{fig:receipt}).
The voter inserts the receipt into the 
newly-chosen envelope 
and marks it
differently from the real one.
Voters can create as many fake credentials as desired.
Officials may inquire if a voter spends excessive time in a booth,
thus imposing an informal, non-deterministic limit.
Once finished,
the voter leaves the booth
and checks out.

\parhead{Check-Out.}
The voter presents any one of their 
credentials to the official, 
who scans the receipt's second QR code
through the envelope's window.
This completes the voter's visit.
To ensure prompt detection and rectification
of any successful impersonation of voters,
the registrar subsequently notifies voters digitally 
and/or by mail of their registration session,
as discussed in \apxref{apx:impersonation}.

\parhead{Activation.}
The voter activates their 
real credential on any device they trust,
whether that is their own or a friend's device.
The voter lifts the receipt one-third out of the envelope
to the \emph{activate} state~(\cref{fig:activate}).
They scan the three visible QR codes:
the receipt's top and bottom QR codes,
and the envelope's QR code.
The voter can now discard this credential,
perhaps shredding it %
if the voter faces a coercion threat.

\parhead{Voting.}
The voter can now use this device to cast their ballot, 
verifying it with existing 
cast-as-intended methods~\cite{benaloh2007BenalohChallenge,
cortier2024BeleniosCaI} to ensure its integrity.
Only the vote cast with the real credential
will count, however.
A voter who loses their device
can re-register for new credentials.
Other options like social key recovery are also possible
as discussed in \apxref{apx:credentials}.

\parhead{Results.}
After each election,
the voter's device downloads, verifies,
and shows the election results to the voter. %

\subsection{Desired System Properties}\label{sec:design:system-properties}

\noindent
We now informally present
\sysname's system
properties.

\noindent
\begin{itemize}[leftmargin=*]
    \item \textbf{Universal Verifiability}
    \cite{juels2010CoercionResistantElections,2021SwissPostProofs}:
    anyone can independently verify that the election
    results accurately reflect the will of the electorate
    as represented by the ballots cast.
    \item \textbf{Individual Verifiability}
    \cite{bernhard2018CHVoteProofs,2021SwissPostProofs}:
    each voter can independently verify that their
    issued real credential casts ballots that are accurately
    recorded and included in the final tally, and that this ballot reflects their intended vote.
    \item \textbf{Coercion Resistance}
    \cite{juels2010CoercionResistantElections}:
        voters can cast their intended vote
        even when faced with coercion because the
        coercer cannot determine whether the voter
        has complied with the coercer's demands, even if the
        voter wishes to comply.
    \item \textbf{Privacy}
    \cite{juels2010CoercionResistantElections}:
        uncompromised voters' ballots stay secret.
    \item \textbf{Usability}~\cite{
    NIST2007UsabilityBenchmarks,
    distler2019EVotingSecurityVisibility,
    kulyk2020HumanFactorsCoercion}:
        voters can understand and use the voting system,
	including for registering and casting ballots,
	and can understand the uses of real and fake credentials.
\end{itemize}

\noindent
\emph{End-to-end verifiability}
requires both individual and universal verifiability,
with voters verifying the integrity of their own real vote,
and the public at large verifying the results.

\section{Registration and voting protocol design}
\label{sec:scheme}
\label{sec:protocol}
This section informally describes the \sysname protocol, with
a particular focus on \name, the system's registration stage.

The core design of \sysname addresses four key challenges:
(1) threat modeling and designing a registration process
in which we trust neither voters \emph{nor} registrars in general;
(2) providing an end-to-end verifiable registration, voting, and tallying process
despite only voters themselves knowing how many credentials they have
or which of them is real;
(3) enabling the registration kiosk to prove interactively to a voter that a credential is real
without the voter being able to prove that fact subsequently to a coercer;
and
(4) materializing this credentialing and interactive proof process in a usable fashion
using only low-cost disposable materials such as paper.
The rest of this section addresses these challenges,
then \cref{sec:protocol:extensions}
summarizes optional extensions to the base design.

\subsection{System model and threat model}
\label{sec:protocol:model}

We summarize \sysname's system and threat models 
here
at a concise intuitive level,
leaving detailed
formal system model and threat models used in our proofs
to \cref{sec:formal-model}.

Consistent with common e-voting system designs,
we assume an \emph{election authority}, or simply \emph{authority},
responsible for setting up and coordinating the election process.
The registration sites or \emph{registrars} at which voters can sign up for e-voting
answer to, or at least coordinate with, the election authority.
The authority provides client-side software that voters need
to activate credentials and cast ballots;
for transparency we assume this software is open source.
Finally, the authority is responsible for providing two logical backend components:
a \emph{ledger} or \emph{public bulletin board}
implementing a tamper-evident log~\cite{
crosby2009TamperEvidentLog},
and a \emph{tally service}.

We assume that these backend services are distributed
across multiple servers,
preferably operated independently,
ensuring that backend services remain secure
provided not all servers are compromised.
Backend trust-splitting is standard in cryptography and e-voting practice
and not novel or unique to \sysname:
\eg the Swiss e-voting system splits backend services
across four ``control components'' maintained by independent teams~\cite{
swisspost2024Architecture}.
For exposition clarity, we leave this trust-splitting implicit for now,
as if the ledger and tally services were trusted third parties.
In contrast with Civitas' expectation
that users interact directly
with multiple registrars~\cite{clarkson2008Civitas},
splitting backend services does not introduce usability issues
because ordinary users need not interact directly with them,
nor even be aware of them.
Users interact with backend services only through the front-end client,
which anyone with sufficient expertise may inspect
but ordinary voters are not required or expected to understand.

One challenge to threat modeling \sysname
is that we wish to trust neither voters \emph{nor} registrars in general,
but if we distrust \emph{all voters} and \emph{all registrars} at once
then we have no foundation on which to build any arguably-secure system.
We address this challenge by specializing the threat model to distinct threat vectors,
which we model as separate adversaries.
To evaluate \sysname's transparency we use an \emph{integrity adversary} $\IA$,
which can cause any or all registrars to misbehave arbitrarily --
attempting to steal voters' real credentials and issue only fake credentials for example --
but which prefers not to be ``caught'' in such misbehavior.
To evaluate \sysname's coercion resistance we use a \emph{coercion adversary} $\CA$,
which can coerce and hence effectively compromise voters,
but which cannot compromise registrars.
The key desirable property this nuanced threat model achieves is that
the majority of voters \emph{not} under coercion need not trust the registrars
in order to verify the end-to-end correctness of the election.
Only the hopefully-few voters \emph{actually under coercion}
must trust the registrar.
This trust appears unavoidable because
in this case we must treat the voter himself as untrustworthy,
by virtue of being coerced or otherwise incentivized to follow the coercer's demands.

Our formal analysis also considers a \emph{privacy adversary}, whose aim 
is to identify and decrypt a voter's real ballot.
Unlike the coercion adversary, however, 
this adversary cannot directly compromise
or exert coercive influence over voters.

\subsection{High-level protocol flow summary}

\begin{figure}[t]
  \centering
  \includegraphics[width=\linewidth]{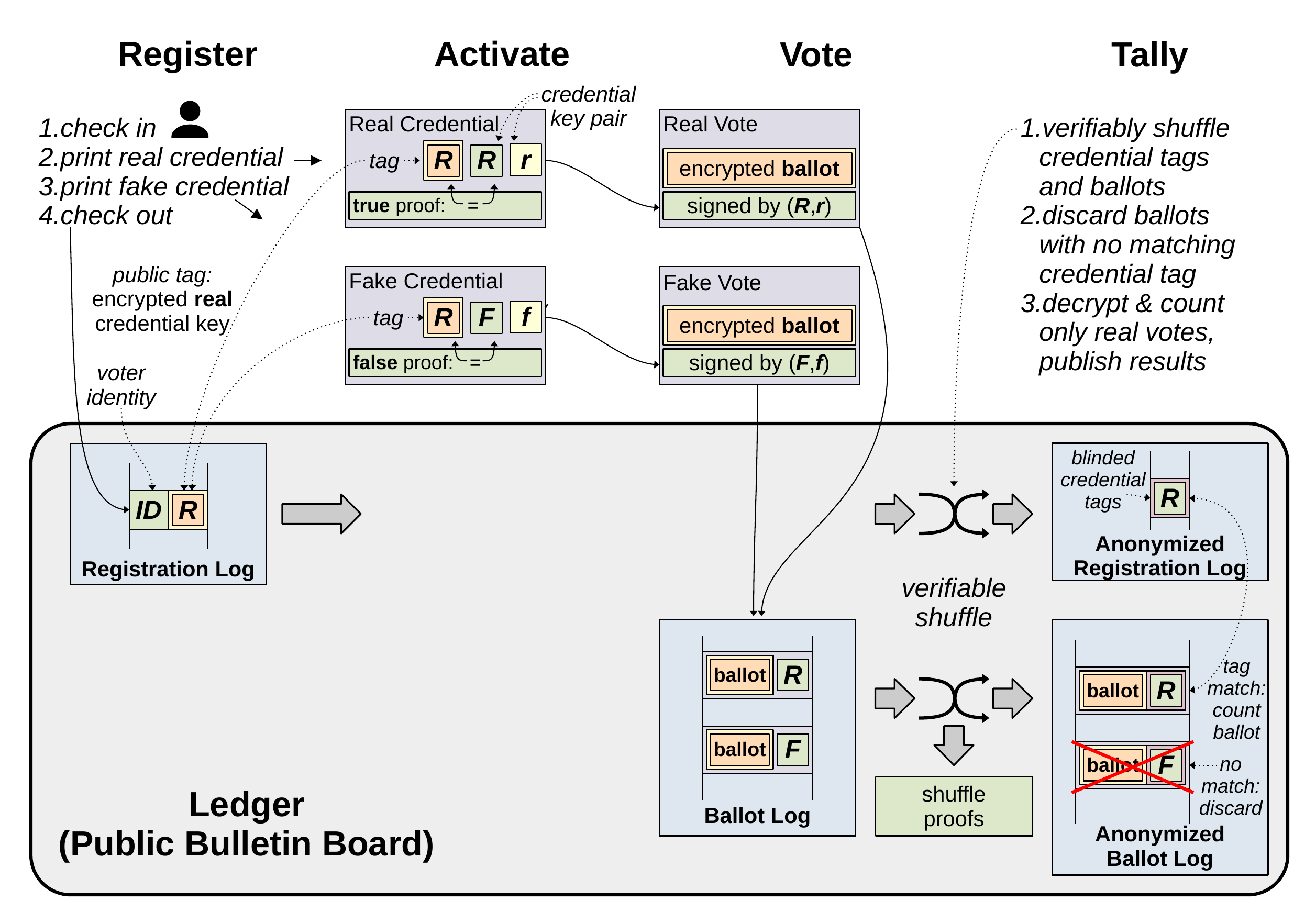}
  \caption{\sysname protocol operation summary}
  \Description{A pipeline diagram summarizes the Votegral protocol across four stages---Register, Activate, Vote, and Tally---flowing left to right above a large rounded rectangle labeled ``Ledger (Public Bulletin Board).'' Register: a numbered list shows check‑in, print real credential, print fake credential, and check‑out. An arrow drops a record into the Registration Log on the ledger that contains the voter identity and a public tag (an encryption of the real credential’s public key). Activate: two small panels depict a Real Credential and a Fake Credential. Both display the same public tag, but the real one carries key pair (R, r) with a ``true proof,'' while the fake shows key pair (F, f) with a ``false proof.'' Vote: two panels show an ``encrypted ballot'' signed by the corresponding key pair—(R, r) for the real vote and (F, f) for the fake vote—feeding into a Ballot Log that lists ballots next to credential keys R and F. Tally: the ledger performs verifiable shuffles, producing an Anonymized Registration Log of blinded credential tags and an Anonymized Ballot Log. Tags and ballots are matched: the ballot whose credential key matches the blinded tag R is counted; the unmatched fake ballot (F) is discarded. A side list summarizes the tally steps: verifiably shuffle tags and ballots, discard non‑matching ballots, decrypt and count only real votes, and publish results.}
  \label{fig:protocol}
\end{figure}

\Cref{fig:protocol} presents an illustration
of \sysname's overall protocol operation,
omitting various details for clarity
and focusing on a single voter, Alice,
who creates one real and one fake credential.
Alice first registers using the voter-facing process above in \cref{sec:design:voter-facing},
then at home activates her real and fake credentials on her voting device(s).
At the next election,
Alice uses each credential (on the same or different devices)
to cast real and fake ballots.
After the election,
the tally service anonymizes all ballots,
discards those cast via fake credentials,
and counts the remaining ballots,
posting the results.

Upon Alice's registration,
her registrar posts a record to the \emph{registration log} on the ledger,
including metadata uniquely identifying Alice as an eligible voter.
This identity metadata might include Alice's name
in districts where voting rosters are considered public,
or might just be a pseudonymous index into the authority's eligible-voter database
if the database is not public.
Each time Alice registers or renews,
the new record supersedes and invalides all prior registration records
for the same voter identity.
Thus, there is always at most one active registration record per voter,
and for transparency
anyone can at least count the total number of active records
and check it against relevant census data,
regardless of how much metadata about each voter may be public.

Each credential in \name
has a unique public/private key pair $(K,k)$.
In this example,
Alice's real credential has key pair $(R,r)$,
and her fake credential has a different key pair $(F,f)$.
The main use of each credential's key pair is to authenticate --
essentially as if by signing --
encrypted ballots submitted using that credential.
Each credential also has a \emph{public credential tag},
which unlike the per-credential key pairs,
is identical across all credentials created in the same registration session.
In all of Alice's new credentials, this public tag is an encryption of
the public key associated with Alice's \emph{real} credential.
This tag goes into Alice's registration record on check-out.
It does not matter which credential Alice shows on check-out
because all have this same %
tag.

After the election deadline,
\sysname's tally service uses verifiable shuffles~\cite{
groth2010VerifiableSecretShuffle}
in a mix cascade~\cite{dingledine2003Mixcascade}
to anonymize the set of public credential tags
associated with currently-active records in the registration log.
The tally service similarly shuffles
the set of all encrypted ballots cast using any credential.
In this shuffling process,
the tally service cryptographically blinds
the public key that each credential was submitted with,
and in parallel,
decrypts while identically blinding all the public credential tags
derived from the registration log.
Because the public tag on all of Alice's credentials
was an encryption of the public key of Alice's \emph{real} credential,
this means that the blinded credential key associated with Alice's real vote
now matches the decrypted and blinded public credential tag
derived from the registration log.
The tally service verifiably counts Alice's real vote
as a result of this tag match.
The credential public keys associated with fake votes, however,
do not match any blinded public credential tag after the shuffle,
and as a result are discarded.%

The design sketched so far mostly achieves universal verifiability.
Anyone may inspect the eligible voter roster on the registration log,
and the encrypted ballots on the ballot log;
anyone can check the shuffle proofs to verify that
both the roster and ballots were shuffled
via a correct but secret permutation;
and, anyone can verify that only ballots authorized
by appropriate registration log entries are counted,
at most one ballot per voter.
We have not yet achieved individual verifiability, however:
how does Alice know or verify which of her credentials is real,
or that any of them are real?

\subsection{Interactively proving real credentials real}
\label{sec:protocol:interactive}

To enable Alice to verify individually
that the ballot she cast with her real credential will be counted,
she needs to ``know'' that the public credential tag
printed on all her credentials and recorded in the registration log
is a correct encryption of her real credential's public key.
This is ultimately the critical single bit of information Alice needs
to complete the end-to-end transparency chain
and verify that her real vote will count.
For coercion resistance, however,
Alice must be unable to prove
that critical information bit to anyone else.

As a straw-man,
the kiosk might give Alice an ordinary non-interactive zero-knowledge proof,
of the kind commonly used in digital signatures,
that her credentials' public tag is an encryption of her real credential.
Alice could then be certain that her real credential is real --
but then she might also take that proof (receipt) 
with her upon leaving the booth and hand it to a 
coercer or vote buyer, who would be equally convinced.
Such a non-interactive proof would therefore support
transparency but undermine coercion resistance.

A key observation is that
we can use \emph{interactive} zero-knowledge proofs or IZKPs
to convince Alice that her real credential is real
without making that information transferable.
\name uses IZKPs
taking the common \emph{$\Sigma$-protocol} form,
which consist of three steps:
\emph{commit}, \emph{challenge}, and \emph{response}.
For such an IZKP to be sound and effectively prove anything,
the protocol must be executed in precisely this order:
\emph{first} the prover chooses a cryptographic commitment;
\emph{then} the verifier chooses a challenge value
previously-unknown to the prover;
\emph{finally} the prover computes the sole valid response
corresponding to the combination of commit and challenge.
If the $\Sigma$-protocol is executed in the wrong order --
in particular, if the prover knows or can guess the verifier's challenge
\emph{before} choosing the commit --
then the IZKP is unsound and the prover can trivially ``prove'' anything.

Leveraging this observation,
\name includes in all credentials, real and fake,
a \emph{transcript} of the three phases of an IZKP,
which \emph{purports} to prove that the public credential tag
contains an encryption of the credential's public key --
\ie that the credential is real.
Anyone can verify the structural validity of these IZKP proof transcripts,
and voters' devices do so automatically when activating credentials.
Such a transcript alone omits one crucial bit of information, however:
was the commit, or the challenge, chosen first?
\name achieves coercion-resistant verifiability
by revealing this one crucial bit of information to the voter
only interactively,
via the sequence of steps that the voter takes in the privacy booth
to print real and fake credentials.
Once the voter observes this distinction in printing steps,
the credentials themselves embody only IZKP transcripts,
which lack the crucial bit of ``voter's-eyes-only'' information,
and are useless to prove to anyone else which credential is real.
Real credentials contain sound IZKPs,
fake credentials contain unsound ones,
and the two are cryptographically indistinguishable once printed.

Prior work has used sound and unsound IZKPs
in similar fashion
for coercion-resistant in-person voting~\cite{
moran2006ReceiptFreeVerifiableEverlastingPrivacy,
neff2006MarkPledge,
adida2009Markpledge2,
joaquim2012MarkPledge3}.
To our knowledge, however,
\name is the first design to use IZKPs in \emph{registration} for e-voting
to achieve verifiability and 
coercion resistance using fake credentials.

\subsection{Physically materializing usable credentials}

Although IZKPs in principle
resolve the conflict between individual verifiability and coercion resistance,
for usability we cannot expect ordinary voters
to understand e-voting systems or cryptography,
or any of the technical details underlying credentials,
or precisely \emph{why} technically Alice should be convinced
that her real credential is real.
In particular,
we need a physically \emph{materialized} credentialing process
that ordinary voters can understand and follow given only minimal training,
and ideally that is cheap and efficient enough to deploy at scale.
These are the considerations that motivate \name's specific design
using paper credentials.

One particular issue is that to ensure that the kiosk
\emph{must} honestly produce sound IZKPs for real credentials,
the voter -- taking the verifier's role in the IZKP --
must choose and give the kiosk a cryptographic challenge,
only \emph{after} the kiosk has printed the IZKP commit.
Choosing and entering a high-entropy random challenge manually
would be tedious and burdensome,
and ordinary users are bad at 
choosing random numbers~\cite{nickerson2002HumanRandomness}.
These considerations motivate \name's choice to use envelopes,
each pre-printed with a unique random QR code,
enabling voters to pick a challenge
by selecting any envelope from a supply provided in the booth.
A compromised registrar could duplicate envelopes
to make challenges more predictable,
but an activation-time check for duplicate challenges
detailed in \cref{apx:security:envelope_analysis}
ensures high-probability detection
if many envelopes are duplicated.

Above and beyond the instructional video,
we would like the credentialing process itself
to help train voters to learn and expect the correct process,
especially for printing real credentials.
To ensure in particular that voters
do not hastily choose or present an envelope too early,
before the kiosk has printed the IZKP commit,
an honest kiosk first chooses one of a few symbols at random
and prints it just above the commit QR code.
The voter must then choose and scan any envelope with the matching symbol.
When interacting with an honest kiosk, therefore,
the voter \emph{must} wait to see the symbol (and hence commit) printed
before picking an envelope;
otherwise the honest kiosk gently rejects a voter's choice of an envelope
with the wrong symbol.
If a voter accustomed to this process later encounters a compromised kiosk
that asks them to present an envelope first
while supposedly printing a ``real'' credential,
we hope that the voter's normal-case training
will make such an irregularity more noticeable.

To create fake credentials, in contrast,
the kiosk asks the voter to choose and present an envelope first.
The kiosk thereby obtains the challenge before choosing its commit,
thus enabling the kiosk to fake a ``proof''
of the credential's ``realness.''
Again, we hope that even voters completely unaware of
the reasons for this curious process
will nevertheless be able to follow it and,
at least with moderate probability,
notice and report if a compromised kiosk ever attempts
to use the fake-credential process to print a ``real'' credential.

\name's envelope and receipt design
secures transport and activation.
Fully inserting the receipt into the envelope
places the credential in the \emph{transport} state
(\cref{fig:transport}).
In this state,
the QR code necessary for check out
appears through the envelope's window,
but the envelope's opaque lower portion hides
the area of the receipt containing the credential's secret key.
Only after the voter transports the credential
to whatever device the voter trusts to activate it on,
the voter pulls the receipt out of the envelope just enough
to reveal the two other QR codes needed for activation,
including the credential's secret key,
as shown in \cref{fig:activate}.

\subsection{Subtleties, design extensions, and limitations}

The above summary omitted
two subtleties of \name's design.

\parhead{Impersonation defenses.}
To ensure prompt detection and remediation of
any successful registration-time impersonation of a voter --
whether by a look-alike or by a corrupt registration official --
\name notifies voters of all registration events.
\apxref{apx:impersonation} discusses in more detail
these threats, defenses,
and their implications for coercion resistance.

\parhead{Credential signing.}
All \name credentials include a signature of the kiosk that produced them,
not shown in \cref{fig:protocol}.
This signature ensures that credentials are authorized
and traceable to a particular registrar and check-in event,
and also counters subtle attacks against related prior approaches~\cite{
smith2005CRLinear}
in which a coercer identifies a voter's real credential
by forging a mathematically-related
fake credential~\cite{clarkson2008Civitas,spycher2012JCJLinearTime,araujo2010CRScheme}.

\label{sec:protocol:extensions}

The design of \sysname supports a few optional extensions,
which might be included or excluded for policy reasons.

\parhead{Voting History.}
Voters can save and view their
voting history on their devices to
further enhance cast-as-intended individual verifiability,
as discussed in \cref{apx:voting-history}.
Coercion resistance remains intact
because casting votes with a fake credential
effectively fabricates a fake voting history.

\parhead{Reducing Credential Exposure.}
Although the protocol above ensures that
theft of and voting with a credential's secret is eventually detectable,
an extension in \cref{apx:credential-exposure}
reduces the ``window of vulnerability''
by ensuring that credential theft is more promptly detectable
\emph{by activation time}.

\parhead{Resisting Extreme Coercion.}
A few voters might face extreme coercion,
such as by being searched immediately after registration.
An extension in \cref{apx:extreme-coercion}
allows voters to delegate their vote within the privacy booth,
\eg to a political party,
leaving the booth
holding only fake credentials.

\label{sec:protocol:limitations}

Like any system, \sysname has important limitations.
\sysname assumes the public roster of eligible voters is correct,
and cannot address the suppression or fakery
of voters,
as discussed in \cref{apx:registration}.
Systematically addressing other threat vectors,
such as voter impersonation (\apxref{apx:impersonation}),
side channels (\apxref{apx:side-channels}),
and rare ``Ramanujan voters''
able to compute cryptographic functions in their heads
(\apxref{apx:devices:voter-limits}),
are also beyond this paper's scope.
\section{Informal security analysis}\label{sec:security}
A voting system is secure
if it resists
attacks
that manipulate
the election outcome,
despite all participants being
untrustworthy to some extent.
This section summarizes how
\sysname achieves
end-to-end verifiability
and coercion resistance,
countering
two of the three
adversaries we model.

We do not cover the
privacy adversary here,
as this paper focuses primarily on
verifiability and coercion resistance.
In brief, \sysname ensures privacy
because decryption of a voter's real ballot 
would require compromising all election authority members,
which is not possible per our threat model.
We present a privacy proof sketch in~\cref{apx:security:privacy}.

\subsection{Individual and universal verifiability}
\label{sec:security:verifiability}

The integrity adversary's goal
is to manipulate the election outcome
\emph{without detection}.
By making each step of the election
process verifiable,
ensuring end-to-end verifiability,
\sysname prevents this adversary from
achieving its goal.

\parhead{Individual Verifiability.}
This notion of verifiability ensures that 
voters can verify that their submitted real ballot 
reflects their intended vote and 
is included in the final tally.
Since this paper focuses on the
registration process (\name),
our security analysis centers on
achieving the latter:
ensuring that the cast ballot
counts.\footnote{
    Numerous prior works~\cite{benaloh2007BenalohChallenge,
    cortier2024BeleniosCaI}
    have explored ways to help voters ensure
    that the ballot they cast contains their
    intended vote.
    \sysname's voting history extension
    (\cref{sec:protocol:extensions})
    offers one such cast-as-intended
    verification method.
}
To achieve this in a system with real
and fake credentials,
the voter must be convinced that
they have obtained their real credential.
The integrity adversary can attempt
to steal the voter's real credential
by either impersonating the voter during
registration or by issuing the voter a
fake credential and claiming it as their
real one.

To counter impersonation,
\sysname publishes
the real credential's public component $c_\pc$
alongside the voter's identity
on the ledger at check-out.
The voter's device then alerts the voter
to any registration event.
If the voter did not initiate this registration,
the attack is detected, and
the voter can re-register to
replace $c_\pc$ and invalidate
the false registration.

To prevent an adversary from claiming a
fake credential as real,
voters are educated through an instructional
video on the four-step process for creating
a real credential and how it differs from
a fake one.
In~\cref{sec:eval:usability},
we present voter performance in this
process from our user study.
Consequently, the
integrity adversary's only
chance of success is to guess the
envelope---the ZKP challenge---that the voter
selects.
While the adversary can
control the number of envelopes in the booth,
and create duplicate envelopes
(envelopes with identical challenges),
they cannot influence the voter's actions,
such as envelope selection
or the number of envelopes consumed
(each credential
consumes one envelope).
This makes the adversary's
success probability minimal,
and it becomes negligible over
repeated attacks against
many voters,
as are usually necessary to
influence an election.

\parhead{Definition (Individual verifiability — informal).}
The integrity adversary interacts with an honest voter's \textit{voter-supporting device} (VSD)
throughout the registration and voting workflow, 
controlling all registrar components.
We fix a target voter $V^\star_{\mathrm{id}}$, 
an \emph{intent} (the public credential and desired vote),
and a conflicting \emph{goal}. 
$\IA$ wins if (1) the VSD's activation‑time checks and cast‑as‑recorded
comparison accept, and (2) the ledger records the adversary's 
goal for $V^\star_{\mathrm{id}}$ (with goal $\neq$ intent). 

\parhead{Theorem (Individual verifiability, sketch).}
Let $n_E$ be the number of envelopes in the booth, 
$n_c$ the voter's chosen number of credentials, 
$D^{c}$ the probability distribution of voters
choosing $n_c$, 
and $k$ the number of envelopes the adversary duplicates with the same challenge. 
Then
the success probability of any 
PPT integrity adversary $\IA$
is at most
\begin{equation*}
    \max_{1\le k\le n_E} \mathbb{E}_{n_c \sim D^{c}} 
    \left[\frac{k}{n_E}\cdot
        \frac{\binom{n_E-k}{n_c-1}}
        {\binom{n_E-1}{n_c-1}}\right]+
        \negl
\end{equation*}

\emph{Proof sketch.}
Tampering \emph{after} correct registration is detected: 
the VSD compares the posted ballot
to the one it formed 
and signed; ElGamal decryption is unique (binding), 
so any plaintext change necessarily affects the ciphertext.
Tampering \emph{at} registration reduces to either 
(a) forging the $\Sigma$‑protocol (negligible under DLP), or 
(b) guessing the voter's chosen challenge in advance. 
The best strategy is to ``stuff'' $k$ envelopes with the 
same challenge $e^\star$ and hope the voter uses one 
such envelope for the real credential (probability $k/n_E$), 
while picking the remaining $n_c-1$ envelopes for fake 
credentials from the honest pool without picking another $e^\star$.
Averaging over $n_c\sim D^{c}$ gives the bound. 
See~\cref{property:iv} for the formal game (Game $\IndVer$),
theorem and proofs.
Across $N$ independent target voters, 
the success probability becomes $p_{\max}^N$ 
(strong iterative IV; \cref{property:iiv}). 

\parhead{Universal Verifiability.}
This enables
anyone to verify the outcome
by ensuring the integrity of the tallying process.
Since this paper focuses on \name, the
registration component of \sysname, 
we omit this proof.
\sysname's voting and tally processes follow the JCJ 
lineage with publicly verifiable shuffles and 
threshold decryption~\cite{juels2010CoercionResistantElections}, 
while using publicly verifiable 
deterministic tags limited to registrar-issued credentials
for linear‑time filtering~\cite{
koenig2011PreventingFloodingVotes}.

\subsection{Coercion resistance}

The coercion adversary aims to influence
election outcomes by pressuring voters
either to cast a specific vote,
or not to vote.
Coercion resistance ensures
this adversary cannot determine
whether a voter complied,
thereby thwarting their objective.
This section summarizes how \sysname achieves
coercion resistance, with formal proofs in~\cref{apx:proofs:coercion-resistance}.

\sysname
achieves coercion resistance by
enabling voters to create fake
credentials,
allowing them to appear compliant
while concealing their real
credential.
These fake credentials
are cryptographically and
visibly indistinguishable from real ones, as discussed
in~\cref{sec:protocol:interactive} on credential issuance.
This indistinguishability also holds when a voter
uses a fake credential
to cast fake votes.
Thus, the coercer cannot
determine voter's compliance
based on credential appearance or
behavior.

Since the coercer cannot
observe the voter creating their
credential---%
the only way to distinguish these
credentials---%
the coercer may attempt, before
registration, to demand
the voter to create a specific
number of fake credentials and
present them along with one
additional credential---%
their real one.
However,
voters can always
generate one more
fake credential, maintaining
the secrecy of their real credential.

Unable to discern compliance through
voter actions, the coercer might turn to
the ledger
to determine whether the voter
complied with their demands.
The ledger reveals
the voter's associated
public component
$c_\pc$, the total number of envelopes used
in the system and the final tally.
Regarding $c_\pc$,
the coercer might attempt to encrypt
the credentials' $c_\pk$ given by
the voter with the
election authority's public key.
Encryptions are cryptographically randomized,
however,
so even if the
adversary possesses the real credential,
the resulting $c_\pc$ will differ from the
one on the ledger.
Similarly, decrypting
$c_\pc$ is infeasible
as the adversary cannot compromise all the tally servers,
as per the threat model,
and thus cannot reconstruct the tally service's private key.
As for the ledger's
disclosure, in aggregate,
of the number of envelopes used
and the tally,
coercion remains ineffective
due to statistical uncertainty
stemming from the actions
of voters the adversary
does not control.
For example, if the
tally shows 5 votes each for candidates A and B,
a coercer demanding a vote for
A cannot determine whether
the coerced voter contributed to A's votes or voted
for B alongside others.

\parhead{Definition (Coercion resistance — informal).}
Compare a real game, 
where coercer $\CA$ interacts with \name 
(choosing a target voter, dictating actions, obtaining controlled voters' credentials, and observing the ledgers), 
with an ideal game
where the attacker sees only
the statistical uncertainty from honest voters' behavior
(distributions $D^{c}$ over number of fake credentials 
and $D^{v}$ over vote choices). The advantage is
\begin{equation*}
    \mathrm{Adv}^{\mathrm{cr}}_{\CA}=\Bigl|\Pr[\text{Real game = 1}]-\Pr[\text{Ideal game = 1}]\Bigr|
\end{equation*}
\noindent
\Cref{apx:proofs:coercion-resistance} defines
games $\cResist$ and $\cResistIdeal$.

\parhead{Theorem (Coercion resistance, sketch).}
Under DDH in $G$,
the $\Sigma$‑protocol's soundness, 
and EUF‑CMA signatures (with NIZK simulations in the random‑oracle model), $\mathrm{Adv}^{\mathrm{cr}}_{\CA}\le \negl$; only uncertainty induced by $D^{c}$ and $D^{v}$ remains.

\emph{Proof sketch.}
Hybrid 1 (Eliminate Voting Ledger View): 
Replace honest ballots by a DDH‑based simulation and program proofs in the ROM; $\CA$'s view becomes independent of honest votes.
Hybrid 2 (Number of Fake Credentials): 
Real and fake credential transcripts are indistinguishable, and
giving $\CA$ the real credential adds no tallying power, 
so only the distribution $D^{c}$ over 
honest voters' fake credentials influences 
$\CA$'s uncertainty. 
Hybrid 3 (Eliminate Registration Ledger View): 
Replace the TRIP roster with a JCJ roster via ElGamal semantic security.
Each hybrid hop is indistinguishable, so the total
distinguishing advantage is negligible.
(See~\cref{apx:proofs:coercion-resistance}.)

\section{Implementation}\label{sec:implementation}
Our full \sysname prototype consists of 9,182 lines of code
as counted by CLOC~\cite{cloc},
broken down further in \apxref{apx:code-size}.
The code is available at \\
\href{github.com/dedis/votegral}{https://github.com/dedis/votegral}.

\name is 2,633 lines of Go~\cite{go-lang}, and
uses \href{https://github.com/dedis/kyber}{dedis/kyber}~\cite{DEDISkyber} 
for cryptographic operations,
\href{https://github.com/makiuchi-d/gozxing}{gozxing}~\cite{gozxing} for QR code processing,
\href{https://github.com/jung-kurt/gofpdf}{gofpdf}~\cite{softwareGitHubgofpdf} and
\href{https://github.com/pdfcpu/pdfcpu}{pdfcpu}~\cite{softwareGitHubpdfcpu} for printing and reading QR codes.
We use Schnorr signatures with SHA‑256 on the edwards25519 curve and 
ElGamal on the same group.

The rest of \sysname comprises 1,816 lines of Go,
utilizing
\href{https://github.com/dedis/kyber}{dedis/kyber}~\cite{DEDISkyber} for
cryptographic operations,
Bayer Groth~\cite{bayergroth2012EfficientShuffle}
for shuffling ElGamal ciphertexts,
and a distributed deterministic tagging
protocol~\cite{weber2007CRLinearDeterministicFingerprint}.
\sysname uses a C implementation
of Bayer Groth for shuffle proofs~\cite{anderspkd2021grothShuffleCLibrary}.
Unlike a complete system,
\sysname simulates each phase of an e-voting system,
focusing on the cryptographic operations as these incur
the highest computational costs.

\section{Experimental evaluation}\label{sec:evaluation}
We evaluate \name's practicality
via these key questions:
\begin{itemize}[noitemsep,topsep=0pt,parsep=0pt,partopsep=0pt,leftmargin=*]
    \item Q1: Is \name fast enough in practice to accommodate voters
        who are willing to spend only a few minutes in a booth?
    \item Q2: How does \name's registration-phase performance compare
        to that of state-of-the-art e-voting systems?
    \item Q3: How does an online voting system using \name perform, and scale with voter population,
	measuring the full ``end-to-end'' (E2E) voting pipeline
	from setup though tallying?
    \item Q4: Can voters effectively register using \name,
        and protect verifiability by identifying a malfunctioning kiosk?
\end{itemize}

\noindent

To evaluate \sysname's computational cost and
answer Q2 and Q3,
we compare \name against three e-voting protocols:
(1) Swiss Post~\cite{2021SwissPostProofs},
a verifiable secret ballot system used in Switzerland;
(2) VoteAgain~\cite{lueks2020VoteAgain},
a coercion-resistant voting system based on deniable re-voting;
and (3) Civitas~\cite{clarkson2008Civitas},
a verifiable and coercion-resistant voting system.
Swiss Post's voting system represents the
state-of-the-art in verifiable secret-ballot systems,
despite lacking coercion resistance.
We chose VoteAgain for its efficient 
tallying process,
despite stronger trust assumptions to
achieve end-to-end verifiability.
We compare against Civitas because it is a well-known
coercion-resistant, end-to-end verifiable system based on
JCJ~\cite{neumann2013SmartCards2,neumann2012CivitasRealWorld,
neto2018CredentialsDistributionUsability,estaji2020UsableCRSmartCards,
koenig2011PreventingFloodingVotes,krips2019CRSummary}.
We omit coercion-resistant systems
that rely on cryptographic primitives
still deemed impractical~\cite{ronne2020CRLinearFHE},
or that we were unable to run despite
our efforts, including Estonia's system based on
deniable re-voting~\cite{EstoniaEVoting}.

\subsection{Experimental setup and benchmarks}\label{subsec:setup}
\begin{figure*}[t]
    \centering
    \begin{subfigure}{0.48\textwidth}
        \centering
        \includegraphics[width=\linewidth]{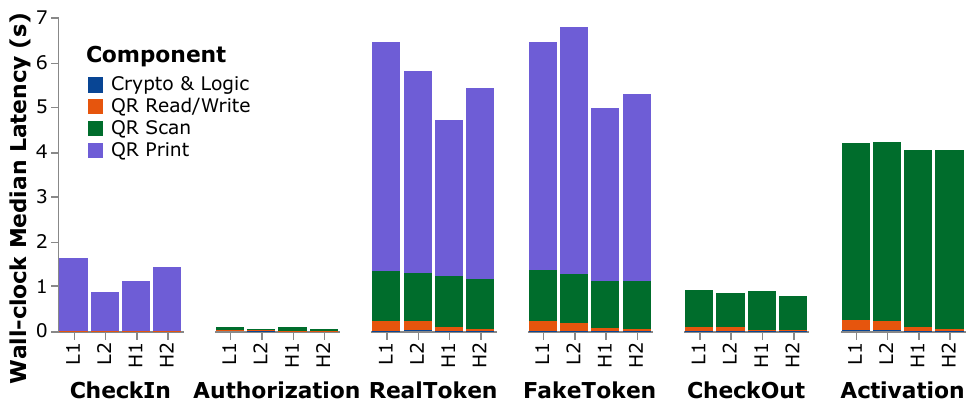}
        \caption{\textbf{Wall-clock time incurred by each sub task in \name}}
        \label{fig:reg_wallclock_time}
    \end{subfigure}
    \begin{subfigure}{0.48\textwidth}
        \includegraphics[width=\linewidth]{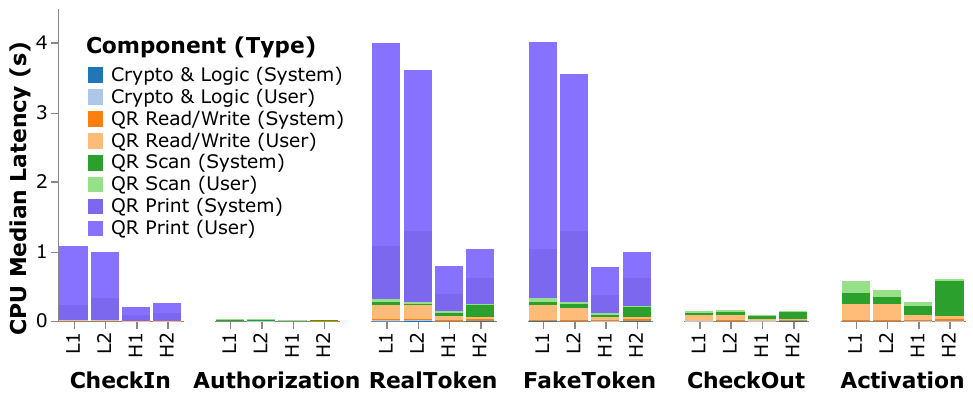}
        \caption{\textbf{CPU time incurred by each sub task of \name}}
        \label{fig:reg_cpu_time}
    \end{subfigure}
    \caption{\textbf{Execution latency}: (L) represents resource-constrained devices while (H) represents resource-abundant devices.}
    \Description{Execution latency of TRIP: (L) represents resource-constrained devices while (H) represents resource-abundant devices.}
    \label{fig:latency-vreg}
\end{figure*}
For~\cref{subsec:eval-latency},
we run \name across four distinct hardware setups:
(L1) a Point-of-Sale Kiosk (Quad-core Cortex-A17, 2GB RAM, Linaro),
(L2) a Raspberry Pi 4 (Quad core Cortex-A72, 4GB RAM, Raspberry Pi),
(H1) a Parallels VM on Macbook Pro (M1 Max 4 cores, 8GB RAM, Ubuntu 22.04.2), and
(H2) a Beelink GTR7 (AMD Ryzen 7840HS, 32GB RAM, Ubuntu 22.04.4).
(L) devices represent resource-constrained systems.

The Macbook Pro serves as our baseline,
while the Raspberry Pi and Beelink are
compact computational units
suitable for integration into devices
with QR printing and scanning capabilities.
The Point-of-Sale Kiosk,
used in our user study~\cite{evoteconscience},
includes both computational elements and
QR peripherals (receipt printer and
barcode/QR scanner).
Due to issues with the kiosk's
built-in receipt printer,
particularly with tearing receipts,
we replaced it with
a dedicated
EPSON TM-T20III printer
with an automatic cutter for easier
receipt handling.
To ensure consistency across all devices,
we equipped each configuration with this printer and a Bluetooth-connected
barcode/QR scanner,
as wired connections were not compatible across all four configurations.

Our end-to-end evaluation
in~\cref{sec:eval:compare-end-to-end}
simulates the main phases of an
election---Registration, Voting and Tally---while focusing
on computational costs.
This focus is because cryptographic primitives usually
incur the most expensive computational costs,
making these costs a standard metric for comparing different
e-voting systems.
For these experiments,
we use Deterlab~\cite{benzel2011Deterlab}
featuring a bare-metal node with
two AMD EPYC 7702 processors for
a total of 128 CPU cores (256 threads) and
256GB RAM.
The latency of the Swiss Post system represents
a realistic benchmark,
as it is a government-approved
e-voting system deployed in Switzerland.

\subsection{Voter-observable registration latencies}
\label{subsec:eval-latency}
\begin{figure*}[t]
    \centering
    \hfill
    \begin{subfigure}[t]{0.45\textwidth}
        \centering
        \includegraphics[width=\linewidth]{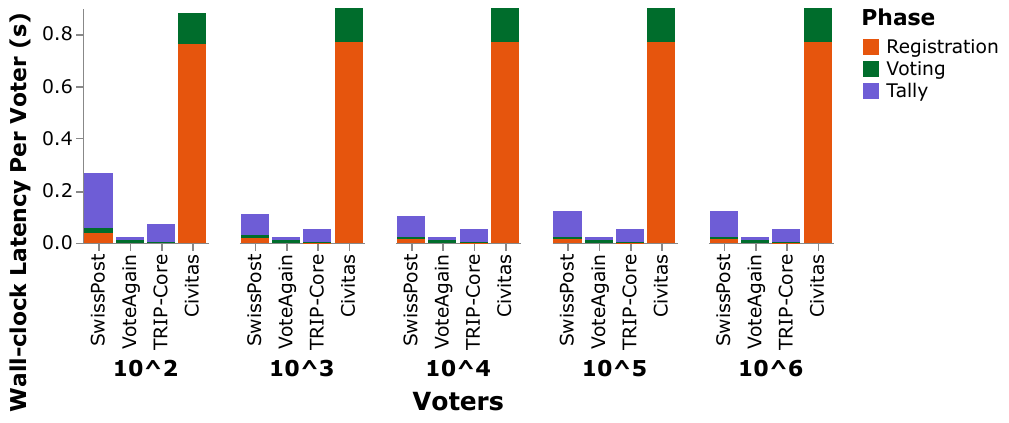}
        \caption{\textbf{Latency per voter for registration, voting, and tallying.} We exclude Civitas tallying to avoid skewing the graph's scale.}
        \label{fig:reg:wallclock:pervoter}
    \end{subfigure}
    \hfill
    \begin{subfigure}[t]{0.48\textwidth}
        \centering
        \includegraphics[width=\linewidth]{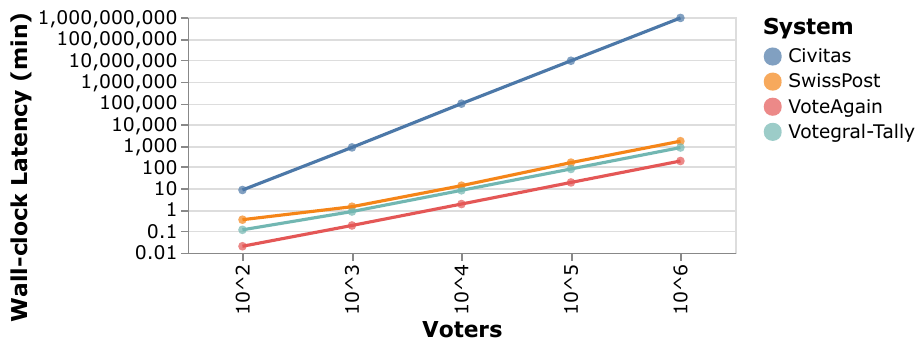}
        \caption{\textbf{Tally phase latency}}
        \label{fig:tally:wallclock}
    \end{subfigure}
    \caption{\textbf{Comparing Phase Execution across Voting Systems.} 
    Swiss Post is end-to-end verifiable but not coercion resistant. 
    VoteAgain is coercion resistant via deniable re-voting. 
    Civitas implements the JCJ system with coercion resistance via fake credentials. 
    Due to its quadratic time complexity, 
    we extrapolate Civitas' latency beyond $10^4$ voters.
    Each system uses four shufflers. 
    As this parameter was not configurable in VoteAgain, we ran its mixing and shuffling
    processes four times.}
    \hfill
\end{figure*} %
This experiment addresses Q1:
whether \name's performance
is adequate for realistic use in environments
where voters typically spend a few minutes each in the privacy booth.
This experiment also explores whether \name
can operate on low-cost, resource-constrained,
or battery-powered devices.

We scripted \name to issue
one real and one fake credential
without human involvement,
measuring all
user-observable wall-clock delays
across all registration phases, including
cryptographic operations,
QR code scanning (``QR Scan''),
encoding/decoding (``QR Read/Write''),
and receipt printing (``QR Print'').
Since QR printing and scanning are mechanical components,
they introduce unique challenges in latency measurement.
For QR printing,
we adapted CUPS~\cite{softwareCUPS}
to capture
CPU and wall-clock latency from job
initiation to completion.
For QR scanning,
we measured the time when \name starts to receive data
to when it captures the terminating character.
We ran \name for 10 consecutive registrations
on each hardware, recording
wall-clock and CPU latencies
for each registration component
and phase (\cref{fig:reg_wallclock_time,fig:reg_cpu_time}).

First, the maximum wall-clock latency perceived by voters
during registration is under 19.7 seconds on all deployments,
with the longest wall-clock latency for any specific registration
process being under 6.5 seconds.
In in-person voting, by comparison,
the duration that a voter stays inside a booth varies greatly,
depending on the complexity of the ballot.
A voting time estimator tool~\cite{VotingTimeEstimator2}
states that marking a single-race ballot---%
where a voter chooses one candidate among several---%
takes between 19 and 44 seconds (90\% confidence interval).
Assuming voters instantly execute the mechanical tasks,
QR printing and scanning, \name meets the lower end of
this time range, even on resource-constrained devices.

Second,
QR-related tasks
significantly contribute to the total wall-clock latency:
QR Printing and scanning account for at least 69.5\% of the time,
with a median overhead of 97.5\%.
In particular,
it takes about 948 milliseconds on average to scan each QR code
across devices.
This duration is primarily due to the time required to transfer
QR data (13-356 bytes)
from the scanner to the computational devices.
We expect that a real deployment
would use a more suitable barcode/QR scanner,
such as the kiosk's embedded scanner, which scans even 300-byte
QR codes almost instantly. %
Using a more efficient scanner could thus
reduce the overall registration latency by about a third, on average,
as each registration run currently takes 7 seconds, on average,
to scan QR codes.

Third,
we find that the resource-constrained devices (L1 and L2)
perform up to 19.8\% slower than higher-end devices (H1 and H2).
Specifically, L1 is the slowest with a wall-clock latency of
19.7 seconds, while H1 is the fastest at 15.8 seconds.
We attribute this difference to the CPU time distribution
shown in~\cref{fig:reg_cpu_time}.
On average, the CPU latency for resource-constrained devices is
260\% higher than that for higher-end devices.
In particular, QR printing on these devices takes 380\% longer
than that on higher-end devices.
Despite these significant increases in CPU latency,
the overall wall-clock latency rises only by an average of 16.5\%.

In answer to Q1, we find that \name appears suitable for environments
on time scales comparable to those applicable to in-person voting,
even on resource-constrained devices.

\subsection{Registration performance across systems}
\label{sec:eval:compare-reg}
We now examine Q2:
How does registration performance in \name
compare to the registration-time costs of other e-voting systems?
This experiment focuses on computational costs,
excluding the I/O-related latencies such as QR code printing and scanning
considered earlier for Q1.
We focus on computation cost for two reasons:
\name's I/O-bound phases lack direct analogs in existing systems,
making ``apples-to-apples'' comparison impractical;
and second, computation cost
is a key metric in e-voting research, as
cryptographic operations,
such as shuffling, decryption, and zero-knowledge proof generation and
verification,
typically dominate and can be prohibitive
for deployment.
We compare the cryptographic performance
between \name, Swiss Post,
VoteAgain, and Civitas, while varying the number of voters.

For \name,
which includes voter-interaction features like QR codes,
we use a configuration termed ``\name-Core'' that omits
all QR-related tasks to isolate the cryptographic operations.
Swiss Post provided us with
their end-to-end election simulator,
enabling us to run their cryptographic functions.
VoteAgain and Civitas did not require any changes
to the code provided by their respective papers.
\Cref{fig:reg:wallclock:pervoter} 
presents our per-voter results,
which also include data for 
the voting and tallying phases,
which is later discussed in~\cref{sec:eval:compare-end-to-end}.

In the 1‑million‑voter configuration, 
the per‑voter registration latency is 
1.2 ms for \name, 13 ms for Swiss Post, 0.1 
ms for VoteAgain, and 771 ms for Civitas (\cref{fig:reg:wallclock:pervoter}). 
Accordingly, 
\name is about two orders of magnitude 
faster than Civitas, about one order 
of magnitude faster than Swiss Post, 
and about one order of magnitude slower 
than VoteAgain. 
Part of the gap comes from group choice: 
Civitas uses large‑modulus primitives, 
whereas \name, 
Swiss Post, and VoteAgain use elliptic curve
cryptography, which is generally faster 
at the same security level.
These results suggest that \name
lies in the same performance range as
modern e-voting systems.

\subsection{End-to-end performance across systems}
\label{sec:eval:compare-end-to-end}

To address Q3,
we broaden our perspective
to the full ``end-to-end'' e-voting pipeline,
comparing
\sysname against Swiss Post, VoteAgain and Civitas.
We used \name-Core as the registration component,
and the voting and tallying schemes detailed
in \apxref{apx:tallying-scheme}.
We measured the execution latency of each phase 
for configurations ranging from 100 to 1 million voters
while maintaining four talliers.
Because tallying in Civitas has quadratic complexity,
we extrapolated its results after 1,000 voters.
\Cref{fig:reg:wallclock:pervoter}
depicts the total execution latency
per voter across each phase
(excluding the tally phase in Civitas),
and \Cref{fig:tally:wallclock}
compares the tally latencies 
of the four systems.

During the voting phase alone,
the per-voter latency for \name, 
Swiss Post, VoteAgain, and Civitas
is 1 ms, 10 ms, 10 ms, and 128 ms, respectively.
Voting latency is independent of
voting population in all systems,
unsurprisingly, since this phase
is ``embarrassingly parallel.''
Voting in \sysname performs an order
of magnitude faster than Swiss Post 
and VoteAgain, and two orders of magnitude
faster than Civitas.

In the tally phase (\cref{fig:tally:wallclock}),
\sysname requires approximately 
14 hours for 1 million ballots,
compared to 3 hours for VoteAgain,
27 hours for Swiss Post and an 
impractical 1,768 \emph{years}
(estimated) for Civitas.
VoteAgain significantly outperforms, 
but it does so under stronger trust assumptions:
it assumes a registration authority that will 
not impersonate voters (\eg not cast votes on
their behalf), and a centralized 
service necessary to preserve 
coercion resistance.
It also inherits the standard revoting
limitation: if a coercer controls the 
voter until the polls close, 
the voter cannot recover.
Civitas' significant tally cost 
stems from the JCJ filtering step~\cite{
juels2010CoercionResistantElections}: 
pairwise plaintext‑equivalence tests (PETs~\cite{jakobsson2000PET}) 
to remove duplicate ballots and 
to test membership against real
credential tags.
While Civitas can improve tally performance 
by partitioning voters into groups and 
tallying each group separately, 
this approach reduces the per‑group 
anonymity set---%
and thus the coercer's
statistical uncertainty---%
which weakens coercion resistance.
\sysname achieves a significant improvement 
over Civitas by leveraging \name's natural 
constraint on issued credentials and 
by restricting valid ballots to 
those cast with registrar‑issued 
credentials (real or fake),
as explained in~\apxref{apx:tallying-scheme}.

In summary, \sysname outperforms Civitas
in ``end-to-end'' performance 
across the pipeline,
and is competitive
with other recent systems such
as Swiss Post and VoteAgain.

\subsection{Usability studies of \name}\label{sec:eval:usability}
To address Q4, whether voters
can use \name and detect a compromised kiosk,
we conducted a usability study with 150 participants in
Boston, Massachusetts.
Our findings, 
detailed in our companion paper~\cite{evoteconscience},
are summarized below.
Our studies were approved
by our institutional review board.

We initially conducted
two preliminary user studies involving 77 participants
to gain insights on our design.\footnote{
The first preliminary study involved 41 PhD students;
the second involved 36 individuals
from varying locations across Boston, Massachusetts.
}
This
led to two enhancements in \name.
First, we replaced a QR code on the check-in ticket with a barcode,
as participants often mistook it for the kiosk's first printed QR code 
used to create real credentials.
Second, to remind participants
to pick and scan an envelope after the first QR,
we added a matching symbol on the receipt and envelopes, 
prompting the voter to select an envelope bearing the same symbol.

In our main study with 150 participants~\cite{evoteconscience},
\name achieved an 83\% success rate
and a System Usability Scale score of 70.4 --
slightly above the industry average of 68.
Participants found \name just as usable
as a simplified one involving only real credentials.
Furthermore, 47\% of participants 
who received security education
could detect and report a misbehaving kiosk,
while 10\% could do so without security education.
Assuming each voter has a 10\% chance
of detecting and reporting a malicious kiosk,
the probability that such a kiosk
could trick 50 voters without detection
is under 1\%.
For 1000 voters, that drops to a
cryptographically negligible $1/2^{152}$.

\section{Related work}\label{sec:related}

\todo{Integrate
However, JCJ and subsequent works~\cite{
juels2010CoercionResistantElections,
spycher2012JCJLinearTime,
araujo2010PracticalSecureCoercionResistant,
benaloh2009CRSingleTV,
weber2007CRLinearDeterministicFingerprint}
primarily focus on the tallying phase,
aiming to achieve the same level of
verifiability and computational complexity
as non-coercion-resistant systems.
In \name,
kiosks are responsible for issuing real credentials
directly to voters in their physical presence.
While this approach might seem counterintuitive---%
since it involves the voter's real credential, and one
might expect the voter to generate it themselves---%
it is essential in preventing coercion.
Allowing voters to generate their own voting credentials
could expose the real credential to the voter's coercer,
by, for example, demanding voters to use the coercer's
device to generate the credential.
One approach taken by other works~\cite{clarkson2008Civitas,
feier2014CRSmartCards,
neumann2013SmartCards2,
estaji2020UsableCRSmartCards}
is for a government agency to provide
each voter with a trusted device (\eg smart card) to
generate the real voting credential.
However, relying on trusted hardware presents numerous challenges,
from security concerns to logistical issues~\todo{cite}.
Therefore,
by having kiosks issue real credentials
on behalf of voters inside the booth,
\name ensures this issuance is completed without
the coercer's access.
}

In JCJ~\cite{juels2010CoercionResistantElections},
the registrar issues a real credential
via an abstract untappable channel.
The registrar and voter are then assumed to protect
the secrecy of this real credential.
\name implements this
untappable channel using a paper-based workflow
utilizing interactive zero-knowledge proofs.

Civitas~\cite{clarkson2008Civitas}
proposes that voters interact with multiple
registration tellers to reduce trust in any one teller.
Asking real-world voters to interact with 
several tellers, however,
incurs significant complexity and convenience costs,
whose practicality has never been tested with 
a usability study.
The election authority also faces the prospect
of explaining this convenience cost to voters
with a justification that risks sounding like:
``You must interact with several registration tellers
because you can't fully trust any of them,
although we hired and trained them and they answer to us.''
Election authorities usually want to and are mandated to promote trust in elections
and electoral processes, not to undermine that very trust!
In \name, coerced voters avoid such a predicament:
they know they must either trust the kiosk (and realistically the
registrar) not to collude with their coercer, or else
comply with the coercer's demands---a simpler binary choice.

In Krivoruchko's work on registration~\cite{
Krivoruchko2007CRVoterRegistration,
krivoruchko2007CRVoterRegistrationMaster},
voters must generate their real credential, encrypt it,
and then provide the encrypted version to the registrar.
This process ensures that the registrar never obtains
the real credential, eliminating the need for the registrar to
prove the credential's integrity.
Nevertheless, this approach has two shortcomings
with respect to coercion resistance:
First, voters must possess a device
prior to registration,
which the coercer could have compromised or confiscated beforehand.
Second,
the scheme lacks a mechanism that proves to the registrar
that the voter's device knows the real credential from
the encrypted version it submits to the registrar.
This is essential to prevent voters from giving the registrar
an encrypted version of a real credential that was
generated by a coercer,
thus rendering the credential inaccessible to the voter.

Prior work used interactive 
zero-knowledge proofs (IZKP)
for receipt-free in-person voting~\cite{
moran2006ReceiptFreeVerifiableEverlastingPrivacy,
chaum2004VoterVerifiableReceipt,
neff2006MarkPledge,
adida2009Markpledge2,
joaquim2012MarkPledge3}.
Moran and Noar's approach~\cite{
moran2006ReceiptFreeVerifiableEverlastingPrivacy}
relied on DRE machines to
place an opaque shield over part of the
receipt and asked voters to enter random words
as a challenge.
In contrast, \name uses IZKPs for registration
rather than voting,
and simplifies the process with a design
that involves selecting an envelope and
scanning its QR code.
TRIP also eliminates the need for a \emph{trusted} party to generate ZKP challenges
for usability~\cite{joaquim2012MarkPledge3},
especially when an adversary targets many voters.
As discussed in \apxref{apx:devices:voter-limits},
we do not deem it necessary to shield
the printed commitment from the voter:
it is hard for ordinary users to interpret QR codes or compute
cryptographic functions without electronic devices,
a much easier unauthorized use of which
is simple recording (\apxref{apx:devices:recording}).

\section{Conclusion}
\revision{
\sysname is the first
coercion-resistant, verifiable, user-studied
online voting system based on fake credentials,
requiring no trusted device in credential
issuance.
We find that \sysname is competitive in performance and efficiency
with today's state-of-the-art e-voting systems.
Formal security proofs
confirm that \name provides individual 
verifiability and coercion resistance.
Finally, usability studies offer evidence
that \sysname is understandable and usable by ordinary voters.
}

\section{Acknowledgments}

The authors would like to thank 
Eric Dubuis, Rolf Haenni, Ari Juels,
Ron Rivest, Peter Ryan,
and the anonymous reviewers
for their valuable feedback
on earlier drafts.
The authors are also grateful to 
the Swiss Post team for their help with running
and evaluating the Swiss Post e-Voting System.

This project was supported in part by 
the Fulbright U.S. Student Program, 
the Swiss Government Excellence 
Scholarships for Foreign Scholars,
armasuisse Science and Technology,
the AXA Research Fund, and 
ONR grant N000141912361.
Simone Colombo's work was partly 
supported by UKRI grant EP/X017524/1, 
and not by armasuisse or ONR grants. 
\bibliographystyle{ACM-Reference-Format}

\bibliography{references}

\clearpage	%
\appendix

\setcounter{tocdepth}{1}

\twocolumn[\section*{\huge Appendices}

The following appendices contain material supplemental to the main paper above.
The authors feel that this material may be useful in understanding
certain subtleties of \sysname and its relation to prior work in more detail.
Readers are advised, however,
that the material in these appendices has not received
the same level of peer review as the main paper.
\\
~]
\addcontentsline{toc}{section}{Appendices}

\tableofcontents

\section{Broader computing systems applications}
\label{apx:broader-app}

Ever since early visionaries such as J.C.R. Licklider~\cite{
licklider68computer},
generations of technology researchers and engineers
have predicted the positive transformative effects
that networked, distributed or decentralized computer systems
would have on the ways people interact socially,
from small groups to entire nations.
The USENET was the first large-scale decentralized platform
that its users called, and widely believed,
would be strongly "democratizing" --
at least by virtue of given everyone with Internet access
an "equal" and "uncensorable" platform for speech, communication,
and coordination with others for myriad purposes~\cite{hauben97netizensFixed}.
Decades later,
after USENET had effectively succumbed to
a heat death from uncontrolled spam
and been replaced by commercialized platforms,
excitement in social media applications in particular
sparked another global wave of optimism
about large-scale computing technologies being "democratizing",
culminating in the short-lived Arab Spring movement~\cite{khondker2011ArabSpringMedia}.

Not long after this movement's collapse, however,
it became increasingly clear that
large-scale computing platforms were just as readily usable
as tools of surveillance and control by anti-democratic actors,
as for support or promotion of democracy or related ideals.
Increasing suspicion has fallen on large-scale platforms
for their potential to interfere with democracy
in nontransparent and unaccountable ways,
whether intentionally --
such as by potentially affecting election outcomes
through search-engine rankings~\cite{epstein2015SearchElectionManipulation} -- 
or promoting sensationalistic and radicalizing content
in the blind algorithmic pursuit of human attention
and associated advertising revenue~\cite{ribeiro2020YoutubeRadicalization} --
or generally increasing the political polarization of a nation by presenting users with information diets consisting mostly of opinions of their friends and others they already 
agree with~\cite{flamino2023TwitterPoliticalPolarization}.
A large segment of the public even in advanced democracies
has increasingly been drawn into a "technology backlash"
against large-scale computing platforms
and the companies and personalities associated with them~\cite{
katsomitros2025TechAntiTrustBreakup}.
In summary,
while the early proponents of large-scale distributed computing technologies tended to see only positive effects on society,
hindsight now paints a much more pessimistic picture.

Even focusing more narrowly on technologies specifically intended to support and improve democracy, 
such as e-voting, boundless optimism has generally turned toward dark pessimism.
Technology-accepting politicians and the public at large regularly express interest in 
and even demand the convenience of voting on their favorite mobile devices, 
while experts in voting practices and security consistently and 
nearly-unanimously agree that using e-voting is extremely risky and 
not generally recommended at least at high-stakes, national scales. 
Worse, large-scale decentralized computing platforms such as 
Decentralized Autonomous Organizations (DAOs) might well be used not 
as tools of democracy but rather as weapons \emph{against} democracy, 
such as to buy votes at large scale while exposing the attacker 
to little or no risk of accountability or deterrence~\cite{
austgen2023DarkDAOPreprint,Ronne2025ToolboxCoercion}.

Despite this pessimistic climate
surrounding computing technology's interaction with democracy,
there nevertheless remains persistent and recurring interest
in the idea of \emph{democratic computing systems} in various forms:
large-scale distributed or decentralized systems designed to
serve and be accountable to their \emph{human} user population,
and to support social and democratic coordination and self-organization
in the ``one person, one vote'' egalitarian characteristic of democracy.
In the recently-burgeoning interest in DAOs, for example,
there has been a significant subcurrent of interest
in \emph{democratic DAOs},
in which decentralized organizations operating entirely
``on the blockchain'' are nevertheless controlled, at least in part,
by deliberative decision-making and voting among their human users.
These aspirations always run into the fundamental problem
of the Sybil attack, however \cite{douceur2002TheSybilAttack} --
namely that our networked computing systems
do not reliably or security "know" \emph{what a real human is}.

This problem in turn led to a subcurrent of interest
in \emph{proof of personhood} protocols and mechanisms --
attempting to distinguish real humans from Sybils or bots,
in various fashions~\cite{ford2020PersonhoodDigitalDemocracy,siddarth2020PoPProtocolsReview}.
Recent lessons in this space have revealed the ultimate importance of not only distinguishing humans from bots, but also of ensuring that participating humans are representing \emph{their own interests} and not those of a coercer, astroturfer, or other vote-buyer.
Such attacks
were found to have gradually taken over the Idena proof-of-personhood system over 
several years, for example~\cite{ohlhaver2025IdenaHistoryCrisis}.
Thus, when a proof of personhood is intended to facilitate democractic self-organization in any fashion, a lack of coercion resistance results in a gaping vulnerability
and invitation to vote-buying or astroturfing, whether at small or large scale.

Beyond merely supporting e-voting in conventional elections, therefore, we see the work represented by this paper as a potential stepping stone or building block for true large-scale democratic computing systems in the future: distributed or decentralized systems governed by, and accountable to, their users collectively, while assuring that their voting participants are actually representing their own interests and not someone else's.  While we leave the details of such democratic-computing applications to future work, one conceivably-workable approach is to combine TRIP's in-person coercion-resistance mechanism with the in-person practice of \emph{pseudonym parties} as a proof-of-personhood mechanism, where people obtain one anonymous participation token each at periodic real-world events~\cite{ford2008OnlineAccountablePseudonyms,BorgeFord2017PoP}.

If this key socio-technical hurdle -- of identifying \emph{real people acting in their own interests} -- can 
be surmounted, then there is already ample evidence that further 
development could enable distributed computing systems to 
improve and enhance democracy in numerous other ways long 
envisioned but as-yet of limited practicality.
Even serious political philosophers and democratic theorists see 
this potential, as exemplified in Hélène Landemore's proposal of 
\emph{open democracy} for example~\cite{helene2020OpenDemocracy}.
Similarly, for over two decades there has been a persistent 
subcurrent of interest in concepts such as \emph{liquid democracy} -- 
enabling large numbers of users to participate more regularly and 
directly in governmental deliberation and voting as their time 
and interest permits, by participating directly in selected 
topics or forums close to their interest while \emph{delegating} 
their vote in other areas to others they trust to represent them, 
individually rather than only in large collectives as in 
traditional representative democracy~\cite{
ford2020LiquidPerspective,blum16liquid}.
This idea took hold for a number of years in the German pirate party, 
which implemented and used this idea for intra-party deliberation 
via their LiquidFeedback platform \cite{kling2015LiquidFeedbackVotingBehavior} -- 
although they never adequately solved the conundrum of wishing to support 
user anonymity and freedom of choice while ensuring that only \emph{real humans} 
were voting, one per person, and only in their own interests. 
Similarly, widespread and increasing interests in \emph{mass online deliberation} 
and scalable participatory \emph{mini-publics}~\cite{helene2020OpenDemocracy}
shows widely-acknowledged promise to help make democracy more effective 
and participatory, but similarly requires as a key building block a 
mechanism to ensure that only real humans are registered and 
participating in their own interests, free from coercion or vote-buying.

Even the recent furor of global interest and research in 
machine learning, artificial intelligence, and large language models 
in particular, often neglects to observe that these technologies 
build on -- and are fundamentally dependent upon -- the \emph{collective intelligence} (CI) 
of all the human users who, intentionally or not, created and provided the 
enormous global datasets of data that the large AI models were trained on.
Further, the inability to distinguish human-generated from 
artificially-generated content presents an increasing threat 
not only to artists and other content produces~\cite{chesney2019deepfakes,amanbay2023AIArtEthics}
but also to the long-term effectiveness and usability of AI systems 
such as LLMs themselves~\cite{adler2025PersonhoodAndAI}.
Without being able to tell what comes from people and 
what comes from bots, Sybils, or AIs, we may lose not 
only the online world's effective anchor on ``truth'' or ``reality'' 
but also the effectiveness of the very AI systems of
such great interest at the moment.

In summary, 
while we fully believe that coercion-resistant e-voting 
along the lines this paper explores is a 
useful, challenging, and interesting systems research topic in its own right, 
we also see it as ultimately far more relevant in the long term as 
a potential enabler for future democratic computing systems and 
human-based collective intelligence platforms, 
which might conceivably serve and be used by collectives 
of all sizes from small groups to entire nations or 
ultimately even the global population of networked computing users. 

\section{Registration and voting channels}
\label{apx:registration}

For clarity of design and exposition,
the main paper simplistically assumes
that there is a single \emph{registration} step that is mandatory for all voters,
and that e-voting is the only voting \emph{channel},
or supported method of ballot casting.

Deploying \sysname or any e-voting architecture in a real nation
would of course require addressing many important considerations
to integrate and adapt e-voting to that nation's
legal, cultural, and practical environment.
These considerations would necessarily include adapting \sysname
to the nation's prevailing voter registration paradigm,
and integrating \sysname properly
with the nation's pre-existing voting channels.
Addressing such challenges in detail for any given nation
is beyond the scope of this paper,
but we briefly outline some of the relevant considerations here,
with a few particular nations as illustrative and contrasting examples.

\subsection{Voting across multiple channels}
\label{apx:registration:channels}

In practice, e-voting is never realistically deployed (at least nationally)
as the \emph{sole} available voting channel.
Instead, e-voting is generally
an opt-in alternative to one or more pre-existing ``base case'' channels,
such as in-person voting with paper ballots and/or postal voting.
We outline these voting channel considerations by briefly examining practices
in three countries: the US, Estonia, and Switzerland.

\paragraph{US:}
Although in-person voting by paper ballot remains
the established baseline voting channel throughout most of the US,
all states allow certain voters to request and submit \emph{absentee} ballots by post,
under conditions varying by state~\cite{MOVEAct09,pildes20accomodate,absenteevoting}.
Some states allow some or all voters to join a
\emph{permanent absentee voting list}
and obtain mail-in ballots automatically before every
election~\cite{ncsl24permanent}.
Eight US states have adopted postal voting as a primary voting channel,
with other states allowing postal voting for all voters
but only in certain districts or for certain elections~\cite{ncsl24mail}.
Although the US has never systematically adopted any remote e-voting system,
many states allow certain citizens, such as military and those with disabilities,
to submit scanned ballots by e-mail~\cite{MOVEAct09}.
As a communication channel that is neither encrypted nor even authenticated by default,
this \emph{ad hoc} adoption of e-mail as the US's standard ``remote e-voting system''
(while carefully never calling it one)
seems to us like a strikingly worst-of-all-worlds solution
to the inevitable demand for remote e-voting.

\paragraph{Estonia:}
\label{apx:registration:channels:estonia}
Estonia introduced e-voting as an opt-in alternative
to in-person voting with paper ballots in 2005,
and its adoption has gradually increased from 2\% to over
50\%~\cite{SmartmaticEstoniaI-Voting}.
Despite e-voting's popularity in Estonia,
casting a paper ballot remains an option.
Further, these two alternative voting channels contribute further
to Estonia's revoting-based approach to coercion resistance.
A paper ballot cast in-person overrides any e-vote cast remotely by the same person,
and the in-person voting deadline is later than the e-voting deadline.
A voter who is coerced (or has their e-voting materials confiscated)
all the way though the e-voting deadline, therefore,
still has the option in principle of casting an uncoerced ballot in person.
The effectiveness of this ``last line of defense'' against coercion
remains limited, however,
by the voter's practical ability to make an in-person trip to a polling station,
in secret without the knowledge of the coercer,
during the short time window between the e-voting and in-person voting deadlines.
For voters living outside Estonia or living full-time with abusive partners, for example,
it seems unlikely that the in-person backup channel can offer an effective defense.

\paragraph{Switzerland:}
Although Switzerland has so far adopted e-voting only in a few cantons,
this adoption came amidst a backdrop of \emph{two}
well-established voting channels already available almost nationwide:
namely both in-person voting and postal voting~\cite{luechinger07impact}.
Since postal voting was already the most popular baseline channel in Switzerland
before the adoption of e-voting,
the general security and privacy properties of this postal baseline
effectively came to define the ``goal posts'' for Switzerland's e-voting program.
This historical development explains in part why coercion resistance
has never been a high priority in Switzerland's e-voting program:
the public and political circles alike had already adopted and accepted postal voting
almost universally,
despite its well-known lack of coercion resistance.
Thus, an e-voting system design without coercion resistance
was at least no \emph{less} coercion-resistant
than the predominant baseline of postal voting.

\subsection{Registering \emph{to vote} versus \emph{for e-voting}}

The e-voting research literature,
and especially that addressing coercion resistance
using fake credentials~\cite{juels2010CoercionResistantElections,clarkson2008Civitas},
typically assumes that all voters must perform a step generically called
``registration.''
In realistic environments supporting multiple voting channels discussed above, however,
the reality of what ``registration'' might mean is more complex.

In particular, we must distinguish in practice between two potential forms of ``registration'':
namely \emph{registering to vote},
and \emph{registering for e-voting}.
``Registering to vote'' means signing up to be allowed
to cast a ballot in an election \emph{at all},
by any voting channel.
``Registering for e-voting'' means signing up to use the e-voting channel in particular,
as distinguished from any other available voting channels.

The \sysname architecture makes an important and essential assumption
that voters are required to register in-person at least \emph{for e-voting},
in particular acquiring real and fake credentials for coercion resistance.
\sysname makes no significant assumptions about
whether or how voters register \emph{to vote}, however.
While we generically use the term ``registration'' in the main paper
for consistency with the e-voting literature,
perhaps a better terms for the in-person step that \sysname requires
might be ``e-voting sign-up'' or ``credentialing'' rather than ``registration.''
The actual requirements for and considerations surrounding
\sysname's ``registration'' stage are more appropriately comparable to that of
obtaining or renewing an identity document at a government office,
and less suitably comparable to ``registering to vote'' in the US for example.
We next briefly examine these contrasting notions of ``registering to vote''
versus ``registering for e-voting'' (credentialing)
in the contrasting contexts of Europe versus the US.

\paragraph{Europe:}
In most European countries,
including the above examples of Estonia and Switzerland,
a standard and accepted responsibility of government
is to maintain an accurate registry of inhabitants:
\ie a database listing who currently lives where,
including basic metadata such as age, origin, and citizenship.
Residents are expected and required to report
when they move into, out of, or between localities,
for various reasons orthogonal to voting,
such as for reliable communication between the resident and the government,
and for determination of the resident's tax obligations.
Determining which current residents of some locality are eligible to vote in which elections
is just one more, of many, standard and required uses of this registry of inhabitants.
Since governments are expected to ``know their inhabitants'' anyway,
including residence-address and voting-eligibility metadata,
``registration to vote'' is generally automatic throughout European countries.
Unlike in the US, voters need not do anything special to ``register to vote'':
instead they receive instructions and materials needed for voting
automatically by virtue of being listed on the appropriate registry of inhabitants.

\sysname's assumption of in-person ``registration'' might therefore seem at first glance
incompatible with Europe's convenient standard of automatic registration to vote.
This is not the case, however,
once we correctly identify \sysname's ``registration'' step as
\emph{registration for e-voting} or \emph{credentialing}.
If \sysname were to be integrated into a European nation's election system, in particular,
it need not and should not affect the European standard of automatic registration to vote.
Instead, in-person credentialing would be a step required only for those voters
wishing to opt-in to the \emph{e-voting channel} in particular,
and need not be required of anyone opting for existing voting channels
such as in-person voting by paper ballot.
Thus, in-person registration for e-voting or credentialing
would remain an occasional convenience cost,
but only one imposed on those desiring the option of using the e-voting channel.
Visiting a government office in-person is often still required occasionally, anyway,
for other security-critical processes such as obtaining or renewing IDs or passports.
European voters could therefore in principle do their in-person credentialing for e-voting
at the same time as they sign up for or periodically renew other IDs.

\paragraph{US:}
In the US, in contrast,
voters must normally register to vote as a separate step,
as a prerequisite to casting a ballot via \emph{any} voting channel.
In states where this mandatory voter registration step
is normally done (or required to be done) in-person anyway,
there is a conceivable opportunity to deploy an e-voting architecture like \sysname
without imposing on voters any ``new'' obligation to visit a government office in most cases.
US voters in principle could opt-in to \sysname
and perform the necessary credentialing for e-voting
immediately after registering to vote, in the same in-person visit.

Despite this potential short-term opportunity, however,
we do not suggest or endorse such an approach as an ideal long-term strategy.
The US's requirement of a separate voter registration step
is not only inconvenient to voters,
but has been regularly abused in practice
as a method of deliberate voter suppression~\cite{ross19passive,hardy20voter}.
The fact that the roster of registered voters (often including party affiliation)
is publicly available in many states further exacerbates this weakness.
In particular, any coercer wishing to ensure that certain ``undesirable voters''
do not vote at all need not monitor the target voters continuously:
the coercer need only watch the public lists of registered voters
and threaten ``consequences'' if any of the targets' names ever appear on it.
Thus,
independent of the coercion-resistance properties of any particular voting \emph{channels},
the US's general voter registration paradigm
has a severe and inherent weakness
to coercion \emph{not to register at all}, at the very least.

The hypothetical adoption of an e-voting architecture like \sysname in the US
would not appear to affect these US-specific registration issues
either for better or worse in any obvious way.
\sysname would not help address the problem of voter suppression at registration time,
but it would not appear to exacerbate the problem either.
Further, realistic solutions to the US's registration-time voter suppression weaknesses
appear to be orthogonal to \sysname and in general to any particular voting channel.
The obvious solution is to adopt the European approach of automatic voter registration:
there is no opportunity for coercers to suppress voters at registration time
if all eligible voters are registered automatically without their taking any specific action.
This solution is clearly feasible and requires no new technology such as e-voting,
and hence we argue is important but orthogonal to and beyond the scope of this paper.

\subsection{Cross-channel coercion considerations}

Any election system supporting multiple channels and desiring coercion resistance
must in general consider coercion-resistance properties not just within but across channels.
We noted about in \cref{apx:registration:channels:estonia}
how Estonia's design consciously uses in-person voting as a backup channel to e-voting,
including for purposes of coercion resistance.

In general, a voting system may be effectively only as coercion-resistant
as its \emph{least-coercion-resistant} voting channel,
unless the system's cross-channel interactions are carefully designed
to avoid this undesirable effect.
As a purely-hypothetical negative example,
suppose that Switzerland were to adopt a coercion-resistant e-voting channel such as \sysname,
without addressing the legacy problem
that Switzerland's currently-predominant postal-voting channel is not coercion-resistant.
Even if both the e-voting and in-person voting channels are perfectly coercion resistant,
a coercer could simply demand that a target voter elect postal voting,
and supervise the filling and mailing of the victim's postal ballots.
In other words, a voter's \emph{choice of voting channel} itself
represents a point of potential vulnerability to coercion.

One way to address this cross-channel issue in general
is to allow votes via less-coercion-resistant channels
to be overridden by votes via more-coercion-resistant channels.
Estonia's cross-channel design is an example of this design principle,
albeit with important time-based constraints due to the revoting approach.

An important challenge with revoting and vote-overriding in general, however,
is ensuring the voting system's end-to-end transparency,
not just in terms of the verifiability each channel provides,
but also verifiability of correct tallying \emph{across} channels.
If Alice revotes several times using Estonia's e-voting channel,
then overrides all of those votes by casting a paper ballot in-person,
how do election observers---or the public at large---%
know that only Alice's in-person vote was counted?
If the government makes publicly-available the fact that Alice cast a ballot in-person,
then this defeats coercion resistance because Alice's coercer can see that information as well.
If the government keeps this information secret, in contrast (as Estonia's must),
then the public must apparently ``just trust'' that Alice's last e-vote and her in-person vote
are not \emph{both} counted,
no matter how transparent the e-voting channel alone might be.
(Estonia's current e-voting channel itself
is not verifiable anyway~\cite{springall2014EstoniaEVotingAnalysis},
so while this cross-channel transparency issue is important in general,
it is moot in Estonia's current practical design.)

In summary, a nation's registration paradigm, and other alternative voting channels available,
represent important environmental issues in any conceivable deployment
of a coercion-resistant e-voting system like \sysname.
Both the details of the environment and the appropriate solutions
are necessarily country-specific, however,
and hence are beyond the scope of this paper.
\section{Extensions to the base \sysname system}\label{apx:extensions}
In this section, we present several extensions
to \sysname that generally improves usability,
verifiability and coercion-resistance.
All of these extensions are optional, however,
and might be omitted from a deployment
due to the complexity they add
or for other technical or policy reasons.

\subsection{Voting history review and verification}\label{apx:voting-history}

The use of fake credentials in principle allows voters
to see a record of how they voted in recent elections
with a particular credential (real or fake).
Being able to see later how one voted is an obvious feature
often desired in e-voting systems,
but is usually not allowed due to receipt-freeness concerns~\cite{
benaloh1994ReceiptfreeSecretBallot}.
In the case of \name, however,
allowing voters to see how they voted in the past does not constitute 
a receipt or compromise coercion resistance,
because the record of votes cast \emph{with a particular credential}
does not leak the crucial single bit of information of
whether this is was a real or fake credential,
and hence whether these votes actually counted or not.

From a cryptographic perspective the ability to view past votes
does not affect \name's security properties such as verifiability:
\eg \name remains verifiable even if the VSD does not allow voters
to view their past votes.
Nevertheless, we feel that allowing voters to view their past votes
can improve the \emph{perception} of transparency in the voting process,
and in this way can provide a useful psychological benefit 
towards the acceptance of an e-voting system,
and the acceptance of the results of any given election.
Such psychological effects of viewing voting history
remain to be studied systematically, however.

Furthermore, this approach actually offers a 
verifiability advantage, 
as each vote can be accompanied by a receipt 
proving the ballot contains that vote. 
Voters could use a second device to verify 
the accuracy of their vote not only for the 
current election but also for all previous ones. 
Additionally, even if a second device is not 
available immediately, a new device purchased 
later can verify the integrity of votes cast
on the previous device(s).

Voters can also request the election authority to 
reveal all previous votes cast with a credential 
stored on their device. 
To accomplish this, the voter's device proves 
ownership of the credential to each election 
authority member and requests verifiable decryption 
shares of the ballots cast with it.
The voter's device then reconstructs the vote 
using the decryption shares while verifying 
its integrity. 
Importantly, 
this process does not disclose the vote to 
any election authority members since the
reconstruction is performed locally, 
on the voter's device.

\subsection{Reducing the credential exposure window}\label{apx:credential-exposure}
The voter's real credential is 
vulnerable in two places:
during transport on the printed
receipt and at the issuing kiosk.
To mitigate transport risks,
envelopes are designed with a hollow section that conceals the
confidential QR codes when the receipt is inserted.
A transparent window reveals only the
non-confidential second QR code,
which contains check-out material,
preventing exposure of the entire receipt.
At the kiosk,
exposure can be addressed using
the signing keypair
to sign votes when they are posted on
the ledger.
Voters' devices monitor
the ledger
to detect if a vote was cast using the signing
keypair it holds.
Coercion resistance is preserved because
$c_\pc$ is a non-deterministic encryption of
$c_\pk$ so releasing $c_\pk$ publicly reveals
nothing about whether a particular $c_\pc$
contains $c_\pk$.

However,
impersonation creates inconvenience for voters,
who must contest fraudulent votes with the election authority.
To avoid this,
another approach enables voters' devices
to sign a newly generated key pair
($\hat{c}_\sk, \hat{c}_\pk$) with their
kiosk-issued key pair.
The device then publicly discloses
the signature, along with the new
public key $\hat{c}_\pk$,
allowing only votes cast with
$\hat{c}_\pk$ to be tallied.
This process effectively transfers
voting rights from the kiosk-generated key
pair to the device-generated one.
Additionally, it enables voters to port
their credentials to new devices,
rendering the old device's credential unusable.
Both approaches apply to fake
credentials since they are
also just signing key pairs.
The tallying process would then use
this intermediary table (post-shuffled)
to link credentials
on the voting log with the credentials on
the registration log.

\subsection{Resisting extreme coercion scenarios}\label{apx:extreme-coercion}
A significant challenge with remote
voting systems is handling voters under
practically constant coercion
(\eg by an abusive spouse).
In \name,
such voters might struggle to hide their
real credential from a coercer who
could, for example, conduct a complete physical search
immediately after registration.
\name can allow these voters to delegate
their voting rights to a well-known entity,
like a political party, while leaving the booth
with only fake credentials.
Thus, when the coercer searches the victim,
all credentials found are fake,
although the victim of course claims
that one of them is real.
This approach
unavoidably requires voters under extreme coercion to
trust the kiosk, however,
as these voters can retain no material
evidence that the kiosk acted honestly.

We envision the kiosk 
asking voters if they believe they 
cannot cast their intended vote 
outside the booth (e.g., lacking a 
device outside a coercer's control).
If the voter acknowledges, the kiosk prompts them 
to delegate their vote by selecting a political 
party they align with. 
The kiosk then encrypts the political party's 
preloaded public key (P), which becomes this voter's 
blinded public credential tag. 
This process does not require the kiosk to have 
the private key for P, eliminating any exposure 
of the party's credential private key to the registrar.
Each political party's vote will then be counted
for each voter who delegated their vote to that
party.

\section{Formal system and threat models}
\label{sec:formal-model}

\subsection{System Model}\label{subsec:systemModel}

\noindent
The \sysname architecture involves the following key actors.

\parhead{Ledger.}
The ledger $\L$ is an
append-only, always available, 
publicly accessible data structure.
It includes three ``sub-ledgers'',
the Registration Ledger $\L_R$,
the Envelope Commitment Ledger $\L_E$,
and the Ballot Ledger $\L_V$.
We idealize the ledger as a
tamper-evident, globally consistent view 
with perfect access: 
All actors observe the same 
authentic ledger state, and any tampering 
(\eg denials, alterations, or view manipulations) 
is detectable with overwhelming probability. 
This assumption abstracts away consensus, 
allowing us to focus on e-voting security 
definitions.
In practice, ledger integrity holds under 
honest-majority assumptions with 
decentralized querying and 
well-decentralized nodes.

\parhead{Registrar.}
The registrar $\R$ enrolls eligible voters
from a given electoral roll $\V$.
The registrar consists of:
(1) kiosks
$\K~=~\{K_1, \dots, K_{n_{K}}\}$,
    each in a privacy booth,
    which issue voters credentials;
(2) envelope printers
$\P = \{P_1, \dots, P_{n_{P}}\}$,
    which issue the envelopes that the voters
    use during credentialing;
and
(3) registration officials
$\O = \{O_1, \dots, O_{n_{O}}\}$,
    represented by their \emph{official supporting device} (\OSD),
    who authenticate voters and
    authorize their credentialing sessions.

\parhead{Voters.}
Voters on the electoral roll
$\V = \set{V_{1}, \dots, V_{n_{V}}}$
who interact with the system to 
register and vote.
Voters obtain their credentials 
in-person at the registrar,
and activate them on their
voter supporting device (\VSD)
to cast ballots on a ledger.
\VSD{}s periodically monitor the ledger
to inform voters of relevant updates,
such as a successful registration session.

\parhead{Authority.}
The election authority
$\A = \set{A_{1}, \dots, A_{n_{A}}}$
consists of $n_A$ members who
jointly process the ballots cast on the ledger to
produce a publicly verifiable tally.
The authority also
manages election logistics, sets policy,
and is ultimately accountable to the public.

\subsection{Threat Model}\label{sec:threat-model}
We define
three adversaries
corresponding to specific
scenarios:
integrity $\IA$, privacy $\PA$ and coercion $\CA$.\footnote{
An availability adversary seeking to deny
service is also relevant,
but we assume this issue is addressed by
the use of high-availability infrastructure.}
\cref{table:adversaries} illustrates which actors,
represented by their device for added granularity,
each adversary may compromise.
We assume that all adversaries are computationally bounded,
cryptographic primitives are secure,
and
communication channels are secure (\eg via TLS),
unless stated otherwise.

\renewcommand\tabularxcolumn[1]{m{#1}}
\newcolumntype{g}[0]{>{\centering\arraybackslash}X}%
\newcolumntype{h}[0]{>{\centering\arraybackslash\hsize=.8\hsize}X}%
\newcolumntype{i}[0]{>{\centering\arraybackslash\hsize=1.3\hsize}X}%
\newcolumntype{j}[0]{>{\centering\arraybackslash\hsize=1.4\hsize}X}%
\newcolumntype{k}[0]{>{\centering\arraybackslash\hsize=.9\hsize}X}%
\newcolumntype{A}[0]{>{\centering\arraybackslash\hsize=1.2\hsize}X}%

\begin{table}[t]
    \setlength\tabcolsep{1pt}
    \def\arraystretch{1.2}
    \footnotesize
    \begin{tabularx}{\linewidth}{|j|k|i|h|h|A|h:h|}
        \hline
        \multirow{2}{=}{\diagbox[width=1.06\hsize]{Adv.}{Device}} & 
        \multirow{2}{=}{\centering Ledger} & 
        \multirow{2}{=}{\centering Authority} & 
        \multirow{2}{=}{\centering \OSD} & 
        \multirow{2}{=}{\centering Kiosk} & 
        \multirow{2}{=}{\centering Envelope Printers} & 
        \multicolumn{2}{c|}{\VSD w/ Cred.} \\
        & {} & {} & {} & {} & {} & Real & Fake \\
        \hline
        \vspace{0.1cm}
        {\centering Integrity} & Yes* & All & Yes\textsuperscript{\dag} & Yes\textsuperscript{\dag} & Yes\textsuperscript{\dag} & No & No \\
        \hline
        {\centering Privacy}& Yes & $n_A-1$ & Yes & Yes & Yes & No & Yes  \\
        \hline
        {\centering Coercion} & 
         Yes & $n_A-1$ & No & No & No & No\textsuperscript{\ddag} & Yes  \\
        \hline
    \end{tabularx}
    \vspace{3mm}
    \caption{\textbf{Threat Model}:
    This table depicts an adversary's ability to 
    compromise a device and the entity or credential it represents. \\ 
    \footnotesize{
    * The ledger can be compromised but since all actors observe the same 
    authentic ledger state, any tampering (e.g., denials, alterations, 
    or view manipulations) is detectable with overwhelming probability. \\
    \textsuperscript{\dag}
    The risk of the adversary compromising voter registration
    (see Section \ref{sec:security}, \cref{apx:proofs:verifiability}),
    is small per registration, 
    and becomes negligible when considering the combined 
    probability across registrations.\\
    \textsuperscript{\ddag}
    If voters cannot conceal their device from a coercer,
    they can entrust their real credential to a trusted
    third party to cast votes on their behalf. \\
    }
    }
    \label{table:adversaries}
\end{table} 
\parhead{Integrity.}\label{sec:threat-model:integrity}
Integrity adversary $\IA$'s goal is
to manipulate the election outcome
without detection,
as any detected interventions could
undermine public confidence in the results.
$\IA$ cannot compromise the electoral roll
nor the \VSD{}s.\footnote{
Works such as~\cite{cortier2024BeleniosCaI,benaloh2007BenalohChallenge}
address the latter problem %
using a cast-as-intended verification mechanism.
}

\parhead{Privacy.}\label{sec:threat-model:privacy}
$\PA$'s goal is to
\emph{reveal} a voter's real vote
to, for example, target specific
voters with personalized political
advertising to influence their vote
in future elections~\cite{rathi2019CambridgeAnalytica}.
$\PA$ can compromise
all but \emph{one} authority member.\footnote{
    This notion is common in electronic
    voting literature~\cite{2021SwissPostProofs,
    juels2010CoercionResistantElections}
    and in privacy-enhancing systems in general~\cite{
    sav2021Poseidon,
    froelicher2017UnLynx}.
}
$\PA$ cannot compromise
the \VSD{}s containing voters' real credentials.

\parhead{Coercion.}\label{sec:threat-model:coercion}
Unlike $\PA$,
the coercion adversary $\CA$
aims to pressure voters overtly into casting
the coercer's vote.
$\CA$'s objective is to determine whether
the targeted voters complied with their demands,
which may involve demanding
voters to cast $\CA$-dictated ballots,
or reveal their real credential.
$\CA$ inherits the same capabilities as $\PA$,
meaning it can compromise all but one authority member,
and cannot compromise the \VSD containing
the voter's real credential.
Additionally, $\CA$ cannot compromise
the registrar nor observe the communication
channel between the voter and the kiosk.
While side-channel attacks are out-of-scope,
we discuss them briefly in~\apxref{apx:side-channels}.
As is typical in coercion-resistant e-voting systems~\cite{
juels2010CoercionResistantElections,clarkson2008Civitas},
we assume that the communication channel used by voters to cast their real vote
is anonymous,
at least to the extent of not being monitored
by the coercer.
Otherwise the coercion adversary could simply demand
that the voter never communicates with the ledger,
and monitor all of the voter's communication
in order to enforce that demand.

\section{Formal \name registration protocol}
\label{sec:formal-scheme}
\renewcommand\arraystretch{1.1}
\begin{table}[t]
\centering
\scriptsize
\begin{tabular}{ c p{0.75\linewidth} }
    \toprule
    Symbol & Description \\
    \midrule
    $G, q, g$ & A cyclic group $G$ of order $q$ with generator $g$ \\
    \hline
    $\A, \O, \K, \P, \V$ &
    Authority, Officials, Kiosks, Envelope Printers, Electoral roll \\
    \hline
    $n_A, n_O, n_K, n_P$ &
    Number of Authorities, Officials, Kiosks, Envelope printers \\
    \hline
    $\L, \L_R, \L_E, \L_V$ &
    Ledger and Registration, Envelope \& Voting (sub-)ledgers \\
    \hline
    $V_{id}, c_\pc, c_\pk, c_\sk$ &
    Voter's identifier, Public Credential, Public \& Private Keys \\
    \hline
    $c, \fc_\pk, \fc_\sk$ &
    A credential, A fake credential's Public and Private Keys \\
    \hline
    $\E, n_E, n_c$ &
    Envelope challenges and number of envelopes \& credentials \\
    \hline
    $\tau, s_{rk}$ &
    \mac~authorization tag, Official \& kiosk shared secret key \\
    \hline
    \OSD, \VSD &
    Registration officials and Voters' supporting devices \\
    \hline
    $t_{in}, t_{out}, q_c, q_r$ &
    Check-In \& Check-Out Tickets, Commit \& Response Codes \\
    \bottomrule
\end{tabular}
\vspace{3mm}
\caption{
\textbf{Scheme Notations.}
}
\label{tab:scheme:notations}
\end{table} %
\begin{figure}[t]
\gameblock[skipfirstln]
{\pcalgostyle{\name} (\secpar, ($G$, $q$, $g$), $V$, $\L$, $\A$, $\O$, $\K$, $\P$, $\E$, $s_{rk}$)}{
    \\
    O, K \gets V(\O, \K) \text{\% Voter-chosen/given official and kiosk} \\
    t_{in} \gets (V_{id}, \_) \gets \CheckIn(O, K, s_{rk}, V_{id}) \\
    c \gets (q_c, t_{ot}, q_r, e) \gets ((V_{id}, c_\pc, Y_c, \sigma_{kc}), (V_{id}, c_\pc, K, \sigma_{k_{ot}}), \pcskipln \\
    \t\t\t (c_\sk, r, K, \sigma_{kr}), (P, e, \sigma_p)) \pcskipln \\
    \t\t\t\t \gets \RealCred(\secpar, (G, q, g), K, \A_\pk, \E, t_{in}) \\
    \label{fig:scheme:line:voter-pick-envelopes} \mathbf{e} \gets \set{e};~\mathbf{c} \gets \set{c}~\text{\% Used challenges; credentials set} \\
    \pcfor 1~\text{to}~(n_c \gets V()) - 1 \pcdo ~\text{\% Voter-chosen \# credentials} \\
    \t \fc \gets
    \FakeCred(\secpar, (G, q, g), K, \A_\pk, \E_{\ominus \mathbf{e}}, t_{ot}) \\
    \t \mathbf{c} \Leftarrow \fc;~\mathbf{e} \Leftarrow \fc[e] \\
    \label{fig:scheme:line:voter-pick-checkout} c_v \gets V(\mathbf{c}) ~\text{\% Voter-chosen credential for check-out} \\
    \L \gets \CheckOut(O, \L, \K, c_v[t_{ot}]) \\
    \pcfor i~\text{to}~n_c \pcdo \\
    \t \L \gets \Activate(\L, V, \mathbf{c}_i)
}
\caption{\textbf{\name} Registration process for a prospective voter.}
\Description{Registration process for a prospective voter.}
\label{fig:scheme}
\end{figure}
This section presents \name,
formally described in \cref{fig:scheme}.

\parhead{Notation.}
For a finite set $S$,
$s \sample S$ denotes that $s$ is sampled
independently and
uniformly at random from $S$.
The symbol $\ominus$ represents exclusion from a
collection of elements.
We denote
$a \Leftarrow b$ as
appending $b$ to $a$,
$a \concat b$ as concatenating $b$ with $a$,
$\mathbf{x}$ as a vector of elements of type $x$, and
$\mathbf{x}[i]$ as the $i$th entries of
the vector $\mathbf{x}$.
We use $\top$ and $\bot$ to indicate success and failure,
respectively.
Variables used throughout the \name scheme are summarized
in~\cref{tab:scheme:notations}.

\parhead{Primitives.}
\name requires
(1) the ElGamal encryption scheme $\elgamal$,
(2) a distributed key generation scheme $\dkg$,
(3) a EUF-CMA signature scheme $\sig$,
(4) a secure hash function $\hash$,
(5) a message authentication code scheme $\mac$, and
(6) an interactive zero-knowledge proof of equality
    of discrete logarithms $\pcalgostyle{ZKPoE}$.
    We define these primitives below.

\subsection{Cryptographic primitives}
\label{sec:crypto_primitives}
\noindent

\parhead{Distributed Key Generation Scheme.}
\name uses a
distributed key generation
protocol
$\dkg$~\cite{froelicher2017UnLynx}
that inputs a group description
$(G, q, g)$---a cyclic group $G$ of order $q$ with
generator $g$---and the number of parties $n$
and creates a private and public key pair
for each party $(P_i^{\sk}, P_i^{\pk})$
and a collective public key $P^\pk$:
\[
  \set{P_i^{\sk}, P_i^{\pk}}, P^\pk \gets \dkg(G, p, g, n),
\]
such that $P_i^{\pk} = g^{P_i^{\sk}}$ where
$P_i^{\sk} \sample \ZZ_q$,
$P^\pk = \prod_{i = 1}^n P_i^\pk$.

\parhead{ElGamal Encryption Scheme.}
This scheme is parameterized by
a cyclic group $G$ of prime order $q$ and
a random generator $g$;
and consists of the following algorithms:
$\pcalgostyle{EG}.\kgen(G, q, g)$
which takes as input the group definition
and outputs a public key $\pk$ along with
a private key $\sk$
such that
$\pk=g^{\sk}$ where
$\sk \sample \ZZ_q$;
a randomized encryption algorithm $\pcalgostyle{EG}.\enc(\pk, m)$
which inputs a public key $\pk$ and a message $m \in G$
and outputs a ciphertext $C = (C_1, C_2) = (g^r, \pk^r m)$
for $r \sample \ZZ_q$;
and a deterministic decryption algorithm
$\pcalgostyle{EG}.\dec(\sk, C)$ which takes as input
a private key $\sk$ and a ciphertext $C$
and outputs a message $m = C_2 (C_1^{\sk})^{-1}$.

\parhead{Signature Scheme.}
\name uses a EUF-CMA signature scheme
defined by the following three algorithms:
a randomized key generation algorithm $\sig.\kgen(\secparam)$
which takes as input the security parameter
and outputs a signing key pair $(\sk, \pk)$;
a signing algorithm $\sig.\sign(\sk, m)$
which inputs a private key and a message $m\in \bin^*$
and outputs a signature $\sigma$;
a signature verification algorithm $\sig.\verify(\pk, m, \sigma)$
which outputs $\top$ if $\sigma$ is a valid signature of $m$ and $\bot$ otherwise;
and an algorithm $\sig.\pubkey(\sk)$ that takes as input a private key $\sk$
and outputs the corresponding public key $\pk$.

\parhead{Hash.}
\name utilizes a cryptographic secure hash function $\hash$,
for which the output is $2\secpar$ bits, for security parameter $\secpar$.

\parhead{Message Authentication Code.}
\name uses a message authentication code scheme
defined by the following two algorithms:
a probabilistic signing algorithm $\mac.\sign(k, m)$
which takes as input a (secret) key $k$ and a message $m$
and outputs an authorization tag $\tau$;
and a deterministic verification algorithm $\mac.\verify(k, m, \tau)$
which takes as input the (secret) key $k$, the message $m$
    and the authorization tag $\tau$
and outputs either $\top$ for accept or $\bot$ for reject.

\parhead{Zero-Knowledge Proof of Equality.}
\name employs an interactive zero-knowledge proof of equality of discrete
logarithms~\cite{chaum1993WalletDatabases}
$\pcalgostyle{ZKPoE}$ so that a prover $\prover$ can convince a verifier $\verifier$
that $\prover$ knows $x$, given messages $y \equiv g_1^x$ (mod $p$) and $z \equiv g_2^x$
(mod $p$) without revealing $x$.
In \emph{interactive} zero knowledge proofs, the verifier $\verifier$ must provide the challenge
only \emph{after} the prover $\prover$ has computed and provided the commit to $\verifier$.

\subsection{Setup.}
\label{scheme:setup}
\begin{figure}[t]

\gameblock[codesize=\small,,skipfirstln,jot=-0.6mm]{$\pcalgostyle{Setup}
(\secpar, \V, (G, p, g), n_A, n_O, n_K, n_P, n_E)$}{%
    \\
    \L \gets \emptyset \\
    \set{A_i^{\sk}, A_i^{\pk}}, \A_\pk \gets \dkg(G, p, g, n_A) \\
    \set{O_i, K_i, P_i \gets \sig.\kgen(\secparam)}_{0 \leq i < n_O, 0 \leq i < n_K, 0 \leq i < n_P} \\
    \L_R \gets \set{\V_i^{id}}_{i \in \V} \\
    \E \gets \{P_j \sample \P; e_i \sample \ZZ_q; \pcskipln \\
    \t \t \, \, \L_E \Leftarrow (P_j^\pk, \hash(e_{i}), \sig.\sign(P_j^\sk, \hash(e_{i})))\}_{0 \leq i < n_E} \\
    \label{line:srk}
    s_{rk} \gets \bin^\secpar
}
\caption{\textbf{Setup} Procedure for the ledger, the authority members,
the officials, the kiosks and the envelope printers
with envelope issuance.
The secret $s_{rk}$ is shared between officials and kiosks.}
\Description{Setup 
procedure for the ledger, the authority members,
the officials, the kiosks and the envelope printers
with envelope issuance.
The secret $s_{rk}$ is shared between officials and kiosks.}
\label{fig:scheme:setup}
\end{figure}
Setup (\cref{fig:scheme:setup})
initializes the core system actors
(Ledger, Authority, and Registrar).
Prior works often include
registration as part of a broader setup process,
but we separate it to delineate registration cleanly:
\begin{itemize}[leftmargin=*]
    \item
        The ledger $\L$ becomes available and made accessible to all
        (including third) parties.
        \revisiontwo{To delineate the different types of recorded events,
        we introduce sub-ledgers: $\L_R$, $\L_E$, and $\L_V$, which are
        dedicated to storing registration sessions, envelope data, and votes,
        respectively.}
    \item
        The authority members $\A$ run \dkg{},
        outputting a private, public keypair for
        each authority member $(A_i^\sk, A_i^\pk)$
        and a collective public key $\A_\pk$
        which is made available to all parties.
        $\A_\pk$ must be a generator of $G_q$.
    \item
        Each registrar actor
        (\OSD{}s, kiosks \& printers)
        generates their own private and public key
        pair using $\sig.\kgen(\secparam)$.
        The registrar uses the electoral roll $\V$
        to populate $\L_R$
        with each voter's unique identifier $V_{i}^{id}$.
        \revision{The printer issues at least $n_E > c|\V| + \lambda_\E|\K|$ envelopes
          \E,
        where constant $c \geq 2$ represents the authority’s estimate of the 
        number of envelopes each voter consumes and $\lambda_\E$ is a security 
        parameter detailed in~\cref{apx:proofs:coercion-resistance}, 
        which represents the minimum number of envelopes 
        required in each booth. If authorities underestimate the average
        consumption of envelopes, $\P$ can issue additional envelopes.}\footnote{
        \revision{
            Unlike paper ballots, envelopes do not expire,
            allowing unused ones to be saved for future registrations.
            This longevity property mitigates the steady-state cost of maintaining an abundant supply of envelopes.%
            }
        }
        Each envelope contains
        the printer's public key $P_i^\pk$,
        a cryptographic nonce $e \sample \ZZ_q$, and
        a signature on this nonce
        $\sigma_{p} \gets \sig.\sign_p(\hash(e))$.
        For each envelope, the printer also publishes
        $(P_i^\pk, \hash(e), \sigma_{p})$ to the ledger $\L_E$.
        \highlightchange{The officials $\O$ and kiosks $\K$
        generate a shared secret key $s_{rk}$
        to create and verify $\mac$ tags that authorize
        voters access to a kiosk.}
\end{itemize}

\subsection{Check-In.}
\begin{figure}[t]

\gameblock[codesize=\small,skipfirstln,jot=-1mm,colspace=0.8cm]
{$\mathsf{CheckIn}(O, K, s_{rk}, V_{id})$}{%
    \pcskipln \\
    \textbf{\OSD}(O, s_{rk}, V_{id}) \< 
    \textbf{Kiosk}(K, s_{rk}, t_{in}) \\
    \tau_{r} \gets \mac.\sign(s_{rk}, V_{id}) \<
    (V_{id}, \tau_{r}) \gets t_{in} \\
    t_{in} \gets (V_{id}, \tau_{r}) \< 
    \mac.\verify(s_{rk}, \tau_{r}, V_{id}) \iseq \top
}
\caption{\textbf{Check-In.} 
The official's device issues check-in ticket and 
kiosk verifies authenticity of check-in ticket.}
\Description{Check-In Procedure: 
The official's device issues check-in ticket and 
kiosk verifies authenticity of check-in ticket.}
\label{fig:scheme:check-in}
\end{figure} %
Upon successful authentication at Check-In
(\cref{fig:scheme:check-in}),
the \OSD{} issues the voter
a check-in ticket, $t_{in}$, consisting of
the voter's identifier, $V_{id}$, and
an authorization tag, $\tau_r$, on $V_{id}$.\footnote{
	\revision{For usability, as discussed in \cref{sec:eval:usability},
we use a barcode instead of a QR code, and due to storage
constraints in a barcode, we use $\mac$ instead of $\sig$.}}

\noindent
The kiosk validates the authorization tag, $\tau_r$,
when the voter presents their ticket $t_{in}$ (\cref{fig:scheme:check-in}).

\begin{figure*}[t]
\centering
\begin{subfigure}[t]{0.51\linewidth}
\gameblock %
[codesize=\scriptsize,jot=-1mm,colspace=0cm,linenumbering,skipfirstln]
{$\pcalgostyle{RealCred}(\secpar, (G, q, g), K, \A_\pk, \E, t_{in})$}{
    \pcskipln \\
    \textbf{\small Kiosk}(\secpar, (G, q, g), K, \A_\pk, \E, t_{in})
    \< \< \textbf{\small Voter} \\
    (V_{id}, \_) \gets t_{in} \<
    \text{\% Check-In Ticket} \\
    (c_\sk, c_\pk) \gets \sig.\kgen(\secparam) \<
    \text{\% Real Credential} \\
    x \sample \ZZ_{q};~X \gets \A_\pk^x \<
    \text{\% ElGamal secret} \< \\
    \label{line:scheme:public-credential}
    c_\pc \gets (C_1, C_2) \gets (g^x, X \cdot c_\pk) \< \text{\% Public Credential} \< \\
    \label{scheme:real:commits}
    y \sample \ZZ_{q}; Y_c \gets (Y_1, Y_2) \gets (g^y, \A_\pk^y) \< \text{\% ZKP commit } \< \\
    \sigma_{k_c} \gets
    \sig.\sign(K^\sk, V_{id} \concat c_\pc \concat Y_c) \<
    \text{\% Commit Signature} \\
    q_c \gets (V_{id}, c_\pc, Y_c, \sigma_{k_c}) \<
    \text{\% Commit} \\
    \< \sendmessage*{->}{length=2.5cm,top={q_c}} \< \\
    \< \< E_{i} \sample \E \\
    \label{scheme:real:envelope_pick}
    \< \sendmessage*{<-}{length=2.5cm,top={E_{i}}} \< \\
    (P^\pk, e, \sigma_{p}) \gets E_{i} \< \text{\% ZKP challenge}\< \\
    r \leftarrow y - ex \< \text{\% ZKP response} \< \\
    \sigma_{k_{ot}} \gets \sig.\sign(K^\sk, V_{id} \concat c_\pc) \<
    \text{\% Check-Out Signature} \\
    \sigma_{k_r} \gets \sig.\sign
    (K^\sk, c_\pk \concat \hash(e \concat r)) \<
    \< \\
    t_{ot} \gets (V_{id}, c_\pc, K^\pk, \sigma_{k_{ot}}) \< \text{\% Check-Out Ticket} \\
    q_r \gets (c_\sk, r, K^\pk, \sigma_{k_r}) \<
    \text{\% Response} \\
    \< \sendmessage*{->}{length=2.5cm,top={t_{ot}}} \< \< \\
    \< \sendmessage*{->}{length=2.5cm,top={q_r}} \< }
    \caption{\textbf{Real Credential Creation Process.} Voter and kiosk follow a sound zero-knowledge proof construction.}
    \label{fig:scheme:cred-creation:real}
\end{subfigure}
\hfill
\begin{subfigure}[t]{0.48\textwidth}
\gameblock %
[codesize=\scriptsize,colspace=-0.3cm,jot=-0.825mm,linenumbering,skipfirstln]
{$\pcalgostyle{FakeCred}(\secpar, (G, q, g), K, \A_\pk, \E_{\ominus \mathbf{e}}, t_{ot})$}{
    \pcskipln \\
    \textbf{\small Kiosk}(\secpar, (G, q, g), K, \A_\pk, \E_{\ominus \mathbf{e}}, t_{ot}) \< \<
    \textbf{\small Voter} \\
    (V_{id}, c_\pc, \_, \_) \gets t_{ot} \<
    \text{\% Unpack Check-out Ticket} \\
    (\fc_\sk, \fc_\pk) \gets \sig.\kgen(\secparam) \<
    \text{\% Fake Credential} \\
    (C_1, C_2) \gets c_\pc \<
    \text{\% Unpack Public Credential} \< \\
    \tilde{X} \gets C_2 / c_\pk \<
    \text{\% Derive ElGamal secret} \< \\
    \< \< E_{z} \sample \E_{\ominus \mathbf{e}} \\
    \< \sendmessage*{<-}{length=2.5cm,top={E_{z}}} \< \\
    (P^\pk, e, \sigma_p) \gets E_z \< \text{\% ZKP challenge} \< \\
    y \sample \ZZ_{q} \< \text{\% ZKP commit } \< \\
    \label{scheme:fake:commits}
    Y_c \gets (Y_1, Y_2) \gets (g^y C_1^e, \A_\pk^y \tilde{X}^e) \<
    \text{\% ZKP commit} \\
    r \leftarrow y \< \text{\% ZKP response} \< \\
    \sigma_{k_c} \gets \sig.\sign(K^\sk, V_{id} \concat c_\pc \concat Y_c) \< \< \\
    \sigma_{k_r} \gets \sig.\sign(K^\sk, \fc_\pk \concat \hash(e \concat r)) \< \< \\
    q_c \gets (V_{id}, c_\pc, Y_c, \sigma_{k_c}) \< \< \\
    q_r \gets (\fc_\sk, r, K^\pk, \sigma_{k_r}) \< \< \\
    \< \sendmessage*{->}{length=2.5cm,top={q_c}} \< \\
    \< \sendmessage*{->}{length=2.5cm,top={t_{ot}}} \< \< \\
    \< \sendmessage*{->}{length=2.5cm,top={q_r}} \< \pcskipln}
     \caption{\textbf{Fake Credential Creation Process.} Voter and kiosk follow an unsound zero-knowledge proof construction. Envelopes cannot be reused.}
     \label{fig:scheme:cred-creation:fake}
\end{subfigure}
\caption{\textbf{Voting Credential Creation Process.}}
\Description{Voting Credential Creation Process.}
\label{fig:scheme:cred-creation}
\end{figure*}
\subsection{Real credential.}
The kiosk now issues the voter their real credential
while proving its correctness~(\cref{fig:scheme:cred-creation:real}).
The kiosk first generates
the voter's real credential's
private and public keys $(c_\sk, c_\pk)$
and ElGamal encrypts $c_\pk$
using the authority's public key $\A_\pk$ to
obtain the voter's public credential $c_\pc$.
To prove that $c_\pc$ encrypts $c_\pk$ without
revealing the ElGamal randomness secret $x$
(to enable the construction of fake credentials later),
the kiosk, as the prover, and the voter, as the verifier,
run an interactive zero-knowledge proof of equality
of discrete logarithms:\footnote{
    For cryptographic purposes,
    we equate the voter with the verifier.
    However, in reality, the voter only observes
    the order in which QR codes are printed
    on the receipt without needing to understand them.
    The voter's device is responsible for
    checking the actual proof transcripts.
 }
    $\pcalgostyle{ZKPoE}_{C_1,X}\{(x):
    C_1 = g^x \land X = \A_\pk^x\}$.
The kiosk first computes the commits
$Y_1 = g^y$, and $Y_2 = \A_\pk^y$
for $y \sample \ZZ_{q}$
and prints the commit $q_c$ containing
the voter's public credential $c_\pc$,
the commits $Y_c = (Y_1, Y_2)$, and
a signature $\sigma_{k_c}$ on
($V_{id} \concat c_\pc \concat Y_c$).
The voter then supplies the kiosk with an envelope
$E_{i} \sample \E$ containing the challenge $e$.
The kiosk finally computes
the response $r = y - ex$,
the signatures $\sigma_{k_{ot}}$ and $\sigma_{k_r}$,
and prints the check-out ticket $t_{ot}$ and response $q_r$.

At this stage,
the voter observes that the process adheres to the
$\Sigma$-protocol sequence: commit, challenge, response.
If the voter detects and reports an anomaly,
we expect the registrar or some other authority
to direct the voter to another kiosk
and inspect the one reported,
as detailed in~\cref{apx:security:envelope_analysis}.
After successful registration,
the voter's device later verifies the computational correctness of the credential.

\subsection{Fake credentials.}
To create a fake credential (\cref{fig:scheme:cred-creation:fake}),
the kiosk generates a new credential $(\fc_\sk, \fc_\pk)$
and falsely proves that the public credential
$c_\pc$ encrypts $\fc_\pk$.
The kiosk first derives
the ``new'' ElGamal secret $\tilde{X} \gets C_2 / \fc_\pk$.
The kiosk has no knowledge of an $\tilde{x}$ that satisfies $\tilde{X} =
\A_\pk^{\tilde{x}}$, requiring one to solve the discrete logarithm problem.
Instead, the kiosk and the voter follow
an incorrect proof construction sequence
that violates soundness without affecting correctness.
In this sequence,
the voter first supplies a new envelope
$E_z \sample \E_{\ominus \mathbf{e}}$ to the kiosk,
where $\mathbf{e}$ are the previously used envelopes/challenges.
Then, the kiosk uses the new challenge $e$ to compute a ZKP commit
$(Y_1, Y_2) \gets (g^y C_1^e, \A_\pk^y \tilde{X}^e)$
for some $y \sample \ZZ_{q}$ and the ZKP response $r \gets y$.
The kiosk finishes
by computing signatures $\sigma_{k_c}$ and $\sigma_{k_r}$
and printing the commit $q_c$, check-out $t_{ot}$, and response $q_r$
sequentially,
where $t_{ot}$ is
identical (both in content and visually)
to the one in the real credential process.
The voter can repeat this process
for any number of desired fake credentials
(within reasonable limits mentioned
in~\cref{sec:design:voter-facing}).

\subsection{Check-Out.}
\label{subsec:checkout}
\begin{figure}[t]
\centering
\gameblock[codesize=\small,colspace=0.1cm,jot=-1mm,linenumbering,skipfirstln]
{$\CheckOut(O, \L, \K, t_{ot})$}{%
    \pcskipln \\
    \textbf{\OSD}(O, \L, \K, t_{ot}) \< \< \\
    (V_{id}, c_\pc, K^{\pk}, \sigma_{k_{ot}}) \gets t_{ot} \< 
    \text{\% Check-Out Ticket} \\
    K^{\pk} \isin \K^\pk; \< 
    \text{\% Authorized?} \\
    \sig.\verify(K^{\pk}, \sigma_{k_{ot}}, V_{id} \concat c_\pc) 
    \iseq \top \< \text{\% Verify Signature} \\
    \sigma_{o} \gets 
    \sig.\sign(O^\sk, V_{id} \concat c_\pc \concat \sigma_{k_{ot}}) \< 
    \text{\% Official Approval} \\
    \L_R[V_{id}] \gets (c_\pc, K^\pk, \sigma_{k_{ot}}, O^\pk, \sigma_o) \< \text{\% Update Ledger} \pcskipln \\
    \textbf{VSD}(\L, V) \\
    \Notify(V_{id}) \< \text{\% Notify Voter} \<
}
\caption{\textbf{Check-Out}.
The official approves the voter's registration session,
publishing on the ledger their signature, the kiosk's signature and
the voter's public credential.
The ledger verifies signatures and 
the \VSD{} monitoring the ledger notifies the voter. }
\Description{Check-Out Protocol: The official approves the voter's registration session,
publishing on the ledger their signature, the kiosk's signature and
the voter's public credential.
The ledger verifies signatures and 
the \VSD{} monitoring the ledger notifies the voter.}
\label{fig:check-out}
\end{figure} %
At check-out (\cref{fig:check-out}),
the official uses their \OSD{}
to scan the credential shown by the voter.
The voter shows this credential in the transport state (\cref{fig:transport}),
which reveals the contents of the check-out ticket $t_{ot}$.

The \OSD first checks the credential's authenticity
by checking the kiosk's public key $K^{\pk} \in \K^\pk$,
and verifying the signature $\sigma_{k_{ot}}$.
The \OSD then provides its stamp of approval with a digital
signature $\sigma_o$ on the voter's identifier $V_{id}$,
the voter's public credential $c_\pc$ and
the kiosk's signature $\sigma_{k_{ot}}$.
Finally, \OSD updates the ledger entry $V_{id}$ with
$(c_\pc, K^\pk, \sigma_{k_{ot}}, O^\pk, \sigma_o)$.
Once updated,
the ledger $\L$ performs the necessary checks
and the \VSD notifies
the voter about their recent registration
with information on how to report any
irregularities.

\subsection{Activation.}
\label{sec:scheme:activate}
\begin{figure}[t]
\centering
\gameblock[codesize=\scriptsize,skipfirstln,jot=-1mm,width=7.5cm]
{$\pcalgostyle{Activate}(\L, V, c)$}{%
    \pcskipln \\
    \textbf{\small VSD}((G, q, g), \L, V, c) \\
    ((V'_{id}, c_\pc, Y_c, \sigma_{k_c}), \_, (c_\sk, r, K^\pk, \sigma_{k_r}),
    \pcskipln \\
    \t (P^\pk, e, \sigma_p)) \gets (q_c, \_, q_r, e) \gets c 
        \< \text{\% Unpack Credential} \\
    c_\pk \gets \sig.\pubkey(c_\sk) 
        \< \text{\% Get Public Key} \\
    \sig.\verify(K^\pk, \sigma_{k_c},  
    V'_{id} \concat c_\pc \concat Y_c) \iseq \top 
        \< \text{\% Receipt Integrity Check 1}\\
    \sig.\verify(K^\pk, \sigma_{k_r}, c_\pk \concat \hash(e \concat r)) \iseq \top 
        \< \text{\% Receipt Integrity Check 2} \\
    \sig.\verify(P^\pk, \sigma_p, H(e)) \iseq \top 
        \< \text{\% Envelope Integrity Check} \\
    (C_1, C_2) \gets c_\pc;~X \gets C_2 / \verifytext{c_\pk} 
        \< \text{\% Derive ElGamal Secret} \\
    (Y_1, Y_2) \gets Y_c \< \text{\% Extract ZKP commitments} \\
    \label{activate:zkp:verifier} Y_1 \iseq g^r C_1^e;~Y_2 \iseq \A_\pk^r X^e 
        \< \text{\% Verify ZKP} \\
    (c_\pc', K^{'\pk}, \sigma_{k_{ot}}, O^\pk, \sigma_{o}) 
    \gets \L_R[V_{id}] 
        \< \text{\% Voter Reg. Session} \\
    c_\pc' \iseq c_\pc \land K^{'\pk} \iseq K^{\pk} \land V_{id}' \iseq V_{id} 
        \< \text{\% Verify Public Cred \& Actors} \\
    \label{activate:envelope:verifier} e \isnotin \L_E[\hash(e)];\t \L_E[\hash(e)] \gets e 
        \< \text{\% Challenge Unused \& Append}
}
\caption{\textbf{Credential Activation.} 
Verifies the integrity of the credential: 
if any procedure with fails, 
then the process aborts.
Upon success, the \VSD{} stores the credential's private key $c_\sk$.
}
\Description{Activation Procedure}
\label{fig:protocol:activate}
\end{figure}
During activation (\cref{fig:protocol:activate}),
the voter uses their \VSD to scan
a credential in the activate state~(\cref{fig:activate}).
This reveals the
commit $q_c$, envelope $e$, and response $q_r$;
the check-out ticket $t_{ot}$ is not visible.
\VSD then verifies the integrity of the credential by
(1) verifying the
    signatures
    ($\sigma_{k_c}, \sigma_{k_r}$, $\sigma_{p}$),
(2) deriving the ElGamal secret $X$ and
    verifying the ZKP,
(3) checking whether the public credential $c_\pc$ matches the
    public credential on the ledger $c_\pc'$, and
(4) checking via the ledger $\L_E$
    that the challenge $e$ has not already been used.
Upon success,
the device publishes the challenge $e$ on $\L_E$
and stores the credential $c_\sk$ for future voting.
The device publishes the envelope challenge
on the ledger for integrity,
ensuring that the challenges are unique.
Upon failure,
the \VSD{} reports the offending actor
that results from the failure check,
and instructs the voter to re-register.

We present the voting and tallying scheme
in~\apxref{apx:tallying-scheme}.
\section{Formal proofs}\label{apx:proofs}
\noindent
We formally prove that \name satisfies
coercion-resistance and individual verifiability,
and provide a proof sketch for the privacy
adversary.
Tables~\ref{tab:scheme:notations}
and~\ref{tab:proof:notations} show our notation
and summarize our variables.

\parhead{\name API.}
\begin{figure}[t]
\begin{mdframed}
\scriptsize
\begin{itemize}[leftmargin=*]
    \item $c_\sk^i, c_\pc^i, P^i, \L_R \gets
    \RealCred(\L_R, \R_\sk, \V^{id}_i, \lambda)$:
    takes as input
        the ledger $\L_R$,
        the registrars' private key $\R_\sk$,
        a voter's identifier $\V^{id}_i$ and
        a security parameter $\lambda$
    and outputs
        the voter's private and public credential $c_\sk^i$ and $c_\pc^i$,
        correctness proofs $P^i$
        and an updated registration ledger $\L_R$.
    For simplicity, \RealCred incorporates the
        \CheckIn and \CheckOut processes required in \name.
    \item $\mathbf{c}^i_f, \mathbf{P}^i \gets
    \FakeCreds(\R_\sk, \V^{id}_i, c_\pc^i, n_f, \lambda)$:
    takes as input
        the registrar's private key $R_\sk$,
        the voter's identifier $\V^{id}_i$,
        the voter's public credential $c_\pc^i$,
        the number of fake credentials $n_f \in \NN$, and
        a security parameter $\lambda$
    and outputs
        $n_f$ fake credentials $\mathbf{c}^i_f$ and
        $n_f$ proofs of correctness $\mathbf{P}^i$.
    \item $out, \L_R \gets
    \Activate(\L_R, c, P)$:
    takes as input
        the registration ledger $\L_R$,
        a credential $c$ and
        the credential's correctness proof $P$
    and outputs
        $out \in \set{\top, \bot}$ and
        an updated registration ledger $\L_R$.
    \item $\L_V \gets
    \Vote(\mathbf{c}_\sk, T_\pk, n_M, D_{n_U,n_C}, \lambda)$:
    takes as input
        the set of credentials $\mathbf{c}_\sk$,
        the talliers' public key $T_\pk$,
        the candidate list $n_M$, and
        a probability distribution $D_{n_H,n_M}$ over the possible (voter, candidate) pairs\footnote{
            \scriptsize
            This distribution captures the uncertainty
            of honest voters' choices, which impairs the
            adversary's ability to detect coercion.}.
    It appends a \name-formatted ballot for
    each credential in $\mathbf{c}_\sk$
    to the ledger $\L_V$.
    \item $(X,P) \gets
    \Tally(T_\sk, \L, n_M, \lambda)$:
    takes as input
        the talliers' private key $T_\sk$,
        the ledger $\L$,
        the candidates $n_M$, and
        a security parameter $\lambda$
    and outputs
        the tally $X$ and
        a proof $P$ showing that the talliers
        computed the tally correctly.
   \item $(X, P) \gets
   \IdealTally(T_\sk, \L, n_M, \lambda)$:
    takes as input
        the talliers' private key $T_\sk$,
        the ledger $\L$,
        the candidates $n_M$, and
        a security parameter $\lambda$
    and outputs
        an ideal-tally $X$
        and a proof $P$.
    This algorithm, as defined in JCJ and only used in the ideal game,
    differs from $\Tally$ by not counting any ballots cast by the adversary
    if the bit $b = 0$.
\end{itemize}
\end{mdframed}
\caption{\parhead{\name API.} }
\label{fig:system-api}
\end{figure}
We first redefine the \name API (\cref{fig:system-api}),
where algorithms append to a ledger instead of submitting to it.
Since the registrar is either all malicious (integrity,
privacy) or all honest for coercion, we denote $\R$
to represent the kiosks, registration officials and the
envelope printers.

\subsection{Coercion resistance}\label{apx:proofs:coercion-resistance}
\label{sec:securityCoercion}
\begin{table}[t]
\centering
\scriptsize
\begin{tabular}{ c p{6.5cm} }
    \toprule
    \emph{Notation} & \emph{Description} \\
    \midrule
    $\R$ &
    Registrar (combines kiosks, envelope printers and officials) \\
    \hline
    $X, P, M$ &
    Tally \& Tally Proofs, Voting options  \\
    \hline
    $D^c, D^v$ &
    Probability distribution of chosen \# credentials and votes \\
    \hline
    $\CA, \mathbf{V}_C, j, \beta$ &
    Coercer, $\CA$-controlled voters, $\CA$-target voter,
    $\CA$-intended ballot \\
    \hline
    $n^c_j, \st_{\CA}$ &
    $\CA$-target number of total credentials, $\CA$ (adversarial) state \\
    \hline
    $n_V, n_\CA, n_M$ &
    Number of voters, controlled voters and voting options \\
    \bottomrule
\end{tabular}
\vspace{3mm}
\caption{\parhead{Proof Notations.}
\label{tab:proof:notations}
}
\end{table}
\begin{figure}[t]
\gameblock[codesize=\scriptsize,,skipfirstln,jot=-0.5mm]{$\mathbf{Game}\ \cResistIdeal^{\CA', b} (\lambda, \V, \R, \A, M, n_\CA)$}{
    \\
    \mathbf{V}_\CA, \set{n_c^i}_{1 \leq i \leq |V_\CA|}, \st_{\CA'} \gets
    \CA'(\V, \text{``Choose controlled voter set''}) \\
    \L^{JCJ}, \L_E, n_V, n_M, n_\CA' \gets
    \emptyset{}, \emptyset{}, |\V|, |M|, |\mathbf{V}_\CA| \\
    \label{fig:resist-ideal:adv-selection}
    (j, \beta, \st_{\CA'}) \gets \CA'(\st_{\CA'},
    \text{``Pick target voter and their parameters''})\\
    \pcif n_\CA' \neq n_\CA~\text{or}~j \notin
    \{1,2,...,n_V\} \setminus \mathbf{V}_C \pcthen
    \pcabort \\
    \pcfor i=1~\mathbf{to}~n_V \pcdo ~\textit{\% All voters --- real credential}  \\
        \t c_\sk^i, c_\pc^i, \_, \L^{JCJ}_R \gets
        \mathsf{RealCred}(\L_R^{JCJ}, \R_\sk, \V_{id}^i, \lambda)\\ %
        \t \pcif i \in \V \setminus (\mathbf{V}_C \cup \{j\}) \pcdo
        ~\textit{\% Honest voters --- fake credentials} \\
            \t\t \mathbf{\fc}_\sk^i, \_ \gets \FakeCreds(\R_\sk, \V_{id}^i, c^i_\pc, D^c_{1, [0, \inf]}, \lambda) \\
            \t\t \_, \L_E \append
            \set{\Activate(\L_E, \fc^i_\sk)}_{
            1 \leq i \leq |\mathbf{\fc}|} \\
    \pcif b = 0 \pcthen
    ~\textit{\% Target voter evades --- casts real vote} \\
        \t \fc_\sk^j, \_ \gets
        \FakeCreds(\R_\sk, \V_{id}^j, c^j_\pc, 1, \lambda) \\
        \t \_, \L_E \append
        \set{\Activate(\L_E, c^j_\sk), \Activate(\L_E, \fc^j_\sk)} \\
        \t \L^{JCJ}_V \Leftarrow
        \Vote(c^j_\sk, \A_\pk, M, \beta, \lambda ) \\
    \label{line:c-resist-ideal:give-credential}
    c_{\CA'} \gets c_\sk^j
    ~\textit{\% Target voter always releases their real credential} \\
    \L^{JCJ}_V \Leftarrow
    \set{\Vote(c^i_\sk, \A_\pk, M, D^v_{n_V - n_C, n_M}, \lambda)}_{
    i \in \V \setminus \set{\mathbf{V}_\CA \cup j} } \\
    \L^{JCJ}_V \Leftarrow
    \CA'(\st_{\CA'}, \L_R^{JCJ}, c_{\CA'},
    \set{c^i_\sk}_{i \in \mathbf{V}_C},
    \lambda, \text{``Coercer casts ballots''} )\\
    \label{fig:c-resist-ideal:ideal-tally}
    (\mathbf{X}, \_) \gets
    \IdealTally(\A_\sk, \L^{JCJ}_V, M, \L^{JCJ}_R, \lambda) \\
    \label{fig:c-resist-ideal:give-ledger}
    b' \gets \CA'(\st_{\CA'}, \mathbf{X}, \L_E, \text{``Guess $b$''})\\
    \pcreturn b'= b
}
\caption{
\textbf{Game C-Resist-Ideal.}
{
The \name ideal game for coercion-resistance,
adapted from JCJ, that considers the
adversary's probabilistic knowledge of
honest voters' fake credentials.
Notations are defined in \cref{tab:scheme:notations,tab:proof:notations},
\name API in \cref{fig:system-api}.
}
}

\label{fig:c-resist-ideal}
\end{figure}
\begin{figure}[t]
\gameblock[codesize=\scriptsize,skipfirstln,jot=-1mm]
{$\mathbf{Game}\ \cResist^{\CA,b}(\lambda, \V, \R, \A, M, n_\CA )$}{
\\
\Vc, \set{n_f^i}_{i \in \Vc}, \st_\CA \gets \CA(\V,
  \text{``Choose controlled voter set''}) \\
\L, n_V, n_M, n_\CA' \gets \emptyset{}, |\V|, |M|, |\Vc| \\
(j, n_f^j, \beta, \st_\CA) \gets \CA(\st_\CA,
\text{``Pick target voter and their parameters''}  \\
\pcif n_\CA' \neq n_C\ \text{or}\ j \notin \{1,2,...,n_V\}
  \setminus \Vc \pcthen \pcabort \pcendif \\
\pcfor i = 1\ \mathbf{to}\ n_V \pcdo \\
    \t c_\sk^i, c_\pc^i, P_c^i, \L_R \gets
    \RealCred(\L_R, \R_\sk, V_i^{id}, \lambda) \\
    \t \pcif i = j\ \text{and}\ b = 0 \pcthen\
      \textit{\% Target voter evades coercion} \\
        \t\t \mathbf{\fc}^j_\sk, \mathbf{\Pfc}_c^j \gets
        \FakeCreds(\R_\sk, \V^j_{id}, c_\pc^j, n_f^j + 1, \lambda) \\
    \t \pcelseif i \in (\Vc \cup j) \pcthen\
      \textit{\% $\CA$-Controlled Voters; target voter submits} \\
        \t\t \mathbf{\fc}^i_\sk, \mathbf{\Pfc}_c^i \gets
          \FakeCreds(\R_\sk, \V^i_{id}, c_\pc^i, n_f^i, \lambda) \\
    \t \pcelse\ \text{\% \textit{Honest voters}} \\
        \t\t \mathbf{\fc}^i_\sk, \mathbf{\Pfc}_c^i \gets
          \FakeCreds(\R_\sk, \V^i_{id}, c_\pc^i, D^c_{1, [0, \inf]}, \lambda) \\
    \t \pcendif \\
    \t \mathbf{c}^i \gets (\mathbf{\fc}^i_\sk, \mathbf{\Pfc}_c^i);\
      \mathbf{c}^i \append (c_\sk^i, P_c^i) \\
\pcendfor \\
\mathbf{c}_\CA \gets \mathbf{c}^i\
  \textit{\% Target voter releases all credentials} \\
\pcif b = 0 \pcthen\ \textit{\% Target voter evades coercion} \\
    \t \_, \L_E \gets \Activate(\L_E, c_\sk^j) \\
    \t \L_V \append \Vote(c_\sk^j, \A_\pk, M, \beta, \lambda) \\
    \t \mathbf{c}_\CA \gets (\mathbf{\fc}^j, \mathbf{\Pfc}^j)\
      \textit{\% Target voter releases only fake credentials} \\
\pcendif \\
\pcfor i \in \V \setminus (\mathbf{V}_\CA \cup \{j\}) \pcdo\
  \textit{\% Honest voters cast vote} \\
    \t \_, \L_R \gets \set{\Activate(\L_R, c_\sk^i)}_{
      1 \leq i \leq |\mathbf{c}^i|} \\
    \t \L_V \Leftarrow
      \set{\Vote(c_\sk^i, \A_\pk, M, D^v_{n_V - n_\CA,
       n_M}, \lambda)}_{1 \leq i \leq |\mathbf{c}^i|} \\
\pcendfor \\
\L \Leftarrow \CA(\st_\CA, \L, \mathbf{c}_\CA,
  \set{\mathbf{c}^i}_{i \in \Vc},
  \lambda, \text{``Activate and cast votes''}) \\
\label{fig:c-resist:tally}
(\mathbf{X}, P_t) \gets \Tally(\A_\sk, \L, n_C, \lambda) \\
b' \gets \CA(\st_\CA, \mathbf{X}, P_t, \L,  \text{``guess $b$''}) \\
\pcreturn b' = b
}
\caption{
\textbf{Game C-Resist.}}
\label{fig:c-resist}
\end{figure}
\noindent
We prove that \name is coercion-resistant
by showing that the difference
between $\CA$'s winning probability
in a real game -- representing the adversary's interactions
with our system --
and in an ideal game --
representing the desired level of
coercion-resistance -- is negligible.
In both games, $\CA$'s goal is to determine whether
the targeted voter evaded coercion and cast a real vote.
Like the total number of votes cast in an election,
we treat the total number of credentials created as public information:
$\CA$ could trivially win if it knew exactly how many (fake) credentials
\emph{all other} voters created.
The adversary's uncertainty about the target voter thus derives from
the other honest voters,
each of whom creates an unknown (to the adversary $\CA$)
number of fake credentials,
which we model as a probability distribution  $D^c$.\footnote{
	In practice, to artificially increase this uncertainty,
	envelope printers can post challenges on the ledger
	\emph{without} printing a corresponding envelope and
	gradually release these values,
	similar to the JCJ option of voting authorities
	or third parties intentionally injecting fake votes
	to add noise.
}
To achieve statistical uncertainty also on the voting choice,
we adopt the same approach for the content of the ballot
with the distribution $D^v$.
This ``anonymity among the honest voters''
mimics the reasoning by which votes themselves
are considered to be (statistically) protected once anonymized.
We present the ideal game in \cref{fig:c-resist-ideal}
to highlight \name's level of coercion-resistance.

In this ideal game,
adapted from JCJ~\cite{juels2010CoercionResistantElections},
the coercer
chooses a target voter $j$ and
a set $\V_C$ of  $n_C < n_V$ controlled voters
who comply with the coercer's demands.
Every voter obtains their real credential,
and each voter may also
create and activate fake credentials.
The envelope ledger $\L_E$'s
(line~\ref{fig:c-resist-ideal:give-ledger})
discloses to $\CA$ only the aggregate number of
credentials created and activated.
A challenge bit $b \in \set{0, 1}$ determines
both the target voter's behavior and which ballot
under the target's real credential is ultimately
tallied by \IdealTally.
If $b=0$ (evasion), the target voter casts a ballot
using the real credential and the adversary also learns
this voter's real credential and may cast ballots
using it.
\IdealTally under $b=0$ then tallies the voter's ballot 
and discards any adversary ballots cast using the 
voter's real credential.
If $b=1$ (compliance), the target voter does not cast
a ballot using the real credential and \IdealTally
tallies any adversary ballots cast under the voter's
real credential.
Ballots from honest voters are sampled from $D^v$
and tallied normally in both cases.
Given the tally, released credentials to the adversary,
and the contents of the ledger, the adversary now guesses
the bit $b$, i.e., whether the target voter has 
cast a ballot.
The degree of coercion‑resistance is 
bounded by the adversary’s residual uncertainty 
about honest voters' fake‑credential behavior $D^c$ 
and voting behavior $D^v$.

We introduce the following three major changes
from JCJ games to model \name's behavior.

\parhead{Change \#1: Voter Registration Algorithms.}
To model information beyond the credential
(\eg proofs of correctness),
we replace JCJ algorithms
$\pcalgostyle{register}$
and $\pcalgostyle{fakekey}$
with $\RealCred$ and $\FakeCreds$,
respectively.
We also use $\Activate$ to
model additional registration data
on the ledger (\eg envelope challenges).
In our C-Resist games,
$\Activate$ consistently returns
$out$ as $\top$ since
the registrar $\R$ is trusted.

\parhead{Change \#2: Fake Credentials Issuance.}
In contrast to the JCJ game where $\CA$ has
no influence over the voter before the voting phase,
\name allows $\CA$ to interact with voters
before registration.
Specifically, $\CA$ can demand voters to generate
$n_f \in \NN$ fake credentials during registration.
We use the probability distribution $D^c$
to model the uncertainty around fake credentials
created and activated by honest voters.

The coercion adversary $\CA$ has no control on the
probability distribution $D^c$ and
we enforce this in the system through the
parameter $\lambda_E$. This
security parameter represents the minimum
number of envelopes that each booth requires
to ensure that coerced voters cannot accurately count
all the envelopes that a booth contains.
This system property---that is,
a minimum number of envelopes per booth---ensures
in practice that the distribution $D^c$ is out
of the coercion adversary's control, as we
assume for the rest of the proof.

\parhead{Change \#3: Ledger Entries.}
While JCJ uses a trusted registration algorithm
to generates a voting roster with each voter's
public credential $c_\pc$, \name incorporates
a more complex protocol to achieve individual
verifiability.
Specifically, during $\Activate$,
the envelope challenge is disclosed on the envelope ledger,
permitting the adversary to discern the total number
of fake credentials created.
Additionally, the registration ledger includes digital
signatures from both the kiosk and officials responsible
for issuing the voter's credential.
Despite these additions,
we show that $\CA$'s winning probability is negligibly
affected, outside of the added probability distribution
of fake credentials represented in the ideal game.
Similar to JCJ, the winning probability of the ideal game
is $\gg 0$ as coercion-resistance
is bounded by the adversary's uncertainty over the
behavior of honest voters.

We present our formal definition for coercion-resistance
under~\cref{def:coercion-resistance},
and use the ideal (\cResistIdeal) and real (\cResist)
games in figures~\ref{fig:c-resist-ideal} and~\ref{fig:c-resist},
respectively.
The coercer wins if they can correctly guess the bit $b$,
representing whether or not the targeted voter gives in to coercion.
While \cResist represents the coercive adversary $\CA$,
to prove security we must compare $\CA$ with another adversary $\CA'$
who plays \cResistIdeal, which embodies the security
we want to achieve against coercion.
We show that the difference between the real and ideal games
is negligible.

\begin{definition}[Coercion-resistance]
A scheme is coercion-resistant
if for all PPT adversaries $\CA$,
all security parameters $\lambda \in \NN$,
and all parameters $\V, \R, \A, M, n_\CA$,
the following holds:
\begin{align*}
    \advantage{\text{coer}}{\CA,\cResist}[(\lambda, \cdot)] = \\
    |\Pr_{\CA,b}[\cResist^{\CA,b}(\cdot)=1] & -\Pr_{\CA',b} [\cResistIdeal^{\CA',b}(\cdot)=1]| \\
    \leq \negl,
\end{align*}
where the probability is computed over all the random
coins used by the algorithms in the scheme.
\label{def:coercion-resistance}
\end{definition}

\begin{theorem}[Coercion-resistance]
The \name registration scheme
(within the JCJ remote electronic voting scheme\footnote{This proof includes voting
and tallying functions as defined in JCJ
to demonstrate the complete
process is coercion-resistant.} \cite{
juels2010CoercionResistantElections})
is coercion-resistant
under the decisional-Diffie-Hellman
assumption in the random oracle model.
\label{theorem:c-resist}
\end{theorem}

\begin{proof}
We use three hybrid games to transition from the real game to
the ideal game, with each game involving a protocol change.

\begin{enumerate}[leftmargin=*]
    \item \textbf{Eliminate Voting Ledger View:}
        Eliminate $\CA$'s access to the \name voting ledger $\L_V$,
        as $\CA$ is incapable of differentiating between a ledger
        filled with honest-voter ballots and a randomly-generated
        set of ballots, assuming the decisional-Diffie-Hellman
        assumption holds.
    \item \textbf{Number of Fake Credentials:}
        $\CA$'s ability to demand voters to create
        a specific number of fake credentials
        is equal to that of an adversary $\CA'$
        who \emph{cannot} demand voters to create
        a specific number of fake credentials,
        given the distribution of the honest voters'
        fake credentials.
    \item \textbf{Eliminate Registration Ledger View}:
        Eliminate $\CA$'s access to \name roster $\L_R$
        by introducing a new ledger $\L^{JCJ}_R$
        where we can decouple $V^{id}_i$ and
        $(c_\pc^i, K, \sigma_k, R, \sigma_r)$ via semantic security:
        $\L_R$: $(V^{id}_i, c_\pc^i, K, \sigma_k, R, \sigma_r)$ \&
        $\L^{JCJ}_R$: $(c_\pc^i)$
\end{enumerate}

\begin{hybrid}
\normalfont
\label{hybrid:c-resist:voting-ledger}

We replace
\Tally (\cref{fig:c-resist}, l.~\ref{fig:c-resist:tally})
with \IdealTally
(\cref{fig:c-resist-ideal}, l.~\ref{fig:c-resist-ideal:ideal-tally})
by proving that $\CA$ no longer has access to $\L_V$.
We use a simulation-based approach to show that
if an adversary with access to $\L_V$ has a non-negligible advantage
over an adversary who does not have access to $\L_V$,
then the decisional-Diffie-Hellman assumption is broken.
The simulator gets as input a tuple of group elements
that are either a Diffie-Hellman tuple or
uniformly random and must output a guess.
It interacts with $\CA$,
simulates the honest parties of the protocol,
and makes a guess based on its output.

The simulator then proceeds as follows:
\begin{enumerate}[leftmargin=*]
\item \textbf{Setup:}
    The simulator receives its 
    DDH challenge tuple $(g, X, Y, Z)$ 
    from an external challenger;
    they do not know the discrete logarithm 
    of any component.
    It sets the system's public key as $\pk = (g, h = X)$
    and broadcasts it.
\item \textbf{Coin Flip}:
    Flip the coin $b$.
\item \textbf{Adversarial Corruption}:
    The adversary chooses the set $\mathbf{V}_C$
    of controlled voters and a target voter $j$.
    For each controlled voter and the target voter,
    the adversary chooses the number of fake
    credentials $n_f$ to create.
    For the target voter, $\CA$ sets the target vote $\beta$.
    If the number of controlled voters is not equal to $n_\CA$,
    or $j$ is not an appropriate index then the simulator aborts.
\item \textbf{Registration}:
    For each voter $i$,
    the simulator runs the \name registration process while
    acting as the registrar.
    The simulator first issues each voter their real credential
    $c^i_{\sk}$ along with their public credential $c^i_{\pc}$.
    The simulator then continues with generating fake credentials
    for each group of voters.
    For each controlled voter $i$ in $\mathbf{V}_C$,
    the simulator issues $n_f^i$ fake credentials as
    specified by the adversary.
    For the target voter, if $b=0$,
    the simulator generates $n_f^i + 1$ fake credentials
    for the target voter;
    otherwise, it generates $n_f^i$ fake credentials.
    For each uncontrolled honest voter,
    the simulator creates fake credentials,
    for which the amount is sampled from
    a probability distribution that
    models the adversary's knowledge of the number of fake credentials
    that these voters intend to create  $D^c$.
    Finally, the simulator carries out check-out for all voters. %
\item \textbf{Credential Release:}
    The simulator gives $\CA$ the real and fake
    credentials of voters in $\mathbf{V}_C$.
    If $b=0$,
    the simulator gives $\CA$ the target voter's
    $n_f^j + 1$ fake credentials;
    otherwise, the simulator gives $\CA$ the voter's
    real credential and $n_f^j$ fake credentials.
\item \textbf{Honest Ballot Casting}:
    For each honest voter $i$,
    the simulator samples a random vote
    $\beta_i \sample D_{n_V - n_C,n_M}$ and
    posts a ballot for this vote on
    the voting ledger $\L_V$.
    The simulator forms the ballot using the
    input DDH tuple $(g, X, Y, Z)$ as follows.
    To compute the encryption of the public credential $c^i_{\pc}$,
    the simulator sets the ciphertext components as:
    $E_1 = (Y, Z \cdot c^i_{\pk})$.
    To compute the encryption of the vote $\beta_i$, the simulator must use an independent set of random values. 
    It can do this by creating a second, re-randomized challenge ciphertext, for instance by picking a random
    $s \in \mathbb{Z}_q$ and computing
    $E_2 = (Y \cdot g^s, Z \cdot X^s \cdot \beta_i)$.
    This way, if the input is Diffie-Hellman,
    the result is a valid ElGamal encryption,
    whereas if the input is random, 
    the ciphertext is a random tuple 
    and contains no information about the vote 
    or the credential used to cast it.
    The simulator must now provide the accompanying 
    NIZK proof that
    $E_1$ correctly encrypts a valid credential.
    Since the simulator does not know the randomness $b$ 
    (from $Y=g^b$) used to form $E_1$, it cannot generate
    the proof honestly.
    Instead, it leverages its control over the 
    random oracle $H$:
    \begin{itemize}
    \item For a given proof 
        (\eg a Schnorr-style proof of knowledge),
        the simulator chooses the final response $z$ 
        and challenge $c$ first, 
        picking them uniformly at random.
    \item It then uses the public verification equation for    
        the proof and works backwards to compute the
        commitment value $A$ that would make the transcript
        $(A, c, z)$ valid.
    \item Finally, the simulator programs the random oracle,
        defining that for the input $(A, E_1, statement)$, 
        the output of $H$ shall be $c$.
    \end{itemize}
    The resulting proof is computationally indistinguishable from a real one for any adversary that can only query, 
    but not control, the random oracle.
    The simulator follows the same procedure for all 
    other required proofs and signatures.

\item \textbf{Adversarial Ballot Posting:}
    The adversary now posts a set of ballots onto $\L$.
\item \textbf{Decryption of Ballots:}
    The simulator can now check the NIZK proofs and
    discard the ballots with incorrect proofs.
    Then, since the simulator plays the role of the honest talliers,
    it can decrypt the ballots to prepare for the tallying process.

\item \textbf{Tallying Simulation:}
    This step is carried out as in JCJ's simulator~\cite{
    juels2010CoercionResistantElections}.
    Namely, the simulator eliminates duplicates, mixes, removes fake votes
    and finally decrypts the remaining real votes:
    \begin{itemize}
    \item \textit{Duplicate elimination:}
    The simulator removes duplicates.
    \item \textit{Mixing:}
    To simulate the MixNets from the real tally protocol,
    the simulator outputs an equal-length list of random ciphertexts.
    \item \textit{Credential Validity:}
    The simulator now checks, for each ballot,
    whether it was cast with a valid credential.
    This check is possible
    since the simulator can
    decrypt all the entries in $\L_R$ and $\L_V$.
    \item \textit{Output final count:}
    The simulator can then use the
    decrypted values to compute the final tally and output it.
    \end{itemize}

\item \textbf{Output Guess}:
    The simulator uses $\CA$'s output to make its guess about
    whether the input was a DDH triplet or a random triplet.
\end{enumerate}
If we can show that $\CA$ has a non-negligible advantage
in the real game over hybrid 1, then this implies that
the simulator can break the DDH assumption.
The key to this argument lies in how the simulator
constructs the ballots in Step 6.
When the input is a DDH tuple, where
$X = g^a$, $Y = g^b$, and $Z = g^{ab}$ for unknown $a, b$,
the public key given to $\CA$ is $h = g^a$.
The ciphertext $E_1$ is $(g^b, g^{ab} \cdot c^i_{\pk})$.
This is a mathematically valid ElGamal encryption of
$c^i_{\pk}$ under the public key $h=g^a$, 
using $b$ as the randomness.
The same holds for $E_2$.
Thus, when the input is DDH, the view of $\CA$
corresponds to the actual game, and the
adversary receives the contents of $\L_V$.
However, if the input tuple is a random one, 
where $Z=g^c$ for a random $c$,
then the ciphertext $E_1$ is $(g^b, g^c \cdot c^i_{\pk})$. 
The second component,
$g^c \cdot c^i_{\pk}$, is a uniformly random 
group element from the adversary's
perspective, as $c$ is unknown and independent of $a$ and $b$.
In this scenario, 
the ciphertext perfectly conceals the credential,
making the adversary's view equivalent to denying them
access to the meaningful encrypted content of the ledger.
Hence,
if the adversary $\CA$ holds a 
significant advantage in distinguishing the
real game from the hybrid,
then the probability that the 
simulator correctly guesses whether its
input was DDH or random will also be significant, 
leading to a contradiction
of the DDH assumption.

\end{hybrid}

\begin{hybrid}
\normalfont
\label{hybrid:c-resist:fake-credentials}
In this Hybrid,
we demonstrate that the adversary's ability to
demand a specific number of fake credentials from voters is lost,
as compared to~\cref{hybrid:c-resist:voting-ledger}.
In \cResist
(\cref{fig:c-resist}, l. 16 and 20),
the voter gives their credentials to the adversary
while in \cResistIdeal,
the adversary always gets the voter's single real credential.
(\cref{fig:c-resist-ideal}, l.~\ref{line:c-resist-ideal:give-credential}).

First, we show that real and fake credentials are
indistinguishable:
both real and fake credentials contain a ZKP transcript,
although the ZKP transcript for the fake credentials
are simulated.
The zero-knowledge property of the proof system
implies indistinguishability.

Next,
from hybrid 1, $\CA$ does not get access to $\L_V$,
thus $\CA$ cannot use any real or fake credentials
to cast valid ballots, or see ballots cast with these.
As a result, since real and fake credentials are
indistinguishable, always giving the adversary the
real credential gives them no advantage since
the credentials
cannot influence the tally outcome with it.
Now the value of $b$ only determines whether or
not the target voter casts a ballot.

For the same reason,
$\CA$ can only use the $n_f$ fake credentials
for determining whether the target voter cast
a ballot or not and to influence the number
of envelope challenges on $\L_E$.

Yet,
the signing key pair $(\fc_{\sk},\fc_{\pk})$
corresponding to each fake credential
is sampled independently from the real pair $(c_\sk,c_\pk)$,
and these cannot be used by the adversary
or the target voter to cast a vote that counts.
Regarding the number of challenges on $\L_E$,
the only difference in the case
where the target voter resists coercion
is that they create an additional fake credential.
As a result,
the number of challenges will only differ by one.
To detect this, the adversary needs
to distinguish between the following distributions:
\begin{itemize}
    \item Number of envelope challenges when target complies:
      $n_C + n_T + n_H + n_f + D^{fake}_{n_H}$
    \item Number of envelope challenges when target resists:
      $n_C + n_T + n_H + n_f + D^{fake}_{n_H} + 1$
\end{itemize}
Since everything but $D^{fake}_{n_H}$ is known to the adversary,
this is equivalent to just distinguishing between $D^{fake}_{n_H}$
and $D^{fake}_{n_H}+1$,
so they get no advantage from requesting
a certain number of fake credentials.
Therefore,
the adversary does not gain an advantage
from specifying the number of fake credentials,
since these fake credentials will not help
them identify whether the voter gave them a
real credential.

\end{hybrid}

\begin{hybrid}
\normalfont
In this hybrid, we replace the \name roster $\L_R$ initialized
on the first line of Figure~\ref{fig:c-resist}
with the JCJ Roster $\L_R'$
from the Ideal Game in Figure~\ref{fig:c-resist-ideal}.

To prove that the advantage of the adversary is negligible between these hybrids,
we show that given a JCJ roster,
a simulator can output a \name roster that is indistinguishable from a real one.
This is possible due to the semantic security of ElGamal.

We describe the simulator: \\
\textbf{Input:} JCJ roster $\L_R'$, List of voter IDs $V_{id}$
\begin{enumerate}
    \item Create a kiosk key pair $(k,K)$ and a registrar key pair $(r,R)$.
    If there are multiple kiosks and registrars, these keys contain
    all the individual keys.
    \item Initialize $\L'_R$ to contain each $V_{id}$ in a different entry,
    along with a random timestamp $d$.
    \item Apply a random permutation to the JCJ roster.
    \item Append one entry $V_e$ of the JCJ roster to each entry of $\L_R$.
    \item For each entry, add the necessary signatures.
    First, use $k$ to simulate the kiosk signature
    $$\sigma_2=\sig.\sign_k(V_{id}^i||d||V_e)$$
    and append $K, \sigma_{k_2}$ to the entry.
    Then, use $r$ to simulate the registrar's check-out signature $$\sigma_r=\sig.\sign_r(V_{id}||d||V_e||\sigma_{k_2})$$
    and append $R,\sigma_r$ to the entry.
    \item Output $(K,R, \L'_R)$. Recall that
      we consider a single logical entity for all the
      registration actors.
\end{enumerate}

As stated earlier,
$L_R'$ is indistinguishable from a real \name roster for the same list of voters
due to the semantic security of ElGamal.
If we start with a real \name roster,
we could create intermediate rosters by swapping two encryptions at a time
and updating the digital signatures until we get a random permutation.
Each of these swaps will yield indistinguishable rosters
by the semantic security of ElGamal.
At the end, the distribution will be the same as that of our simulator.

Recall that the view of $\CA$ is the list of valid verification keys
for the kiosks and registrars, along with the \name roster.
Since we have shown that the additional elements contained in the \name
roster
but not in the JCJ roster can be simulated,
this means that the advantage of $\CA$ is negligible.

With this, we have reached the Ideal JCJ game
and have thus shown that the advantage of $\CA$ in the $\cResist$ game
is negligible over the advantage in $\cResistIdeal$.
\end{hybrid}

\end{proof}

\subsection{Privacy}\label{apx:security:privacy}
The privacy adversary seeks to uncover
a voter's vote by identifying
and decrypting the voter's real ballot.
See our threat model for the capabilities
of such an adversary (\cref{sec:threat-model:privacy}).
Unlike the coercion adversary,
this adversary cannot directly interact
with or influence voters;
it must achieve its objective solely
via electronic means.
A voter's ballot is electronically
accessible in three locations:
on the voter's device, on the voting sub-ledger $\L_V$,
and in the final tally.
In what follows, we informally argue that
the privacy adversary is unable to
decrypt the voter's real ballot in these
three locations.

\parhead{Voter's device.}
Per our threat model (\cref{sec:threat-model:privacy}),
this adversary cannot compromise the
device containing the voter's real credential.
Compromising devices that hold
fake credentials
offers no advantage because ballots cast
from these devices are not counted:
the adversary cannot deduce from them
the voter's real vote.

\parhead{Voting sub-ledger $\L_V$.}
The voter's real ballot
on the voting sub-ledger is encrypted, so
the adversary has access only to the real ballot's ciphertext.
The adversary can identify the voter's
real ballot on the ledger because it can
compromise the registrar, including the kiosk
that issued the target voter's credentials.
Decrypting this ballot, however, would require
the adversary to obtain the election authority's
private key, which is impossible since the
adversary can only compromise all but one
authority member.
Secret keys from compromised members
provide no information about the secret
key of the
member that remains uncompromised,
thus preventing the adversary from reconstructing
the election authority's collective private key.

\parhead{Final tally.}
In the final tally,
ballots are revealed in decrypted form,
but only through a mixing process that unlinks
ballots from voters.
While the adversary can acquire the
mixing permutation from all the compromised
authority members, it cannot obtain it from
the one member it cannot compromise.
Each authority member generates and keeps
their permutation secret, leaving no
information for the adversary to deduce
the permutation of the uncompromised member.

The privacy adversary is thus unable to achieve
its goal of revealing any voter's vote,
thereby preserving privacy.

\parhead{Excluded attack.}
As in the Swiss Post e-voting system~\cite[Section 18.2]{2021SwissPostProofs},
we exclude trivial attacks where all voters
vote for the same option, which would indicate
to the adversary what the targeted voter voted.
Also, we do not consider the adversarial strategy
to delete from the voting sub-ledger and final tally
all but one vote, to learn the targeted ballot from
the tally's result. We assume this attack to be readily detectable
and thus inapplicable to the privacy adversary.

\subsection{Individual verifiability}
\label{property:iv}
\label{apx:proofs:verifiability}
In this section,
we provide a formal security analysis of the 
individual verifiability (IV) property for \name.
Informally, IV ensures that an honest voter can
detect any adversarial attempt to tamper with either
(1) the credential produced during registration or (2)
the ballot that will eventually be tallied.
\Cref{property:iv:threatmodel} revisits the
threat model tailored to individual
verifiability.
\Cref{property:iv:definition} formalizes
the IV security game.
\Cref{property:iv:instantiation} maps \name's
procedures to the formal model.
\Cref{property:iv:proof} states the main security
theorem followed by its proof.
\Cref{property:iv:implications} presents the
implications of the adversary's advantage.
\Cref{property:iiv} presents our new security
definitions, iterative individual verifiability,
and strong iterative individual verifiability,
which take into account the adversary's campaign
of targeting more than one voter to 
alter the election outcome.
We also show that \name satisfies strong 
iterative individual verifiability.

\subsubsection{Threat model}
\label{property:iv:threatmodel}
The adversary's objective is to cause the final
registration or ballot associated with an honest
voter on the ledger to be inconsistent with that
voter's actions, without this inconsistency
being detected by the voter or their \VSD.

The honest parties are the voter and their
personal device, \VSD, which follow the
protocol exactly.
This includes the voter following the
procedural steps required during in-person
registration, such as completing the correct
interactive sequence of the $\Sigma$-protocol.

The integrity adversary $\IA$ is computationally
bounded and controls every actor other than the 
voter and their \VSD.
This includes other voters, registration kiosks, printers,
officials, the election authority and
the public ledger.
In essence, $\IA$ has complete, active
control over the entire election
infrastructure, and consequently
possesses all long-term secret keys.
The adversary, however, cannot interfere
with the physically printed envelopes 
once the voter has entered the booth for
their registration session without being
detected.

The public ledger $\L$ is modeled as an ideal,
append-only functionality accessible to all
parties.
The adversary $\IA$ may append arbitrary entries but cannot
delete, modify, or reorder existing entries.

\subsubsection{Formal definition.}
\label{property:iv:definition}
We now present our formal definition,
building on prior
work such as Swiss Post E-Voting System~\cite{2021SwissPostProofs},
and CH-Vote~\cite{bernhard2018CHVoteProofs}.
We first define the syntax of a generic voting
scheme $\Pi$ --- an election scheme without
tallying --- and then use it to construct our security
game, $\IndVer$.

A voting scheme $\Pi$ is a tuple of four algorithms: 
\begin{itemize}[leftmargin=*]
    \item $\Setup_\IA(\params) \rightarrow (\pp, \L, \V, \st_{\IA})$:
        An algorithm run by the adversary $\IA$ that        %
        takes as input the game's public parameters $\params$ which are assumed to
        be honestly generated and hardcoded into the game.
        It outputs scheme-specific
        public parameters $\pp$, the initial ledger state $\L$,
        the list of the eligible voters $\V$ and the adversary's
        state $\st_{\IA}$ containing any secret keys (like the 
        authority's secret key $\A_{\sk}$ to determine the 
        game's outcome).
    \item $\VoterInteraction(V \leftrightarrow \infra)$: 
        An interactive protocol between the honest voter's
        agency (represented by an oracle, \emph{V}) and 
        the malicious infrastructure $\infra$ controlled
        by the adversary $\IA$.
        This protocol encapsulates
        registration and voting, 
        resulting in a voter receipt $\pi$,
        and a possibly modified ledger $\L$.
    \item $\VoterVerify(\params, \pp, \L, \pi) \rightarrow \{\top, \bot\}$:
        A deterministic algorithm executed on \VSD.
        It takes as input the public parameters \params,
        the scheme-specific parameters $\pp$,
        the final public ledger $\L$ to be used for tallying,
        and the voter's receipt $\pi$.
        It outputs $\top$ (accept) if the voter's interactions
        are correctly reflected on the ledger, and $\bot$ (reject)
        otherwise.
    \item $\Extract(\params, \pp, \L, \st_{\IA}, V_{id}) \rightarrow \goal$:
        A deterministic algorithm used only inside the security game
        that takes as input the public parameters \params, 
        scheme-specific parameters $\pp$,
        the final ledger $\L$, the adversary's state $\st_{\IA}$ 
        and a voter identifier $V_{id}$.
        It outputs the recorded information
        associated with that voter $V_{id}$ on the ledger or $\bot$ 
        if no valid record exists.
\end{itemize}

\begin{figure}[t]
    \centering
    \gameblock[bodylinesep=1mm]
    {$\mathbf{Game}~\IndVer_{\Pi, \IA}(\params)$}{%
        (\pp, \L, \V, \st_{\IA}) \gets \pcalgostyle{\Pi.Setup}_\IA(\params) \\
        (V_{\text{id}}^\star, \intent, \goal) \gets \IA(\text{Choose Target \& Goal})\\
        (\pi, \L) \gets \IA^{V(V_{\text{id}}^\star, \intent)}(\L) \\
        \happy \gets \pcalgostyle{\Pi.VoterVerify}(\params, \pp, \L, \pi) \\
        \advoutcome \gets (\Extract(\params, \pp, \L, \st_{\IA}, V_{\text{id}}^\star) = \goal) \\
        \pcreturn 1~\pcif (\happy \land \advoutcome \land (\goal \neq \intent))~\pcelse 0
    }
    \caption{\textbf{Individual Verifiability (Ind-Ver).}
    An adversary wins if it can make the ledger reflect its malicious \goal for an honest voter, 
    without the voter's verification procedures detecting the manipulation.}
    \label{fig:definition:ind-ver}
\end{figure}

We formalize individual verifiability using the
security game $\text{Game}~\IndVer$,
shown in~\cref{fig:definition:ind-ver}.
The game begins with the adversary $\IA$ running $\Setup$.
Following this, $\IA$ chooses a target voter $V_{id}^\star$, 
an honest \intent for that voter, and a malicious
\goal it wishes to have recorded on the ledger,
where $\goal \neq \intent$.

The core of the game is the interaction phase,
where $\IA$ is given oracle access to 
$V(V_{id}^\star, \intent)$ which perfectly
simulates the honest voter's actions.
At the end of the interaction,
$\IA$ outputs the voter's receipt $\pi$ and the
final ledger $\L$.

The adversary wins if its malicious \goal is recorded
on the ledger (\Extract returns \goal) \emph{and}
the voter's local verification on \VSD passes
(\VoterVerify returns $\top$).
We define the advantage of an adversary $\IA$ against 
individual verifiability of scheme $\Pi$ as:
$$Adv_{\Pi, \IA}^{\IndVer} = 
\Pr[\text{Game}~\IndVer_{\Pi, I} = 1]$$

\subsubsection{Protocol instantiation.}
\label{property:iv:instantiation}

We now map the abstract syntax to the concrete
procedures of $\Pi_{\name}$.

The global public parameters are $\params = (G, q, g, \; H)$,
where $(G, q, g)$ is a cyclic group of prime order $q$ where
the DDH problem is hard, and $H \colon \{0,1\}^\ast \to \mathbb Z_q$
is a cryptographically secure hash function. These are assumed to be honestly set
to prevent the adversary from winning vacuously.

\Setup is executed by $\IA$, taking $\params$ as input.
$\IA$ may generate arbitrary keys for the election authority,
officials, kiosks, and printers, and may populate the ledger $\L$ 
and voter list $\V$ in any way. It outputs the scheme-specific
parameters $\pp$ which would include the public keys of all
parties.

The $\VoterInteraction$ abstract protocol is instantiated 
by two concrete interactions: the in-person 
\name registration procedure (\cref{fig:scheme}, lines 1-9),
and the subsequent ``at-home'' \Vote procedure (\apxref{fig:scheme:vote}).
The adversary $\IA$ controls the infrastructure side
of both interactions.
For the registration procedure,
the voter --- not their \VSD --- performs the interactive
choices, such as selecting the number of credentials $n_c$ 
(line~\ref{fig:scheme:line:voter-pick-envelopes}), and the final
credential for check-out $c_v$ (line~\ref{fig:scheme:line:voter-pick-checkout}).
The pick of $n_c$ envelopes is a sequential process,
where the voter picks one envelope for $\RealCred$ and then
$n_c - 1$ envelopes for $\FakeCred$.
The voter aborts the game, returning $0$, if the adversary deviates
from the $\Sigma$-protocol during $\RealCred$.
For the voting procedure,
\VSD also keeps the computed ballot $B$ in memory
for the $\VoterVerify$ procedure.

The voter receipt $\pi$ corresponds to the full set of
data contained in the $n_c$ physical paper credentials
(QR codes) the voter takes home after registration.

The abstract $\VoterVerify(\params, \pp, \L, \pi)$ protocol
is instantiated as a procedure run on the voter's trusted \VSD,
taking the final public ledger $\L$ to be tallied
and the registration receipt $\pi$ as input.
It returns $\top$ if and only if both of the following checks
succeed:
\begin{itemize}[leftmargin=*]
    \item \textbf{Registration verification}:
        \VSD runs the \Activate procedure for
        every credential contained in the receipt $\pi$
        (\cref{fig:scheme}, lines 10-11).
        This includes verifying the correctness of the
        zero-knowledge proof constructed during $\RealCred$,
        and the first-time redemption of the envelope's 
        nonce against $\L_E$.
        This check succeeds only if this loop completes
        for all credentials created without aborting.
    \item \textbf{Vote verification}:
        \VSD retrieves the locally stored ballot $B_{\local}$ that
        it cast during the \Vote procedure.
        It queries the ledger $\L$ for the ballot associated with the
        voter's credential $c_\pk$. Let this be $B_{\posted}$.
        This check succeeds only if the ballot posted on the ledger
        $B_{\posted}$ is bit-for-bit identical to $B_{\local}$.
        If multiple ballots are posted for $c_\pk$ it retrieves
        the one that will be tallied --- based on some policy,
        which for this instantiation, we set as the last ballot cast counts.
\end{itemize}

The game's $\Extract$ algorithm determines the final outcome
for a voter as recorded on the ledger.
It first parses the registration ledger $\L_R[V_{id}]$
to find the public credential $c_\pc$.
It then decrypts $c_\pc$ using the authority key $\A_\sk$
from $\st_{\IA}$ to recover the associated credential public
key, $c'_{\pk}$.
It continues by finding the ballot that counts (\ie last ballot cast)
$B_{\posted}$ on the voting ledger $\L_V$ associated with
$c'_{\pk}$.
Finally, it decrypts $B_{\posted}$ using $\A_{\sk}$ to recover
the vote message $m'$.
If any of the steps fail, it returns $\bot$. Otherwise, it outputs
the tuple $(c'_\pk, m')$.
The \intent and \goal are thus defined as tuples of the voter's
credential public key and the choice of vote.

\subsubsection{Security Theorem.}
\label{property:iv:theorem}

We now state our main theorem regarding the individual verifiability 
of our protocol.

\begin{theorem}
    Let $\Pi_{\name}$ be instantiated as described 
    in~\cref{property:iv:instantiation}.
    If the underlying $\Sigma$-protocol is computationally sound
    (which relies on the hardness of the Discrete Logarithm Problem),
    and the ElGamal scheme is perfectly binding,
    then, for any PPT adversary $\IA$ playing $\text{Game}~\IndVer$,
    its advantage $Adv_{\Pi_{\name}, \IA}^{\IndVer}(\lambda)$ is bounded by:
    $$Adv_{\Pi_{\name}, \IA}^{\IndVer}(\lambda) \leq \max_{1\leq k \leq n_E} 
        E_{n_c \sim D_C}\left[ \frac{k}{n_E} \cdot \frac{\binom{n_E - k}{n_c-1}}{\binom{n_E - 1}{n_c-1}} \right] + \negl$$
    where $n_E$ is the total number of physical envelopes present in
    the booth during the voter's registration,
    $k$ is the number of those envelopes maliciously prepared by $\IA$,
    $n_c$ is the number of credentials the voter chose to create,
    and $D_C$ is the distribution of the voter's choice for $n_c$.
    \label{theorem:ind-ver}
\end{theorem}

\parhead{Proof.}
\label{property:iv:proof}
Let \Win be the event that $\IA$ wins the Game~\IndVer.
In other words, \Win means that \VoterVerify returns $\top$
and \Extract returns the adversary's \goal.
The adversary can win by tampering with registration or
by tampering with the vote.
Let \RegSub be the event that the adversary's \goal for
the credential key, $\goal.c_{\pk}$, differs from the honest
intent, $\intent.c_{\pk}$: $\goal.c_{\pk} \neq \intent.c_{\pk}$.
Let \VoteSub be the event that $\goal.m \neq \intent.m$ where
$m$ represents the vote.

For the adversary to win, the \goal must differ from \intent,
which means either \RegSub or \VoteSub (or even both)
must occur.
Therefore, the \Win event is a sub-event of $(\RegSub \lor \VoteSub)$,
allowing us to partition the \Win event into two disjoint cases
based on whether registration was tampered.
By the law of total probability, we get:
\begin{align*}
\Pr[\Win] &= \Pr[\Win \land (\RegSub \lor \VoteSub)] \\
&= \Pr[\Win \land \RegSub] \\
&\quad + \Pr[\Win \land \neg\RegSub \land \VoteSub]
\end{align*}
We bound each term in a separate lemma,
\cref{thm:ind-ver:reg} and \cref{thm:ind-ver:vote},
respectively.

\begin{lemma}
    The probability of adversary $\IA$ winning \text{Game}~\IndVer
    via registration tampering is bounded by:
    \begin{multline*}
        \Pr[\Win \land \RegSub] \leq \\
        \max_{1\leq k \leq n_E} E_{n_c \sim D_C}
        \left[ \frac{k}{n_E} \cdot 
        \frac{\binom{n_E - k}{n_c-1}}{\binom{n_E - 1}{n_c-1}} \right] + \negl
    \end{multline*}
    \label{thm:ind-ver:reg}
\end{lemma}

\parhead{Proof.}
The proof proceeds via a game hop argument.
Let $\Win_i$ be the event that the adversary wins $\text{Hybrid}~i$.

\setcounter{hybrid}{-1}
\begin{hybrid}
\label{thm:ind-ver:regsub:real}
\normalfont

This is the real Game~\IndVer.
The adversary $\IA$ interacts with the game
as defined in the \name protocol.
The adversary's success probability is
$\Pr[\Win_0 \land \RegSub]$.
\end{hybrid}

\begin{hybrid}
\label{thm:ind-ver:regsub:binding}
\normalfont

This hybrid is identical to \cref{thm:ind-ver:regsub:real} 
except that it aborts if the adversary $\IA$ 
produces two different plaintexts (\eg credential public
keys) that are valid openings to the same ElGamal ciphertext
$c_\pc$.
Since ElGamal is perfectly binding, this event is
information-theoretically impossible.
Thus,
$$|\Pr[\Win_0 \land \RegSub] - \Pr[\Win_1 \land \RegSub]| = 0$$
\end{hybrid}

\begin{hybrid}
\label{thm:ind-ver:regsub:zkp}
\normalfont

This hybrid is identical to \cref{thm:ind-ver:regsub:binding}
except we replace the $\Sigma$-protocol with an idealized proof system that aborts if $\IA$ tries to produce a proof for a false statement.
The statement being proven in $\RealCred$ is knowledge of the
randomness $x$ used to create $c_\pc$ as the encryption of
$c_\pk$.
$$\pcalgostyle{ZKPoE}_{C_1,X}\{(x):
    C_1 = g^{x} \land X = \A_\pk^{x}\}$$
where $c_\pc = (C_1, C_2)$ and $c_\pk = \frac{C_2}{X}$.
An adversary attempting to tamper with registration wants 
to get on the ledger $c_\pc$ that encrypts some other $c_\pk'$ ---
known only to $\IA$ --- instead of the voter's intended public
key $c_\pk$.
This is a false statement, so the probability that the adversary wins in Hybrid 2 is:
\[ 
    \Pr[\Win_2 \wedge \RegSub] = 0 
\]
We now wish to bound the advantage of $\IA$ between Hybrid 1 and Hybrid 2.
To do this, we must adapt the soundness analysis of the Schnorr $\Sigma$-protocol to our setting where adversarially chosen envelopes are used for the challenge instead of a uniformly random group element.
The adversary can break soundness by either guessing the envelope challenge or finding the corresponding discrete logarithm that allows them to produce the desired ciphertext.
This second case is oblivious to how the challenge is chosen, and puts us back in the standard computational soundness analysis of the Schnorr $\Sigma$-protocol.
By reduction to the hardness of the Discrete Logarithm Problem,
the probability of generating a valid proof in this case is negligible.

Now we must bound the probability that $\IA$ is able 
to guess the challenge $e$ that will be selected by the voter.
Recall that our threat model allows $\IA$ to adversarially select
the set of envelopes available to the voter.
We first show that the best strategy that $\IA$ can adopt is
to have a malicious set of credentials $\mathcal{M}$ that 
contains the same repeated nonce, and an honest set $\mathcal{H}$ 
consisting of envelopes with a unique nonce.

Consider an arbitrary strategy where the adversary prepares $n_E$ envelopes consisting of a malicious set $\mathcal{M}$ and an honest set $\mathcal{H}$.
We partition $\mathcal{M}$ into subsets $\mathcal{M}_1, \cdots \mathcal{M}_k$ where $\mathcal{M}_i$ contains various envelopes with the same nonce $e_i^*$.
Now, when guessing the nonce the voter will choose, let $p_i$ be the probability that the adversary guesses $e_i^*$.
Then, let $P=\sum p_i$, so the probability that the adversary tries to guess an envelope from $\mathcal{H}$ is $1-P$.
The adversary wins if it correctly guesses the envelope picked by the voter for their real credential \textit{and} the voter only picks envelopes with unique nonces.
We show that this probability is maximal when all the envelopes in $\mathcal{M}$ have the same repeated nonce.

Now assume the voter creates $n_c$ total credentials.
We can upper bound the probability that the adversary correctly guesses the voter's real credential without getting caught as follows.
Let $\EnvGuess$ be the event where the adversary correctly guesses the nonce and let $\DupCatch$ be the event where the voter finds duplicate nonces.
First, we lower bound the probability that the adversary gets caught.
Note that for all $\mathcal{M}_i$, this probability is at least the probability that it gets caught on a specific $\mathcal{M}_i$.
This tells us that
\begin{multline*}
    \Pr[\DupCatch \mid \EnvGuess] \geq \\ \max_{i} \Pr[\IA \text{ caught on } \mathcal{M}_i \mid \EnvGuess],
\end{multline*}
which implies that
\begin{multline*}
    \Pr[\neg \DupCatch \mid \EnvGuess] \leq \\ 1-\max_{i} \Pr[\IA \text{ caught on } \mathcal{M}_i \mid \EnvGuess].
\end{multline*}

Then, the probability of success is
$$\Pr[\neg \DupCatch \wedge \EnvGuess]$$
$$\leq (1-\max_{i} \Pr[\IA \text{ caught on } \mathcal{M}_i\mid \EnvGuess]) \cdot \Pr[\EnvGuess]$$
By Union Bound:
$$\leq (1-\max_{i} \Pr[\IA \text{ caught on } \mathcal{M}_i \mid \EnvGuess]) \cdot \left( \sum_{i=1}^k p_i \frac{|\mathcal{M}_i|}{n_E} \right)$$
$$\leq (1-\max_{i} \Pr[\IA \text{ caught on } \mathcal{M}_i \mid \EnvGuess]) \cdot \left( \sum_{i=1}^k \frac{|\mathcal{M}_i|}{n_E} \right)$$
$$\leq (1-\max_{i} \Pr[\IA \text{ caught on } \mathcal{M}_i \mid \EnvGuess]) \cdot \frac{|\mathcal{M}|}{n_E}.$$
Now, note that if $\mathcal{M}$ consists of envelopes with a single repeated nonce, then this actually becomes an equality. 
We can therefore conclude that an adversary gains nothing from using various repeated nonces.

Now, we analyze the optimal strategy, which is:

\begin{enumerate}[leftmargin=*]
    \item $\IA$ prepares $n_E$ envelopes composed of
    two sets:
    a malicious set $\mathcal{M}$ of $k$ envelopes containing
    the same, pre-determined nonce $e^\star$,
    and an honest set $\mathcal{H}$ of $n_E - k$ envelopes
    containing unique nonces.
    \item During the \RealCred interaction, $\IA$ must commit
    to $Y_c$ \emph{before} seeing the voter's challenge $e$.
    Therefore, to prove a false statement that $c_\pc$ encrypts
    $\intent.c_\pk$ when it actually encrypts $\goal.c_\pk$,
    $\IA$ must guess the challenge $e$ that the voter will choose.
    Thus, $\IA$ hopes that the voter chooses an envelope from
    $\mathcal{M}$ (containing $e^\star$) for this step.
    \item The \VoterVerify procedure checks for reused nonces
    across all envelopes posted on the ledger $\L_E$.
    Therefore, to pass this check, $\IA$ must hope that the
    voter during $\FakeCred$ picks only envelopes from the
    honest set $\mathcal{H}$.
\end{enumerate}

The adversary wins if the voter chooses an envelope from
$\mathcal{M}$ for $\RealCred$ --- probability of $\frac{k}{n_E}$
\emph{and} chooses the subsequent $n_c-1$ envelopes for
$\FakeCred$ entirely from $\mathcal{H}$.
Given the first choice was from $\mathcal{M}$,
there are $n_E-1$ envelopes left, of which $n_E-k$
are in $\mathcal{H}$.
The probability of the second event is then:
$$\frac{\binom{n_E-k}{n_c-1}}{\binom{n_E-1}{n_c-1}}.$$

Given that these are two distinct and ordered probabilistic events,
the total advantage of the adversary in Hybrid 1 over Hybrid 2 (and thus its probability of success),
maximizing over $\IA$'s choice of $k$ and averaging over the voter's
choice of $n_c$, is:
\begin{multline*}
    \Pr[\Win_1 \land \RegSub] = \\ \text{max}_{1 \leq k \leq n_E} E_{n_c \sim D_C} 
    \left[ \frac{k}{n_E} \cdot 
        \frac{\binom{n_E - k}{n_c-1}}{\binom{n_E - 1}{n_c-1}} \right] + \negl[].
\end{multline*}

This probability represents the adversary's optimal strategy in a game of uncertainty.
The adversary must choose the number of malicious envelopes, $k$, without knowing how
many credentials, $n_c$, the voter will choose to create.
A greedy strategy of setting $k = n_c$ (all envelopes being malicious) would guarantee
a win if the adversary knew that $n_c = 1$, as the nonce-reuse check in \VoterVerify
would never detect duplicates assuming the adversary issues new envelopes after each
registration session.
However, this strategy fails if $n_c > 1$, as the \VSD would immediately
detect a nonce-reuse on the activation of the second credential.
Since $\IA$ does not know $n_c$, it must balance the potential reward of subverting
the \RealCred interaction against the risk of detection during \FakeCred procedures.
This formula captures the maximum expected success for $\IA$, averaged over the
voter's behavioral distribution $D_C$.
\end{hybrid}

Summing this statistical term with the negligible terms from the 
game hops concludes the proof of this lemma.
\qed

\begin{lemma}
    Given that registration was not subverted (indicated by $\neg \RegSub$),
    the probability of an adversary winning via vote manipulation
    is $0$, i.e., 
    $\Pr[\Win \land \neg \RegSub \land \VoteSub] = 0$
    
    \label{thm:ind-ver:vote}
\end{lemma}
\parhead{Proof.}
We proceed by direct argument.
The non-subversion condition $\neg \RegSub$ implies that the public credential
$c_\pc$ on the ledger correctly encrypts the honest credential
key $c_\pk$ that the voter possesses.
The adversary $\IA$, however, knows the credential's secret key
as it generated the key pair $(c_\sk, c_\pk)$ on the kiosk during
$\RealCred$.
The adversary's goal is to have the final tallied vote be
$\goal.m$ while the voter's intent was $\intent.m$, without
the \VoterVerify check failing.
We consider all possible strategies for $\IA$.

The first strategy is to submit a ballot on the ledger
given $\IA$ knowledge of $c_\sk$.
$\IA$ constructs and signs a ballot $B_{\IA}$ for the
vote $\goal.m$.
$\IA$ then sends this ballot to ledger $\L_V$ such that it
will be the one extracted for the tally.
The game requires $\IA$ to output the final ledger $\L$
\emph{before} \VoterVerify is run, as a means of mimicking
continuous checking by \VSD.
If $\IA$ posts only its ballot $B_{\IA}$,
the voter's \VSD will detect a ballot $B$ associated with
$c_\pk$ on $L_V$ when no vote was cast and fail.
If the voter's ballot is posted $B_V$, and then $\IA$
posts $B_{\IA}$, \VoterVerify will compare the last ballot
$B_{\IA}$ with its locally stored ballot $B_V$.
Since $B_V \neq B_{\IA}$, the check will fail.
If $\IA$ posts $B_{\IA}$ and then the voter's ballot is posted
$B_V$, the last ballot will be $B_V$, and the \Extract function
will find $\intent.m$, not the adversary's $\goal.m$.
Therefore, in all cases $\IA$ does not win.

The other strategy is to modify the decryption of the ballot.
$\IA$ intercepts or appends a new ballot based on
the voter's honestly generated ballot $B_V$
(containing an encryption of $\intent.m$) and attempts to
modify it into $B_{\IA}$ such that $B_{\IA}$ decrypts to $\goal.m$.
For \VoterVerify to pass, $B_\IA$ must be bit-for-bit 
identical to the original $B_V$ that the \VSD stored locally.
However, since the ElGamal encryption is perfectly binding,
$\IA$ cannot get the ciphertext to decrypt to anything but
the voter's plaintext $\intent.m$.
It is information-theoretically impossible to find a different
message $\goal.m$ that is a valid opening of the same ciphertext.
Therefore, any modification that changes the underlying vote
will necessarily change the ciphertext, causing the bit-for-bit
check in \VoterVerify to fail.
The probability of success for this strategy is $0$.

In all scenarios, the combination of the \VSD's bit-for-bit check
and the perfect binding property of the encryption scheme ensures
that any attempt by $\IA$ to change the vote after a successful
registration will be detected. The probability of success is therefore
$0$. \qed

We now combine the results of \Cref{thm:ind-ver:reg} and \Cref{thm:ind-ver:vote}
to complete the proof.
The total advantage of the adversary $\IA$ is the sum of its advantages in
the two disjoint scenarios we analyzed:

\begin{align*}
    \text{Adv}_{\Pi_{\name}, \IA}^{\IndVer} &= \Pr[\Win] \\
    &= \Pr[\Win \land \RegSub] \\
    &\qquad\qquad + \Pr[\Win \land \neg \RegSub \land \VoteSub] \\
    &\leq \left( \max_{1\leq k \leq n_E} \mathbb{E}_{n_c \sim D_C}\left[ \frac{k}{n_E} \cdot \frac{\binom{n_E - k}{n_c-1}}{\binom{n_E - 1}{n_c-1}} \right] + \negl \right) %
\end{align*}

This concludes the proof of the theorem. \qed

\subsubsection{Implications of the security bound}\label{apx:security:envelope_analysis}
\label{property:iv:implications}
\begin{figure}[t]
    \centering
    \begin{subfigure}[t]{.49\linewidth}
        \centering
        \includegraphics[width=\linewidth]{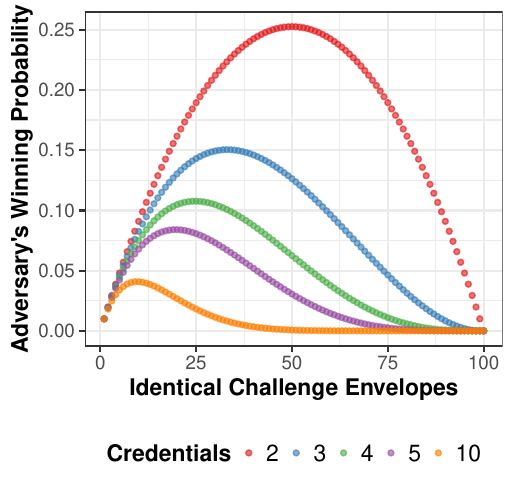}
        \caption{Adversary's Winning Probability based on the
        $k$ identical envelopes \& $n_c$ voter-chosen credentials}
        \label{fig:envelope_analysis_envelopes}
    \end{subfigure}
    \hfill
    \begin{subfigure}[t]{.49\linewidth}
        \centering
        \includegraphics[width=\linewidth]{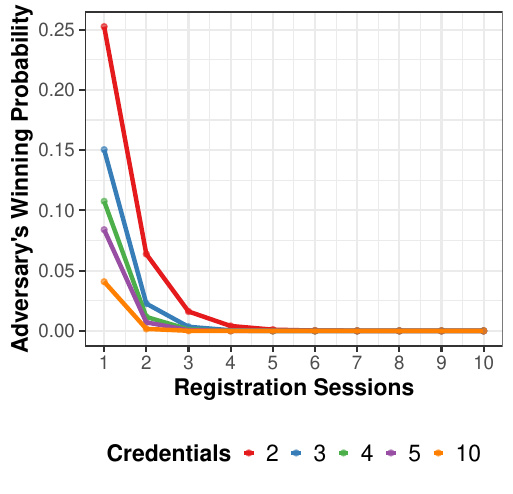}
        \caption{Adversary's Winning Probability across Voter Registration Sessions \& Number of Credentials Voter
        Creates.}
        \label{fig:envelope_analysis_voters}
    \end{subfigure}
    \caption{\textbf{Adversary's probability of winning for 100 dnvelopes.}}
    \label{fig:envelope_analysis}
\end{figure} 
Beyond the proof, we can build intuition for the adversary's
security bound by analyzing their strategic options.
To subvert a voter's registration 
without being cryptographically detected, 
the adversary must forge a zero-knowledge proof
during the $\RealCred$ procedure.
This requires knowing the voter's ZKP challenge \emph{before}
committing to the proof, as done for $\FakeCred$.
The adversary's only viable strategy to achieve this is
to ``poison'' the deck of envelopes in such a way that they
can predict which challenge the voter will select.

\parhead{The adversary's dilemma.}
The core of the adversary's strategy is to place multiple envelopes
containing the same, pre-selected nonce $e^\star$ in the booth:
the malicious set $\mathcal{M}$.
Their hope is that the voter, when creating their real credential,
will pick up one of these malicious envelopes.
This strategy, however, presents a fundamental dilemma, a trade-off
between maximizing the chance of success and minimizing the risk
of immediate detection.
This is precisely the dynamic captured by the security bound
in \cref{theorem:ind-ver} and visualized in \Cref{fig:envelope_analysis}.

Consider the adversary's choice of how many malicious envelopes, $k$,
to place in the booth our of a total of $n_E$.
As an adversary increases $k$, the probability that the voter selects
a malicious envelope for their real credential, $\Pr[e^\star \in \mathcal{M}]$,
increases linearly $k/n_E$. This is the adversary's path to a successful
tampering.
The voter's primary defense is the creation of fake credentials ---
not just individual verifiability, but also for coercion-resistance!
For each fake credential, the voter picks another envelope from the remaining
pool and if \emph{any} of these additional draws also comes from the malicious
set, then \VSD will detect a reused nonce during activation and the attack will
fail.
As $k$ increases, so does the probability
that the \VSD detects the attack.

This tension is illustrated in \Cref{fig:envelope_analysis_envelopes}.
For any given number of credentials the voter creates,
the adversary's winning probability initially rises with $k$,
but then decreases as the risk of detection increases.
The peak of each curve represents the adversary's optimal choice of $k$,
but the adversary does not know which curve they are on when selecting $k$,
as this depends on the voter's choice of $n_c$.
The expectation over the voter's choice distribution $D_X$ captures this
uncertainty and significantly limits the adversary's real-world advantage.
The user study (\cite{evoteconscience})
demonstrates the soundness of this notion:
76\% of participants
chose to create at least one fake credential,
with 53\% stating they
would create fake credentials if such a system existed today.

\parhead{Large-scale fraud.}
The adversary's difficulty in winning compounds dramatically when attempting
to influence an entire election.
To succeed, they must win the registration gamble not just for one targeted voter,
but for \emph{every} single voter they target, as a single detected failure
could expose the entire operation.
If the maximum probability of deceiving a single voter is $p_{\max}$,
the probability of deceiving $N$ independent voters is $p_{\max}^N$.

This exponential decay is visualized in \cref{fig:envelope_analysis_voters}.
Even with an optimistic success probability of $\sim 25\%$ for a single voter
(the peak for $n_c = 2$), the chance of successfully subverting just three
such voters undetectably drops to less than 2\%.
This shows that the success probability makes large-scale fraud infeasible
against a population of even moderately cautious voters.
This fact, where security against one is amplified into security against
many, is what motivates the concept of \emph{iterative individual
verifiability} (\cref{property:iiv}).

\parhead{Additional adversarial behaviors.}
An adversary might exploit human biases --- humans being non-random ---
by, for example, placing their malicious envelopes at the top of the stack.
However, even if this could double or triple the success rate against one
voter, it provides only a constant-factor advantage against an exponential
problem when targeting many voters.

The adversary could also program the kiosk to abort if the voter
selects an ``un-guessed'' envelope.
This, however, is not an undetectable subversion but rather an
observable and undeniable failure.
A single voter reporting such behavior can be sufficient to reveal misconduct.

The authorities are capable of challenging the kiosk
to reveal the ZKP commitment secret $y$ for the 
$Y_c$ it printed in the first QR code
(\cref{fig:scheme:cred-creation:real}, line~\ref{scheme:real:commits}) 
--- a computationally infeasible task if it cheated.
If the kiosk
refuses to reveal $y$ or provides an incorrect value,
authorities gain unequivocal proof of misconduct.

Ultimately, while the adversary has a non-negligible advantage 
of winning against a single voter, this advantage becomes negligible
when attempting to scale the attack.

\subsubsection{Iterative individual verifiability}
\label{property:iiv}

While the $\IndVer$ game establishes security for a
single voter, the adversary's true objective is not to
deceive one voter but to deceive enough voters to 
alter an election's final outcome.
This requires a model of security that accounts for
an adversarial campaign across a population of voters.
To change an election outcome,
they must successfully and undetectably subvert a 
number of votes $\lambda_v$ equal to or greater than 
the election's margin of victory.

We explore this requirement through two
complementary views.
First, we discuss a general model applicable
to any voting protocol,
resulting in a standard binomial security bound.
Second, we leverage a specific property of
\name --- the ability to produce non-repudiable
proof of misbehavior --- to define a
``high-stakes'' game, \IterIndVerStrong, which yields
a much stronger exponential security bound
for \name.

\begin{figure}[t]
    \centering
    \gameblock[bodylinesep=1mm]
    {$\mathbf{Game}~\IterIndVer^{\lambda_v}_{\Pi, \IA}(\params)$}{%
        \V \gets \IA(\text{\% Compile electoral roll}) \\
        \wins \gets 0 \\
        \pcfor V_{id}~\text{in}~\V \pcdo \\
            \t \wins \mathrel{+}= \text{Game}~\IndVerTarget_{\Pi, \IA}(\params, V_{id}) \\
        \pcendfor \\
        \pcreturn 1~\pcif (\wins \geq \lambda_v)~\pcelse 0
    }
    \caption{\textbf{Iterative Individual Verifiability ($\IterIndVer$)}
    Definition of iterative individual verifiability to account
    for an adversary's goal of altering an election outcome.}
    \label{fig:definition:iter-ind-ver}
\end{figure}

\parhead{Formal definition: iterative individual verifiability.}
We define the $\IterIndVer$ property using the security
game $\text{Game}~\IterIndVer$ shown in~\cref{fig:definition:iter-ind-ver}.
This game is a wrapper that repeatedly calls a voter-targeted
version of our base \IndVer~game, which we now define formally.

The \IterIndVer~game requires a sub-routine that forces the adversary
to attack a specific, pre-determined voter.
We denote this game as Game~\IndVerTarget. It is identical in
all respects to Game~\IndVer~from \Cref{fig:definition:ind-ver}
but with two modifications.
First, the game now takes as an additional input parameter $V_{id}$ 
and hardcodes the target voter as $V_{id}^\star \gets V_{id}$.
Then, the step $$(V_{id}^\star, \intent, \goal) \gets \IA(\text{Choose Target \& Goal})$$
becomes
$$(\intent, \goal) \gets \IA(\text{Choose Goal}).$$

With this sub-game defined, we now present the main game
in~\cref{fig:definition:iter-ind-ver}.
In this game, the adversary first commits to an electoral roll $\V$.
The challenger then executes $|\V|$ independent rounds
of $\IndVerTarget$, one for each voter $V_{id} \in \V$.
The adversary wins if they accumulate at least $\lambda_v$
successful attacks.

\parhead{General model for iterative attacks.}
For any voting scheme where a detected failure
might be deniably attributed to a ``systems fault'', 
an adversary's campaign can be modeled as a 
sequence of Bernoulli trials.
An adversary might therefore attack $n_V > \lambda_v$ voters,
tolerating some failed attempts to maximize their
chance of reaching $\lambda_v$ wins.
Each attempt against a single voter
is a trial with maximum success probability
$p_{\max}$.
The adversary's success in this scenario
is bounded by the tail property of the
binomial distribution $B(n_V, p_{\max})$:
$$ \Pr[\wins \geq \lambda_v] \leq \sum_{i=\lambda_v}^{n_V} \binom{n_V}{i}(p_{\max})^i(1-p_{\max})^{n_V-i}$$
This bound reflects a weaker security
guarantee, as an adversary can trade
a higher number of attempts for a greater
probability of success.

\parhead{Strong iterative individual verifiability.}
\name provides a stronger guarantee that allows us
to move beyond the general binomial model.
Specifically, any detected attempt at tampering provides
non-repudiable proof of the adversary's malicious actions.
We show this in the following lemma.

\begin{lemma}
    Let $\bot$ be the event that \VoterVerify returns reject
    for an honest voter's interaction in \name.
    This event constitutes non-repudiable cryptographic proof of
    misbehavior, which an
    adversary can only deny culpability for this event
    by breaking the hardness of the Discrete Logarithm Problem
    or the EUF-CMA security of the underlying signature scheme.
    \label{lemma:proof-of-guilt}
\end{lemma}

\parhead{\Cref{lemma:proof-of-guilt} proof sketch.}
A rejection by $\VoterVerify (\bot)$ during the registration
phase of \name arises from a failed attempt to
deceive the voter.
We can disregard failures during the voting phase,
as \cref{thm:ind-ver:vote} of \cref{theorem:ind-ver} 
shows that the adversary's success probability is 0
in that phase.
The registration failures can be categorized into two main types:
ZKP soundness failure and nonce-reuse detection.

As established in \cref{theorem:ind-ver},
to cheat the $\RealCred$ ZKP, the adversary must
commit to $Y_c$ before seeing the voter's challenge $e$.
This forces the kiosk to guess a challenge $e^\star$.
If the voter selects an envelope with a different challenge
$e \neq e^\star$, the adversary has two options:

First, it may abort the protocol.
The voter is left with an incomplete but signed receipt
(\ie the first QR containing the commitment $Y_c$).
Upon reporting the voter reporting the kiosk's malfunction, 
authorities can challenge the kiosk to reveal the ZKP secret $y$
corresponding to $Y_c$.
A malicious kiosk, having computed $Y_c$ to cheat, will be unable
to produce a valid $y$ that satisfies the ZKP commitment equation:
$Y_c = (Y_1, Y_2) = (g^{y}, \A_\pk^{y})$,
which is a failure reducible to the hardness of the DLP.

The adversary may also complete the protocol with an invalid proof.
If the adversary provides an invalid ZKP response $r$ on the receipt,
the voter's \VSD will detect this during the \Activate procedure,
triggering the $\bot$ event.
The voter now possesses, at the minimum, a receipt containing a signed
ZKP commitment in the first QR code and an invalid ZKP response $r$
in the third QR code.
This is also non-repudiable proof that the kiosk acted maliciously
unless the adversary can shift the blame to a third party by demonstrating
that they can break the
EUF-CMA security of the signature scheme to claim forgery.

The second class of attacks is the nonce-reuse detection.
Suppose the adversary successfully guesses
the challenge for $\RealCred$
(\ie the voter picks an envelope with $e^\star$).
They must still pass the $n_c-1~\FakeCred$ checks.
If the voter selects another malicious envelope containing $e^\star$,
the kiosk similarly has two options: abort or complete the procedure.

If the kiosk aborts during $\FakeCred$,
the authorities check that each envelope used has a
unique challenge.
Furthermore, for additional verification, 
the authorities can perform the same ZKP soundness 
check as described earlier using the voter's stated real
credential.
An honest kiosk can always comply with the challenges faced
by the authorities by retaining the secrets produced
during their interaction with the voter until the 
end of that interaction.

If the kiosk completes the protocol,
the \VSD will later detect a nonce re-use during 
the \Activate procedure, triggering a $\bot$.
This detection is non-repudiable because each physical envelope
is supposed to contain a unique challenge, signed by 
an (adversary-controlled) envelope printer.
The voter can now present these distinct envelopes with the 
same challenge to the authorities, providing a direct proof 
of malicious behavior.\footnote{
    Our threat model assumes voters are honest and 
    will not duplicate envelopes to cause disruption.
    To mitigate such an attack in practice, 
    the envelope printer could employ 
    physical security measures like unique watermarks 
    on the envelopes.
}

In all scenarios,
a $\bot$ event provides the voter and the authorities with
cryptographic evidence of misbehavior.
Therefore, the adversary's failure to tamper is non-repudiable,
bounded only by the negligible probability of breaking DLP
and the EUF-CMA property of the underlying signature scheme.\qed

\cref{lemma:proof-of-guilt} allows us to model
a game where the entire campaign fails upon a single detection.
Furthermore, it implies that a rational adversary has no incentive
to continue attacking once their goal of $\lambda_v$ wins is met,
as any further attempts only add risk for any $p_{\max} < 1$.
We formalize this with the game $\IterIndVerStrong$.

\parhead{Formal Definition: strong iterative individual verifiability.}
We now formalize our notion of strong iterative individual
verifiability using the following game $\IterIndVerStrong$
shown in~\cref{fig:definition:strong-iter-ind-ver}.
This game models a sequential and adaptive attack where
the adversary $\IA$ must achieve a target,
$\lambda_v$ wins, without a single detection.

To define the outcome of each round in $\IndVer$,
we define another sub-game $\IndVerTargetStrong$,
which adopts the same modifications as $\IndVerTarget$
with the one additional modification: $\IndVer$
returns three values: 1 for win, $\bot$ for detection,
and 0 for benign loss.
Concretely, for detection ($\bot$), 
this means that $\goal \neq \intent$,
$\advoutcome$ is $\top$ and $\happy$ (\VoterVerify) 
is $\bot$.

\begin{figure}[t]
    \centering
    \gameblock[bodylinesep=1mm]
    {$\mathbf{Game}~\IterIndVerStrong^{\lambda_v}_{\Pi, \IA}(\params)$}{%
        \V \gets \IA(\text{\% Compile electoral roll}) \\
        \wins \gets 0 \\
        \pcfor V_{id}~\text{in}~\V \pcdo \\
            \t r = \text{Game}~\IndVerTargetStrong_{\Pi, \IA}(\params, V_{id}) \\
            \t \pcif r = \bot \pcthen \pcreturn 0 \quad \text{\% Adversary caught} \\
            \t \wins \mathrel{+}= r \\
        \pcendfor \\
        \pcreturn 1~\pcif (\wins \geq \lambda_v)~\pcelse 0
    }
    \caption{\textbf{Strong iterative individual verifiability ($\IterIndVerStrong$)}
    Definition of strong iterative individual verifiability to account
    for an adversary's immediate loss if caught.}
    \label{fig:definition:strong-iter-ind-ver}
\end{figure}

\begin{theorem}
    Let $\Pi_{\name}$ be instantiated as described in~\cref{property:iv:instantiation},
    and $p_{\max} = \text{Adv}^\IndVer_{\Pi_{\name}, \IA}$ be the maximum advantage in
    the $\IndVer$ game.
    Under the hardness of the Discrete Logarithm problem and the EUF-CMA security of
    the signature scheme, for any PPT adversary $\IA$ 
    playing the $\IterIndVerStrong$ game with a success threshold of $\lambda_v$, 
    its advantage is bounded by:
    $$ \text{Adv}_{\Pi_{\name}, \IA}^\IterIndVer (\lambda_v, \lambda) \leq (p_{\max})^{\lambda_v} + \negl$$
    \label{thm:iter-ind-ver}
\end{theorem}

\parhead{Proof.}
Let $\Win$ be the event that Game~\IterIndVerStrong~returns $1$.
Our goal is to bound $\Pr[\Win]$.
Let $R_i \in \{1, 0, \bot\}$ be the random variable for the outcome
of the $\IndVerTargetStrong$ game against voter $V_i$.
The adversary wins if 
$$\left(\bigwedge_{i=1}^{|\mathcal{V}|} R_i \neq \bot\right) \land \left(\sum_{i=1}^{|\mathcal{V}|} \mathbf{1}_{R_i=1} \geq \lambda_v\right)$$
The proof proceeds via two lemmas.

\begin{lemma}
    For any round $i \in \{1, \cdots, |\mathcal{V}|\}$ and for any PPT adversary $\IA$,
    the probability of success in that round is bounded by $p_{\max}$,
    conditioned on the event that no prior round resulted in a detection,
    and on the full transcript $\mathsf{T}_{1}^{i-1}$. Formally,
    $$ \Pr[R_i = 1 | \wedge_{i=1}^{i-1} (R_i \neq \bot) \land \mathsf{T}_{1}^{i-1} \leq p_{\max} + \negl.$$
    \label{lemma:iter:conditional}
\end{lemma}

\parhead{\cref{lemma:iter:conditional} proof.}
We proceed via a standard reduction.
Assume for contradiction that there exists a PPT adversary $\IA$ for
which this inequality does not hold for some round $i$.
That is, for some history $\mathsf{T}_{i-1} = \wedge_{j < i} S_j \neq \bot \land \mathsf{T}_{i-1}$,
$\IA$'s success probability $p_i = \Pr[S_i = 1 | \mathsf{T}_{i-1}]$ is greater than $p_{\max} + \negl$.
We construct an adversary $B$ that uses $\IA$ to win $\IndVer$ game with probability
$p_i$, contradicting the definition of $p_{\max}$.

$B$ receives the parameters of $\IndVerTargetStrong$ game targeting an honest voter oracle $V$.
$B$ internally runs $\IA$ and perfectly \emph{simulates} the history $\mathsf{T}_{i-1}$ by playing
the role of the honest voters for rounds $\{1, \dots, i-1\}$ 
(\ie creates a fake history for all previous voters).
At round $i$, $B$ connects $\IA$ to the voter oracle $V$ in the $\IndVerTargetStrong$ game.
From $\IA$'s perspective, this interaction is indistinguishable from
a real round $i$ of $\IterIndVerStrong$, as the sources of randomness
(the voter's choice for $n_c$ from the distribution $D_C$) are fresh and 
independent of past transcripts.
$B$ then forwards $\IA$'s output to the challenger.

$B$'s advantage in its $\IndVerTargetStrong$ game is exactly $p_i$.
Our assumption $p_i > p_{\max} + \negl$ thus implies $B$ 
can break the underlying cryptography underpinning the
$\IndVerTargetStrong$ game, a contradiction. Therefore, the lemma holds. \qed

\begin{lemma}
    For any PPT adversary $\IA$, $\Pr[\Win]$ is bounded by $(p_{\max})^{\lambda_v} + \negl$.
    \label{lemma:iter:product-bound}
\end{lemma}

\parhead{\cref{lemma:iter:product-bound} proof.}
For the \Win event to occur,
the adversary must achieve $\lambda_v$ wins
before any $\bot$ outcome occurs, at which point the game terminates.
Let us analyze the adversary's probability of success.
The adversary's strategy involves choosing which voters
to attack.
Let the sequence of attacked voters be $V_{i_1}, V_{i_2}, \dots$.
To win, the the adversary must succeed in the first $\lambda_v$
of these attacks.

Let $\EventE$ be the event that the adversary succeeds in its
first $\lambda_v$ attempted attacks.
The $\Win$ event is a sub-event of $\EventE$,
so $\Pr[\Win] \leq \Pr[\EventE]$.
We can express $\Pr[\EventE]$ using the chain rule of probability:
\begin{align*}
    \Pr[\EventE] &= \Pr[R_{i_1} = 1] \cdot \Pr[R_{i_2} = 1 |R_{i_1} = 1] \cdot \dots \\
    &\; \; \cdot \Pr[R_{i_{\lambda_v}} = 1| \wedge_{k=1}^{\lambda_v - 1} R_{i_k}] \\
    &=\prod_{k=1}^{\lambda_v} \Pr[R_{i_k} = 1| \wedge_{l=1}^{k-1} R_{i_l} = 1].
\end{align*}
For each term in this product, the conditioning event
$(\wedge_{l=1}^{k-1} R_{i_l} = 1)$ implies no prior detections
occurred.
By \cref{lemma:iter:conditional},
each term is bounded by $p_{\max} + \negl[]$.

\begin{align*}
\Pr[\Win] \leq \Pr[\EventE] &\leq \prod_{k=1}^{\lambda_v} (p_{\max} + \negl[]) \\
&= (p_{\max} + \negl[])^{\lambda_v} \\
&\leq (p_{\max})^{\lambda_v} + \negl[].
\end{align*}
\qed

We now conclude the proof.
\Cref{lemma:iter:conditional} establishes a bound on the adversary's
per-round success probability, showing that past interactions do not
increase its chances beyond $p_{\max}$.
\Cref{lemma:iter:product-bound} then leverages this per-round bound
to analyze the probability of the global \Win event.
By substituting the final result of \cref{lemma:iter:product-bound},
we have:
$$\text{Adv}_{\Pi_{\name}, \IA}^{\lambda_v} = \Pr[\Win] \leq (p_{\max})^{\lambda_v} + \negl[\lambda].$$
\qed

This result demonstrates
that with \name's non-repudiable proofs of malicious behavior, 
the security against scalable
attacks on individual verifiability scales exponentially with the number
of votes the adversary must alter, as shown in \Cref{fig:envelope_analysis_voters}.

\iftoggle{longpaper}{
\section{The paper baseline versus digitalization}
\label{apx:baseline}

The only approach to national voting in democracies today
enjoying near-universal acceptance by experts
is in-person voting with paper ballots.
Marking paper ballots in a privacy booth at a controlled polling site
helps protect ballot secrecy against coercion or vote buying,
which were widespread problems historically
before the Australian secret ballot's
introduction in 1856 and its subsequent global acceptance~\cite{crook2011ModernSecretBallot}.
Supporting transparency,
voters normally deposit their marked ballots into an on-site ballot box themselves,
and independent election observers ensure ``end-to-end'' transparency
by monitoring the subsequent handling and traditionally-manual counting of these paper ballots.

The rise and ubiquitous cultural acceptance of digital technologies, however,
has inevitably raised a variety of complex issues
surrounding the adoption of digital technologies
into the voting process.
One side of this issue is the desire of voters
to bring and use their own personal devices while voting.
The other side is the desire of governments
to ``improve'' the voting process
by incorporating digital technologies.

\subsection{Pressures for voter-driven digitalization}

In the first case,
voters may have understandable desires
to use their own personal devices
even during paper-based voting in a ballot booth.
Conscientious voters may wish to refer back
to digital notes they made earlier,
for example,
while deciding (and perhaps discussing with friends)
how to vote in a complex election
involving numerous different races, candidate choices,
popular initiatives and referenda, etc.
Referring to such digital notes is arguably non-problematic
in itself as long as it remains the voter's free choice --
in the privacy booth --
\emph{which} of their digital notes or past discussions
to reflect on their actual ballot and which to disregard.

A similarly-understandable but more-problematic practice,
however,
is that of taking \emph{ballot selfies}:
photos that voters take of themselves in the ballot booth
with their marked ballot visible~\cite{
Benaloh2013RethinkingCoercion}.
Voters may be innocently tempted to take ballot selfies
in order to show their civic engagement
or their enthusiasm for particular parties or candidates,
for example.
The problem is that ballot selfies and other recordings
can double as \emph{receipts} usable
for participating in coercion or vote-buying schemes.
The proximity of officials arguably remains
at least a psychological deterrent
against fraudulent voting behaviors on-site,
but the strength of this deterrence is limited,
its effectiveness depending on the strength
of coercive threats.
Thus, even the question of whether to \emph{allow}
voters to use personal devices in privacy booths
is a delicate policy issue,
as further discussed later in \cref{apx:devices}.

\subsection{Pressures for government-driven digitalization}

From the perspective of democratic governments
and their election authorities,
there are appeals to the prospect
of digitalizing their official voting processes.
Incorporating digital technologies
anywhere in the voting process is problematic,
however,
due in particular to the fundamental difficulty
of ensuring that all of the system's critical digital components
are running the correct software,
and that this software is bug free~\cite{thompson84reflectionsFixed,feldman2007VotingMachineAnalysis}.

The high stakes of national elections~\cite{2024MoldovanVoteBuying,Poteriaieva2024ElectionPolymarket}
create enormous incentives
for nefarious actors to manipulate elections,
and the general insecurity of today's devices
creates endless potential opportunities to do so.

One proposed safety test
for the incorporation of digital technologies is
\emph{software independence}~\cite{rivest2006SoftwareIndependence}.
Briefly,
anyone should be able to verify independently
that an election's announced outcome is correct,
even if all of the official software 
used in the election process
is incorrect (\eg buggy or compromised).
Even if we accept this principle, however,
applying it throughout the voting pipeline
is nontrivial.

We can categorize the main attractions
toward voting-process digitalization --
together with the attendant risks of each --
according to the stage of voting they primarily affect:
ballot \emph{marking},
ballot \emph{casting},
and ballot \emph{counting}.

\subsubsection{Digitally-assisted ballot marking.}
Incorporating digital technologies into ballot marking
can make it easier and more reliable
for voters to express their choices.
Electronic \emph{ballot marking devices} or BMDs
deployed in voting booths,
for example,
can improve usability in numerous ways
over hand-marked paper ballots~\cite{
bernhard2020VotersDetectManipulation,chapman2024NISTVoterVerificationReview}.
BMDs can guide a voter through the choices step-by-step,
making relevant instructions and background information available on demand.
Similarly, BMDs can assist voters with disabilities,
for example by displaying choices in large fonts
or audibly reading them for those with vision impairments.
By validating the voter's choices as they are entered,
BMDs can proactively help ensure
that cast ballots are valid and not accidentally spoiled.
This automatic validation can prevent numerous common errors
that are prevalent on hand-marked paper ballots,
such as accidentally or ambiguously marking
two candidate choices when only one choice is legally allowed.\footnote{
    When hand-marking a ballot by filling in ovals 
    with a pen or pencil,
    for example,
    voters may start filling in the oval for one candidate,
    then change their mind and try to ``cross out'' their initial choice,
 and then fill in another candidate's oval.
    Deciding whether to attempt to interpret the ``voter's intent''
    in such cases or merely disqualify that choice
    (or even the whole ballot)
 is a tricky and potentially-costly policy choice.}

In widely-accepted best practices,
voters still use BMDs alone in a privacy booth 
to preserve ballot secrecy.
Well-designed BMDs maintain transparency
by printing a paper ballot
that the voter can -- and arguably is \emph{expected} to --
inspect and verify for correctness before casting.
Even if a relatively small percentage of voters
actually do check their printed ballots carefully
before casting them,
a compromised BMD that attempted to manipulate many ballots --
as would be required to affect most elections --
is statistically almost certain to be detected
by at least some voters.
BMDs, therefore, arguably satisfy the software-independence test,
but leave unchanged the usual requirement that voters vote \emph{in person} and \emph{in private},
only at official polling sites.

\subsubsection{Digitally-assisted ballot counting.}
Another attractive and partly-accepted digital advance
is automated ballot scanning and counting.
Machine-counting paper ballots is clearly much faster
and less costly
than manual counting,
and arguably less prone to the obvious human-error risks
of hand counting.
Counterbalancing these perceived advantages, however,
is the risk of digital scanners
miscounting ballots or misreporting the results,
for countless potential reasons including
mechanical failures, software bugs, or malicious compromise.
Accepted best practice therefore dictates that
automated ballot-counting processes must at the very least
leave a tamper-evident ``paper trail''
of paper ballots that could be recounted by hand if needed.

An assurance that machine-counted ballots
\emph{could in principle} be recounted manually, however,
does not ensure that evidence of an irregularity
will actually emerge to trigger such a recount --
or even that the ballots will actually be recounted
even if such evidence does emerge.
Unlike the case with BMDs for ballot marking,
individual voters suspicious of an election's outcome
cannot unilaterally decide to recount the ballots:
only the government can decide to do so.
The genuinely-high cost of a manual recount,
which obviously cancels the cost-savings of machine counting,
incentivizes governments
to impose high evidentiary demands to trigger a recount.
This high cost is also temptingly easy to cite
as a reason not to recount
even in the face of controversy over an election's results --
especially if the authorities in power
are already happy with the machine-counted results.

For these reasons,
the mere existence of a paper trail
appears insufficient to ensure software independence.
There must also be some process to verify
machine-counted results \emph{proactively},
ideally during every election,
before and independently of any evidence of irregularity.
Risk-limiting audits,
which can statistically check an election's results
without incurring the full cost of a manual recount,
represent one such mechanism
that is gradually gaining acceptance~\cite{
bernhard2021RiskLimitingAudits}.

\subsubsection{Digitally-assisted ballot casting.}

Neither of the above tentatively-accepted digital advances,
of course,
affect the best-practice expectation
that voters must normally still vote in person,
in the privacy of a ballot booth in a controlled environment.
The inconvenience and sometimes impracticality of in-person voting, however,
creates pressure on governments to adopt remote processes,
such as postal voting,
which allow voters to vote from anywhere.
Because the election authority cannot control arbitrary locations, however,
remote voting of any kind normally compromises ballot secrecy
and hence resistance to coercion or vote-buying.

This weakness is not purely theoretical:
for example, absentee ballot fraud in North
Carolina was found to have altered election
outcomes~\cite{graff2021NorthCarolinaAbsenteeBallots,2022NCAbsenteeFraud}.
Despite this weakness,
convenience and voter-turnout considerations have led a few countries,
such as Switzerland,
to adopt remote processes such as postal voting
as the normal-case approach~\cite{luechinger07impact}.
A few Swiss cantons still use the practice of
\emph{landsgemeinde},
physically gathering in the town square
and raising hands to vote,
an approach similarly lacking protection
against coercion or vote-buying~\cite{osullivan2025GlarusPublicVoting,
durst04landsgemeinde}.
Maintaining this cultural indifference
to the coercion and vote-buying problem,
despite its recognition as a central voting-security issue
in most other countries,
coercion resistance was likewise omitted from the goals of
Switzerland's e-voting program~\cite{post2023VotesNowPossible,swisspost2024Architecture}.

Even without coercion resistance,
achieving end-to-end verifiability and software independence
is fundamentally difficult in e-voting
because digital threats affect
every stage of the voting pipeline.
Switzerland's e-voting system, for example,
is carefully designed
to provide \emph{cast-as-intended} verifiability,
ensuring that a voter's compromised personal device
cannot silently alter or suppress the voter's choices
without detection.
Both Switzerland's and Estonia's e-voting systems
arguably fail the software independence test
in their backend services,
however,
because voters must just trust and cannot independently verify
that the backend infrastructure is securely running the correct
software~\cite{ford22auditing}.

Although Switzerland's system admirably splits
the critical backend functions
across four \emph{control components}
run by separate operational teams,
and these control components produce
cryptographic zero-knowledge proofs of the
correct handling of encrypted ballots,
these proofs are not made publicly available,
so only one designated ``insider'' authority can verify them.\footnote{
    On the basis of private communication with the Swiss e-voting authorities,
    the perceived threat of future quantum computers being able to decrypt past ballots and deanonymize voters in past elections is one key motivation
    for this deliberate limitation of the system's verifiability properties.
}
Switzerland's e-voting system in addition depends on
a trusted \emph{printing authority}
responsible for printing and mailing security-critical
\emph{return codes} to voters;
any leak of these codes could potentially enable
an adversary to impersonate many voters remotely at scale.

Between the facts that today's deployed e-voting systems
address coercion only weakly (Estonia)
or not at all (Switzerland),
the significant design and deployment challenges
of ensuring end-to-end verifiability,
and the pragmatic difficulty of ensuring that
security-critical code is sufficiently bug free,
it is understandable why voting-security experts globally
still consider paper-based voting the standard,
accept limited digital enhancements to in-person voting
only with abundant caution,
and remain highly skeptical about --
if not categorically opposed to --
remote e-voting.

\section{Credential lifetime and surprise coercion}
\label{apx:lifetime}

Any deployment of a system like \sysname
faces a policy choice of how long e-voting credentials should remain valid
before expiring and requiring renewal.
While making this policy choice in a given environment
is beyond the scope of this paper,
we briefly examine here a few credential-lifetime considerations in general,
and their important interactions with coercion resistance in particular.

\subsection{Potential principles for credential lifetimes}

Usability considerations alone might suggest
allowing credentials to be used indefinitely without ever expiring.
Many of the multitude of reasons
that identity documents such as passports usually have expiration dates,
however,
apply similarly to the e-voting credentials that \name produces.
Like any important security token or identity proxy,
\name credentials might fall into the wrong hands for numerous reasons,
their security might erode due to weaknesses gradually discovered
in the cryptographic algorithms they depend on, etc.
Imposing an expiration date on e-voting credentials
is simply good security practice,
for standard reasons not in any way specific or unique to \name.

A more arguably-reasonable choice of credential lifetime, therefore,
might be to coincide with the lifetimes
of other government-issued identity documents
that a citizen may need to renew occasionally in person,
such as passports, driver's licenses, or other ID cards.
Such an alignment of lifetimes across identity documents
could in principle save voters' time and reduce convenience costs
by allowing voters to renew multiple documents at once
during the same government office visit,
in the common case.
This benefit makes the important presumption, of course,
that the government is sufficiently well-organized
to ensure that a single government office
can offer these distinct renewal services --
or at least, can offer them at multiple offices geographically colocated at the same site.

\subsection{Surprise coercion threats}

An important issue with long credential lifetimes, however,
is the threat of \emph{surprise coercion}.
An average voter may arguably be unlikely
to understand or appreciate the risk of coercion sufficiently
to have taken adequate precautions against this threat,
before being ``surprised'' by a coercer who suddenly appears
and demands immediate compliance under some threat
(or offering financial rewards for compliance).

An obvious surprise-coercion scenario in the case of \name
is that a voter printed only one real credential and no fake credentials
at the time of last registration,
and thus has no fake credentials prepared to give or sell to the coercer.
If the coercer guesses, perhaps accurately,
that most voters are likely in precisely this boat --
at least before they have ever actually faced a coercion threat --
then the coercer might simply demand their real credential
or whatever voting device it is installed on,
and take it as ``likely enough''
that this is indeed the voter's real credential.
After surprising the voter this way,
the coercer might successfully forbid the voter from re-registering
at least until the voter's current credentials expire,
since we consider it infeasible to hide registration events from coercers
for reasons detailed in \cref{apx:impersonation}.

Long credential lifetimes thus risk increasing the amount of time
a successful surprise-coercion attack succeeds
in allowing the coercer to control or vote on behalf of the victim,
before a government-imposed credential-renewal obligation
forces the coercer either to permit (or require)
the voter to re-register in order to continue e-voting at all.
Shorter credential lifetimes are thus preferable
from a coercion-resistance perspective,
in order to reduce this window of vulnerability
that many ordinary voters might have to surprise coercion.

A long-term coercer could of course forbid a victim 
from \emph{ever} registering even to renew expired e-voting credentials.
In a realistic environment with multiple voting channels
properly designed to ensure coercion resistance jointly
and not just individually,
as detailed in \cref{apx:registration},
forbidding the victim from ever renewing e-voting credentials
should effectively leave a fallback path to a backup channel
such as traditional in-person voting --
at least if any backup voting channel is effectively available to the victim.
In a worst-case scenario where a coercer can successfully forbid a victim
from \emph{ever} visiting a government office,
whether to register for e-voting or to vote in person,
we consider this tantamount to long-term imprisonment of the victim.
This scenario may unfortunately occur in extreme circumstances,
but probably no realistic approach to coercion resistance can address it.

\subsection{Random or deniable renewal obligations}

Although we cannot realistically hide registration events
for the reasons discussed in \cref{apx:impersonation},
a potential defense against surprise coercion
might be to make credential lifetimes \emph{unpredictable}.
In addition to imposing a fixed upper bound on credential lifetime,
for example,
the government might periodically select a subset of voters
by random lottery,
notifying them that their e-voting credentials
will expire within a few months
and must be renewed if they wish to continue using the e-voting channel.
Even a voter with brand-new credentials would be subject to this lottery.

While such a proactive credential-expiry lottery
would of course inconvenience the unlucky selected voters,
it would also have the global benefit of providing ``cover''
for the government to force early renewal of e-voting credentials
for other reasons as well:
such as due to
an anonymous tip-off from a coercion victim or the victim's friend
that the victim needs a plausibly-deniable,
ostensibly \emph{government-initiated} reason
to visit the registration office, re-register,
and perhaps report the coercion incident in relative safety.
Such an ``anonymous tip-off'' mechanism
might in principle even be built into client-side voting software.
Such a mechanism would be
analogous to the way that some banks currently issue customers with \emph{duress PINs}
that they can use to inform the bank silently
if they are coerced to withdraw money at an ATM~\cite{hameed2013SafePass}.

\subsection{On-demand fakery versus preparation}

Some prior coercion-resistant e-voting schemes
such as Civitas~\cite{clarkson2008Civitas}
have the attractive property that a voter can, at least in theory,
cryptographically create new fake credentials at any time.
This \emph{on-demand} creation of fake credentials
is sometimes cited as a reason to prefer such designs
over a design like \name,
which requires voters to ``prepare in advance''
by obtaining fake credentials at registration time before coercion occurs,
and otherwise leaves voters vulnerable to surprise coercion as discussed above.

Closer analysis of realistic suprise-coercion scenarios
in the context of realistic, arguably-usable deployments of such systems,
however,
reveal this apparent advantage of Civitas-like designs
to be more theoretical than practical.
The only practical embodiments of Civitas-like designs
ever subject to a user study~\cite{neto2018CredentialsDistributionUsability}
rely on trusted hardware to keep the voter's real credential safe
and issue fake credentials on demand,
\eg on entry of incorrect PINs
to the trusted smart card~\cite{estaji2020UsableCRSmartCards}.
In such a deployment context,
we may consider it a near certainty that an unprepared voter
subject to surprise coercion
will have obtained only one such trusted smart card,
which the coercer can simply demand and confiscate.

Without the assistance of \emph{some} electronic device
that holds the voter's real Civitas credential,
and which we must assume the voter has somehow hidden and protected
from the coercer despite being entirely unprepared for coercion,
the theoretical possibility of on-demand creation of Civitas credentials
essentially relies on the voter himself being a cryptographer.
This hypothetical voter must in fact be
not just an average cryptographer,
but a genius in the league of the legendary Indian mathematician Ramanujan,
who could perform complex mathematical calculations on large numbers
in his head without the aid of any electronic device.
Needless to say, we find it implausible
that this ``Ramanujan voter'' profile fits
any significant percentage of real voters anywhere on Earth.

Because the trusted hardware
assumed by usable Civitas-like schemes is undoubtably
much more expensive than the paper credentials that \name uses,
it seems likely that typical voters unprepared for coercion
will be much more likely in practice to have only one smart card
in a Civitas-like design
than a typical \name user is to have obtained
only one paper credential.
\name's user-facing design, in fact,
uses the term \emph{test credentials} in place of \emph{fake credentials},
and informs voters that test credentials may be legitimately used
for purposes other than coercion --
such as for casual testing of the voting system
and for educational purposes,
such as to show children or friends not yet eligible to vote
how the e-voting system works.
By making fake/test credentials not only inexpensive
but also in principle worth the trouble of creating
for reasons entirely orthogonal to worries about coercion,
\name's design attempts to increase the chance that
many or even most voters will create one or more fake credentials
when registering,
``just for fun'' if for no other reason.

\section{Personal devices and usage policies}
\label{apx:devices}

Even in accepted in-person voting processes using paper ballots,
personal devices capable of recording
can already compromise the coercion resistance
that privacy booths are intended to provide at official polling sites.
The use of recording devices can similarly compromise the effective coercion resistance
of \sysname's credentialing process in the registration booth,
in essentially the same fashion.
Just as a coercer can already demand that a voter photograph themselves
with their ballot marked according to the coercer's preferences in a voting booth today,
a coercer could similarly demand that a voter record themselves
creating their real and fake credentials in \name's privacy booth,
showing the coercer in particular how the voter marked their real credential.

The fundamental threat that recording devices impose on registration in \name
thus appear largely equivalent to the corresponding threat of recording devices
used in privacy booths during in-person voting:
it appears \sysname neither solves this problem nor makes it worse.
Nevertheless,
it is worth exploring the background and current prevailing situation
surrounding this challenge for in-person voting,
then outlining potential policy and enforcement options that might be applicable
to e-voting registration using \name.

\subsection{Ballot selfies and recording prohibitions for in-person voting}
\label{apx:devices:recording}

A common practice is to take a \emph{ballot selfie} in the voting booth
showing the voter's marked ballot,
then perhaps posting it to social media.
This practice, while usually intended innocently to demonstrate civic engagement,
could easily double as a ``proof'' of how one voted
to satisfy a coercer's demands or to participate in a vote-buying scheme.
As a result, ballot selfies and related photography of marked ballots
is at least legally prohibited in some countries
as well as a dozen US states~\cite{demsas2020BallotSelfieUS,
AP2014BrazilNoBallotSelfie,kelly2015IrelandNoBallotSelfie}.

Even where ballot selfies are legally prohibited,
the level enforcement and potentially-applicable penalties vary.
Many jurisdictions do not enforce their ballot selfie-ban at all
unless there is evidence of coercion, vote buying, or other fraudulent activity.
In most US states that ban ballot selfies,
being caught taking one is a misdemeanor punishable by a fine,
although in Illinois it is a felony punishable by 1--3 years in
prison~\cite{illinoiselectioncode}.

Even where ballot selfies are illegal and even strongly punishable,
that does not of course mean that voters cannot or will not secretly take them anyway --
if an abusive spouse, local gang member, or other strong coercer threatens violence
or other significant repercussions if the voter does not, for example.
The deterrent represented by \emph{potential} punishment by the government,
only if participation in coercion or vote buying is actually caught,
does not necessarily outweigh \emph{likely} or \emph{certain} punishment by a strong coercer
whose demands go unfulfilled.
Thus, today's accepted coercion-resistance practices
appear to go only so far and are clearly weak against stronger forms of coercion,
and unfortunately we know of no systematic evidence of how prevalent such strong coercion might be.
We hope it is extremely rare,
but it is also possible that it is more common but just rarely
\emph{reported}---%
because in such cases the coercer usually ``wins''
and successfully intimidates the voter not only into following the coercer's demands,
but also into never revealing the incident.

In principle,
in-person polling sites could address stronger coercion \emph{proactively}
by requiring all voters to check personal devices into lockers
before entering the voting area containing the privacy booths.
Polling sites could even enforce such a recording device ban
by requiring voters to enter the voting area through airport-style metal detectors
after checking all personal devices,
just as visitors often must already do at many high-security government sites
such as embassies and military bases.
Such strong measures would incur high costs, however:
financially, logistically, in terms of convenience and perceived friendliness to voters,
and potentially in terms of negative impact on voter turnout.
The cost to all voters and to the election authority thus seems unlikely to be justified
by the incremental protection such measures might provide against
apparently-rare incidents of actual coercion or vote buying.
As a result,
we are not aware of any government that has actually imposed such extreme measures
at ordinary polling sites in the normal case for in-person voting.

\subsection{Anti-recording considerations for \name}
\label{apx:devices:anti-recording-trip}

A government considering the adoption of
a coercion-resistant e-voting system like \sysname
would need to decide whether to impose a ban on the use of recording devices
during a \name registration session in a privacy booth,
and if so, to what degree to enforce such a ban.
We consider this policy decision to be important
but orthogonal to the technical focus of this paper.
Many of the considerations and tradeoffs outlined above for in-person voting
seem likely to translate directly
to corresponding tradeoffs for \name privacy booths.

It is relatively inexpensive and practical simply to forbid the use of recording devices
in \name privacy booths,
to inform voters of this ban in educational materials before they enter the booth,
and to impose penalties as a deterrent to noncompliance with the recording ban.
As with ballot-selfie bans for in-person voting,
such prohibitions may be effective in deterring casual violations,
but may not be effective against stronger forms of coercion
such as the abusive partner or gang member
demanding a secret recording ``or
else.''

A government concerned with strong coercion
could of course take stronger proactive measures,
such as requiring voters to check personal devices and pass through metal detectors
before entering a privacy booth for \name credentialing.
For the same reasons such measures are already rarely if ever taken
for in-person voting, however,
we do not expect such measures to be common in a practical deployment of \sysname.
Regardless, this important consideration remains a policy decision rather than a technical one.

We envision there may be situations in which stronger anti-recording measures
may be cost-effective, however.
One prominent use-case for remote e-voting is for expatriates or military members living overseas.
Expatriates often need to visit an embassy or consulate of their nationality
at least once every few years anyway to renew their passport,
and embassies are typically considered high-security establishments
that often require visitors to check personal devices and enter through metal detectors anyway.
If a government were to deploy \sysname
so as to make \name privacy booths and kiosks available to expatriates
at embassies and consulates,
then there may be little if any extra cost to imposing stronger anti-recording measures
at these sites.

Similarly, overseas military members are often stationed at, or in the general vicinity of,
a military base that has strongly-protected areas
for other othogonal reasons anyway,
which could likewise double as secure sites for \name registration for e-voting
at little to no incremental cost.

\subsection{Assumed limitations of ordinary voters}
\label{apx:devices:voter-limits}

Our assumption above that the use of personal devices
is at least forbidden in, if not proactively excluded from,
registration booths,
affects \name's design for coercion resistance in other more subtle ways.

In particular,
\name's design assumes that ordinary voters
\emph{without the help of an unauthorized personal device}
cannot readily perform cryptographic calculations in their heads.
(If a coerced voter \emph{does} have and is successfully coerced to use
an unauthorized device,
we expect that a recording attack as discussed above
is the simplest and most likely coercion threat vector.)

A rare genius to whom this assumption about the limitations of ordinary voters
does \emph{not} apply,
we call a \emph{Ramanujan voter},
in reference to the legendary Indian mathematician Srinivasa Ramanujan,
who may be one of the few historical examples
of a real person potentially capable of such feats.

If a victim of coercion is such a hypothetical Ramanujan voter,
then the coercer could readily defeat \name's coercion resistance
by demanding that the voter mentally decode the QR code 
containing the real credential's IZKP commit once it is printed,
then compute a cryptographic hash of that commit,
and use the resulting hash in selecting an envelope
from the stack of those available in the privacy booth:
\eg such that the cryptographic challenge matches a certain number
of prefix bits from the envelope's cryptographic hash.
The coercer could still not be certain,
but could obtain some probabilistic confidence,
that the revealed ``real credential'' is indeed real,
because when printing fake credentials
the Ramanujan voter cannot see the kiosk's commit
(in order to hash it mentally)
before having to choose an envelope.

This theoretical attack has inspired related prior proposals,
using IZKPs for in-person voting,
to conceal the real credential's printed commit
until the voter's challenge
is chosen~\cite{moran2006ReceiptFreeVerifiableEverlastingPrivacy}.
While \name kiosks could certainly be designed
with such mechanical protections,
we do not consider such a defense to be necessary
or cost-justified in practice.
Ramanujan voters capable of performing such feats
seem likely to be vanishingly rare in practice --
especially considering that this attack only conceivably matters
to voters \emph{actually under coercion},
a hopefully-rare event in the first place.

\section{Impersonation attacks and defenses}
\label{apx:impersonation}

Any government process that involves identity checking,
such as \name's in-person e-voting registration process,
is potentially vulnerable to \emph{impersonation} attacks.
Anyone able to impersonate a voter in registering for e-voting
can potentially steal the target voter's real credential
and subsequently cast real votes in place the victim.

While we do not expect it is feasible
to prevent impersonation attacks entirely in most practical situations,
our pragmatic goal is to ensure that successful impersonation
is promptly \emph{detectable} by the genuine voter,
who can then report and rectify the successful impersonation
by re-reg\-is\-ter\-ing to obtain new credentials in person.
Ensuring that impersonation is detectable
implies that \name must follow common security practices
by \emph{notifying} voters of registration events.
This notification requirement in turn implies
that we cannot realistically keep the \emph{act of registration}
secret from potential coercers,
as we might wish to in order to maximize coercion resistance.
We discuss these issues and the resulting tradeoffs
in the rest of this section.

\subsection{Some impersonation scenarios of concern}

One situation in which it is hard to prevent impersonation attacks
is when an impersonator
happens to look ``enough'' like the victim to pass as the victim
when the registration official checks the voter's identity
during \name check-in.
Identical twins are ideally positioned for such impersonation attacks.
Many people have and occasionally encounter
proximate ``look-alikes'' to themselves
in the real world, however.
Such a look-alike who identifies and stalks a target voter sufficiently
to learn their name and other basic demographics
might be able to trick a registration official's identity check.
We do not expect either of these attacks to be common,
but we do expect them to occur (and succeed) occasionally.

A variation on the above scenario is if an impersonator
does \emph{not} look enough like the victim to pass an identity check,
but instead presents a forged identity to the registrar,
who does not check the identity closely enough to detect the forgery.
We again do not expect this situation to be common in practice,
as being caught presenting a false identity
in an important government process is strongly punishable almost everywhere,
representing a strong deterrent to such uses of fake identity documents,
even if they might be fairly likely to go uncaught.

A third scenario we are concerned with, however,
is if a particular \emph{registration official} is corrupt,
and deliberately accepts faked (or no) identity documents
in collusion with an impersonator.
The corrupt official might indeed \emph{be} the impersonator.
A corrupt registration official may in principle be ideally positioned
to impersonate not just one but many voters, in fact,
choosing some time other non-colluding officals are absent --
\eg outside the registration office's opening hours --
to ``register'' for e-voting on behalf of any number of voters
who the corrupt official believes are unlikely to notice the impersonation.
Even if such a corrupt registration official
faces severe consequences if caught,
the potential scalability of such impersonation attacks,
and the potentially-scalable financial rewards (\eg bribes)
for participating in such attacks --
or alternatively, potentially-severe punishments
for \emph{not} participating under threats of violence
by local criminal organizations for example --
make such scalable impersonation attacks plausible and perhaps even likely
in certain environments.

\subsection{Detecting impersonation via notifications}

Scenarios like those above lead us to conclude
that it is unrealistic and unwise to assume
that impersonation attacks will never occur, or never succeed.
The common and widely-accepted security ``best practice''
of \emph{notifying} users of security-critical events, however,
is applicable to \name as a measure that can help ensure that
impersonation attacks, when they occasionally do succeed,
are at least promptly detected by the victim.
On successful detection of such an impersonation attack,
the victim can report the incident and re-register in-person for e-voting,
thereby invalidating the impersonator's illegitimate voting credentials.
In the process of reporting the incident,
the victim also hopefully places (non-corrupt)
local registration officials on alert
for any future such impersonation attempts.

Since governments typically have at least a mailing address
for all eligible registered voters,
one obvious default notification channel is by mail.
We envision the voter receiving a letter or post card
a few days after any successful registration,
informing them that the registration occurred,
and advising them to contact the registration office immediately
in the hopefully-rare event it was not the voter who registered.

An alternative notification channel is digitally,
via an E-mail or push notification that the registrar's backend services
automatically sends to any voting device(s)
the voter has recently registered with or used for voting
using any credentials (real or fake) associated with that voter.
Ideally, a successful registration for e-voting
would notify voters of the registration via both, or all,
available notification channels.

A potential weakness with these and other notification channels
is if a corrupt registration official can control or suppress this notifcation,
or can collude with other insiders capable of doing so.
We expect such attacks involving impersonation
\emph{coordinated with} notification suppression
should be significantly harder and riskier to pull off
even than ``lone'' corrupt official attacks suggested above,
since in most cases the official sitting at \name's check-in desk
to verify voters' identities
is unlikely to be the same official (or have the same required expertise)
as the officials who control the notification channels (\eg IT workers).
Nevertheless, collusion is not inconceivable,
especially in the event of bribery or threats
by powerful organized crime in the area,
and both postal and digital notification channels
tend to be single points of failure/compromise
that might conceivably be susceptible to suppression.

\subsection{Detection via periodic ledger checking}

Provided the election authority's backend services
are split across several independent servers run by independent teams,
as discussed in \cref{sec:protocol:model},
and provided not \emph{all} of these servers or teams are compromised
as \sysname's threat model assumes,
there remains at least one further means by which voters
might learn of successful impersonation attacks reasonably promptly,
even if proactive notification channels such as those above
fail or are successfully suppressed.

Each voter's client-side device normally ``watches''
the ledger (public bulletin board) anyway,
periodically checking it for any updates relevant to the voter,
whenever the voting application is launched or left running.
Thus, if the voter launches the voting application
some time after a successful impersonation attack --
perhaps to check how they voted in the past
as discussed in \cref{apx:voting-history},
perhaps to vote in an upcoming election,
or perhaps for other reasons such as to obtain information
or demonstrate the voting system to a family member or friend --
this (or any) launch of the voting application represents
an opportunity for the client software to ``notice''
a new registration of the voter appearing on the ledger's registration log,
and notify the voter of that successful registration.

If the voting application is run on a device such as a desktop
that allows the application to maintain a background ``daemon''
that occasionally wakes up and proactively checks the ledger
(even if only once per day or once per week, for example),
then such periodic checks could reduce the duration with which
successful impersonation might go undetected,
even if proactive notifications are successfully suppressed.

\subsection{Coercion resistance implications}

All of the above defenses to impersonation attacks, however,
underline an important property of \sysname's threat model:
we assume that \emph{the act of registering for e-voting} --
or re-registering for any purpose,
such as to renew credentials or to recover from an impersonation attack --
is public information,
which \sysname's design makes no attempt to keep secret.

This threat-model assumption unfortunately implies,
at a fairly fundamental level,
that we cannot realistically hide \emph{the act of registration itself}
from potential coercers.
This means for example that if a coercer
has successfully obtained a victim's real credential,
and has forbidden the victim from re-registering,
the coercer can see and punish a victim who disobeys this demand.
The successfully-coerced victim will have the opportunity
to recover from this coercion
the next time the victim's credentials expire by government policy
and must be renewed --
but the victim may \emph{not} have an opportunity to recover until then,
if the coercion threat persists for the long term
(\eg in an abusive-partner scenario).

Any attempt to target registration notifications more narrowly --
\eg sending push notifications only to voting devices
on which \emph{real} credentials have been activated --
risks failing to notify the genuine victim
if an impersonator has taken over the victim's real credential,
and also risks undermining coercion resistance
by creating a new way a coercer might distinguish fake credentials from real.
Thus, we do not see any realistic middle ground
between \emph{always notifying} and \emph{never notifying} a voter
of successful registration events,
via any and all available notification channels.

The obvious potential design alternative, then,
is \emph{never} to notify voters of registration events,
and hence to hide such events from voters and coercers alike.
While conceivable,
such a design choice appears to have severe negative consequences
for transparency in general,
and in particular for ensuring that voters have any realistic defense
for the impersonation attack scenarios discussed above.
In particular,
such a secret-registration design
places registration events in a fundamentally non-transparent context
in which there appears to be no conceivable way
for the voter, or anyone else,
to notice that an improper registration event has occurred.
Voters must individually, and at large,
simply trust that registration officials never
attempt to register on behalf of voters
in order to steal their real credentials and silently manipulate the vote,
potentially at scale.

While we are obviously concerned with coercion resistance,
that being the focus of this paper,
we do not believe the threat of coercion is common or important enough,
in almost any conceivably-realistic election environment,
to justify the transparency cost of making it even potentially feasible
for a corrupt registration authority
to impersonate their own voters at scale without detection.

\subsubsection{Coercion to abstain from e-voting}

Given the impracticality of hiding e-voting registration events
as discussed above,
this leaves us with the issue that a long-term coercer
can demand that a voter refrain from registering for e-voting at all,
and simply watch the public registration log to ensure the voter complies.
This perhaps-unavoidable weakness
does not mean that a long-term coercer
can force the voter to abstain from \emph{voting}, however.

As discussed in \cref{apx:registration},
in most realistic scenarios,
e-voting via \sysname would normally be deployed alongside
at least one other ``baseline'' voting channel such as in-person voting.
With proper coercion-resistant cross-channel integration,
a long-term coercer who successfully forbids a voter
from ever registering for e-voting
denies the voter only \emph{the convenience of e-voting},
forcing the voter instead to vote in-person for example.
This can still be a problem if the coerced voter
cannot readily travel to an in-person voting site
any time within the voting period,
of course --
but this is an existing problem that e-voting at least does not make worse.

In a hypothetical future government wishing to go ``all-digital'' --
as Estonia seems to aspire to --
we might envision e-voting at some point
being the \emph{only} available voting channel.
In such an environment,
successful coercion to abstain from registering for e-voting
would indeed by default translate into coercion to abstain from voting at all.
In such an all-digital environment, however,
one potentially-realistic mitigation would be
to make e-voting registration mandatory.
Citizens would be expected and required to maintain valid e-voting credentials,
regardless of whether or how they use them to vote.
Periodic mandatory renewal of e-voting credentials might in principle be
simply part of an occasional mandatory in-person ``citizenship checkup'':
\eg an opportunity for the government to verify that the voter is still alive,
at the very least,
to provide a safe opportunity to report
coercion attempts or other issues that might arise,
and to renew other identity documents at the same time.
If a governmental call to renew e-voting credentials is treated
similarly to a call to mandatory jury duty,
then even a long-term coercer must allow the voter to go renew occasionally,
or else risk government officials coming to check on the voter
and detecting the long-term coercion in the process. 
\section{Management and storage of credentials}
\label{apx:credentials}

The focus of this paper is on \name,
the e-voting registration stage in \sysname:
in particular the printing of real and fake paper credentials
to be activated later on any device(s) the voter trusts.
Once a credential has been activated on a particular device,
the long-term storage and management of that credential,
potentially across several subsequent elections until the credential expires,
is an important security and usability challenge
beyond the scope of this paper.
Nevertheless,
we briefly discuss here some general considerations
for the secure storage and lifecycle management of credentials.

\subsection{Credential storage lifecycle}
\label{apx:credentials:lifecycle}
The advantage of using voting credentials that are stored
on the voter's device is convenience and cost reduction,
for both the registrar and the voter, compared to issuing
devices, such as smart cards.
However, we cannot presume that voters will keep the
same device for extended periods (\eg years), unlike
other government materials such as passports.
Fortunately, since the private voting materials are digitally
stored on the voter's device, they can easily be transferred
over to a new device using NFC or QR codes.
This raises an obvious question, however: what if the voter loses their
device?

\parhead{Re-registration.}
The simplest solution is for voters to re-register for a
new set of credentials.
This approach obviously imposes some inconvenience on voters, however.

\parhead{Cloud Storage.}
Another approach is to back up all voting data onto a cloud
service~\cite{puttaswamy2011silverline}.
Voters would then only need to authenticate with the cloud service
to retrieve their voting credential.
This method is vulnerable to single points of failures,
however,
including the potential compromise of the cloud provider or
phishing attacks targeting the voter.

\parhead{Threshold solutions.}
A third approach involves using threshold solutions
via secret sharing~\cite{shamir1979SecretSharingFixed} 
to entrust voting materials to a set of trustees:
\ie people trusted by the voter, such as friends and family.
This method allows a voter to back up their
credentials without depending on any single third party.
To retrieve their backed-up credential, the voter must
collaborate with a threshold of their trustees
to gather the necessary shares on a new device.
No single trustee can act maliciously, 
as each trustee holds only a share of the voter's voting materials,
which is completely useless in itself.
However, this solution is not as convenient compared to
using a cloud provider and may face reliability 
challenges~\cite{schechter2009SocialAccountRecovery,
javed2014secure}.
User studies on social recovery mechanisms
therefore represent important future work,
which could
provide insight into whether social recovery is
suitable for this purpose.

\subsection{Management of multiple credentials}
\label{apx:credentials:one-device}

If a voter decides to store
multiple credentials on the same device,
this raises the question of how the user can securely and
properly access their intended credential.

\parhead{Security Questions.}
One possible method for securing credentials is the use of
security questions.
A user could link each credential to a specific security question,
simplifying access without the need to remember complex passwords.
Alternatively, different answers to the same security question
could also reveal a specific credential.
However, security question-based solutions are known to be
susceptible to social engineering attacks~\cite{bonneau2015secrets}.

\parhead{Passwords.}
Another approach is to link a credential 
with a password.
While passwords are generally high-entropy alphanumeric strings,
remembering multiple complex passwords can be challenging for voters 
and may negatively impact the usability of the system~\cite{gao2018forgetting}.

\parhead{PINs.}
PINs are also possible but are composed of a small number of digits, 
making them lower in entropy and, therefore, less secure than passwords.
Furthermore, having to remember several PINs could still 
be cumbersome for voters~\cite{
estaji2020UsableCRSmartCards}.
Some applications like Signal require users to re-enter
their PIN periodically to keep them from forgetting it.

\section{Side-channel attacks and defenses}\label{apx:side-channels}

While side-channel attacks are beyond this paper's scope
and threat model in general,
we acknowledge that they represent
important and potentially-realistic threats in practice.

The most privacy-sensitive stage of \name's registration process
is of course the time the voter spends in the privacy booth
to print their real and fake credentials.
A variety of side-channel attacks
could compromise the privacy that the booth is intended to afford.
We briefly discuss a few such potential side-channel attack vectors below,
together with some potential mitigations.
This list is by no means comprehensive, however,
and any systematic treatment of side-channel attacks
we leave to future work.

In general,
a coercer need only learn the precise \emph{number of credentials}
that the coercer's victim creates in the privacy booth.
Given precise knowledge of this number,
the coercer can simply demand that the victim
hand over exactly that number of credentials after registration,
and will know that one of these credentials
must be the victim's real credential,
even if it is not obvious which one that is.
The coercer can simply vote the same way with all of the victim's credentials,
knowing that whichever one is real will cast a vote that counts.
Some of the discussion below therefore focuses
on side-channel attacks that might give a coercer
precise knowledge about the number
of credentials created by a voter inside the booth.

For further reference,
a broader survey on side channels may be found
in~\cite{lavaud2021SideChannelSurvey}.

\parhead{Recording.}
Perhaps the most-realistic side-channel attack
against the privacy of \name credential printing
is the direct analog of the ``ballot selfie'' attack
against privacy booths used for in-person voting:
\ie voters might be coerced to use recording devices in the privacy booth.
We discuss this threat vector and potential mitigations
more extensively in \cref{apx:devices}.

\parhead{Timing.}
A coercion adversary
$\CA$ might be able to infer the number of credentials
a voter creates by analyzing the duration spent in the
booth.
Kiosks might introduce a random, artificial delay
during credentialing to counteract this threat, making
the adversary's estimations more challenging.
This delay might appear during screen changes, or
a static ``Please wait...'' message without any underlying
activity.
Further,
kiosks could enforce, for each voter,
a minimum random number of fake credentials.

\parhead{Sound.}
The noise generated by \name's credential printer 
might expose the of precise number of credentials a voter creates,
if the coercer is near enough to hear the printer noise.
To mitigate this side-channel threat,
we suggest minimizing printer sounds
using a low-noise printer.
Designing privacy booths as separate rooms with a degree of soundproofing
could provide a stronger defense against sound-based attacks,
at a significantly higher construction cost of course.
Additionally, introducing a noise simulation
device might help mask the origin of the sound.

\parhead{Electromagnetics.}
A coercer located nearby might use more sophisticated scanning technology
to detect electromagnetic leakage from the kiosk the victim is using,
or to ``see through'' the privacy booth's protection in other ways.
Privacy booths could in principle be TEMPEST-shielded
to protect against such attacks~\cite{faulkner2022TEMPEST},
but we doubt the cost of such defenses will be seen
as justified by hopefully-rare coercion threats in most cases.
As pointed out in \cref{apx:devices:anti-recording-trip}, however,
some sites where we envision \name might be deployed --
such as at embassies or military bases
for the use of expatriates and military personnel, for example --
might already have suitable areas with stronger side-channel protections,
thus reducing the incremental cost of stronger defenses.

\section{Formal voting and tallying protocol}
\label{apx:tallying-scheme}

In this section,
we present the voting and linear-time 
tallying protocol used within \sysname. 
The key idea follows Koenig et al.'s 
work~\cite{koenig2011PreventingFloodingVotes}
to admit only ballots 
cast with registrar-issued credentials,
which enables the removal of fake ballots in linear time 
using deterministic tags computed after 
a publicly verifiable shuffle. 
This design is a natural fit with \name: 
the kiosk issues a finite set of credentials per voter, 
and issued credentials are generated independently, 
which allows us to retain the performance benefits 
of deterministic tags while preserving coercion-resistance
against algebraic-relation attacks~\cite{
smith2005CRLinear,
weber2007CRLinearDeterministicFingerprint,
koenig2011PreventingFloodingVotes,
pfitzmann1995BreakingAnonChannel}.

\subsection{Overview}

Many JCJ-style schemes remove duplicates---ballots cast with
the same credential---and fake ballots 
using plain\-text-e\-quiv\-a\-lence tests (PETs) once the 
voting phase is complete~\cite{juels2010CoercionResistantElections,
clarkson2008Civitas,
schlapfer2012VoteAuthorization,
achenbach2015JCJDeniableRevoting}.
As a result of this design choice, 
removing duplicates in these schemes is quadratic in the number of cast ballots, 
and real-credential membership testing is linear in the 
number of voters per remaining ballot. 
These costs are evident in Civitas~\cite{clarkson2008Civitas},
where tallying latency quickly becomes impractical
as the ballot count increases (\cref{sec:eval:compare-end-to-end}).

In \sysname, 
since voters obtain their finite set of real and fake 
credentials during registration, 
we can enforce a rule that only registrar-issued credentials 
are admissible for casting ballots.
Concretely, the ledger maintains a 
credential sub-ledger $\L_C$ of all kiosk-issued 
credential public keys;
voters' \VSD{}s can easily verify that their credential's
public key $c_\pk$ is present on $\L_C$.
A ballot is accepted only if $c_\pk\in\L_C$,
includes a valid signature under the corresponding secret key 
and is the first ballot posted under $c_\pk$ for the election. 
After voting closes, the election authority runs a publicly 
verifiable shuffle~\cite{neff2001VerifiableShuffling}, 
computes deterministic tags on the encrypted credential 
components from both the registration roster $\L_R$ and 
the accepted ballots on $\L_V$, and then performs a 
set-join on tags, thereby retaining exactly those 
ballots whose credentials appear on the roster, 
achieving linear-time filtering~\cite{
smith2005CRLinear,
weber2007CRLinearDeterministicFingerprint,
koenig2011PreventingFloodingVotes}.

Deterministic tags are vulnerable to coercion if an adversary 
can inject algebraically-related credentials 
(\eg $\sigma,\sigma^{a}$), allowing the adversary to 
test later for the same relation post-shuffle~\cite{
pfitzmann1995BreakingAnonChannel}. 
In \sysname, however, all admissible credentials are 
issued by the kiosk and independently sampled,
and the kiosk is trusted for coercion resistance,
so an adversary cannot cast admissible ballots 
using specially-related keys
not authorized by the election authority's kiosks.

\subsection{Voting Scheme}
\label{apx:tallying-scheme:voting}

\begin{figure}[t]
\gameblock[codesize=\scriptsize,skipfirstln,jot=-0.6mm]{$\pcalgostyle{Vote}(\L, c, \A_{pk}, \beta, M)$}{%
    \pcskipln \\
    \textbf{\small VSD}((G, q, g), \L, \A^\pk, V_{id}, c, M, \beta) \\
    c_{\pc}' \gets L_R[V_{id}]; c_{\pc}' \iseq c_{\pc} \< 
        \text{\% Latest registration match?}\\
    c_\pk \in \L_C 
        \< \text{\% Voter credential listed?} \\
    E_{vt} \gets \elgamal.\enc(\A^{\pk}, \beta) 
        \< \text{\% Encrypt vote option}\\
    E_{cr} \gets \elgamal.\enc(\A^{\pk}, c_{\pk}) 
        \< \text{\% Encrypt credential}\\
    \pi_{vt} \gets \nizk\{ 
        \beta: E_{vt} = \elgamal.\enc(\A^\pk, \beta) \< \text{\% Well-formed vote option} \\
    \t \t \t \t \t \t \land \beta \in M
        \} \< \\
    \pi_{cr} \gets \nizk\{
        c_\sk : E_{cr} = \elgamal.\enc(\A^\pk, g^{c_\sk}) \< \text{\% Credential binding}\\
    \t \t \t \t \t \t \land c_\pk = g^{c_\sk}
        \} \< \\
    \sigma_{vt} \gets \sig.\sign(
        c_\sk, E_{vt} \concat E_{cr} \concat \pi_{vt} \concat \pi_{cr}
        )
        \< \text{\% Ballot authentication} \\
    B \gets (
        c_\pk, E_{vt}, E_{cr}, \pi_{vt}, \pi_{cr}, \sigma_{vt})
        \< \text{\% Store ballot} \\
    \L_V \append B \< \text{\% Post ballot}   
}
\caption{\textbf{Voting}
The voter encrypts their choice and their credential,
proves the encryption of their choice and their credential 
is well-formed, and signs and posts the ballot.}
\Description{Voting Procedure:
The voter encrypts their choice and their credential,
proves the encryption of their choice and their credential 
is well-formed, and signs and posts the ballot.}
\label{fig:scheme:vote}
\end{figure} %
When casting a vote (\cref{fig:scheme:vote}), 
the \VSD{} first verifies that the public
credential on the device still remains the
voter's public credential on the ledger;
this confirms that the credential is part
of the voter's latest registration session.
If successful,
it prepares the ballot by 
ElGamal encrypting the voter's chosen vote $\beta$
along with the credential selected by the voter $c_\pk$.
Furthermore, the \VSD{} constructs $\nizk$ proofs
to prove correctness for both ElGamal encryptions.
Finally, the \VSD{} digitally signs the ballot
with the credential's corresponding private key
to form the ballot.
When the ledger receives this ballot,
it verifies that $c_\pk \in \L_C$, 
the digital signature is valid,
the $\nizk$ proofs are valid and
that this is the first ballot 
containing $c_\pk$ in this election.

\subsection{Tally.}
\label{apx:tallying-scheme:tally}
We adopt $\Tally$ from the \name API and focus on 
computing the tally $X$ and publishing a proof bundle $P$.

\parhead{Inputs.}
From $\L_R$, $\A$ collects roster ciphertexts 
$$\mathbf{C}^r=(c_{\pc,1},\ldots,c_{\pc,n_V})$$
where each $c_{\pc,i}$ encrypts a public key 
$c_{\pk,i}$---the real credential.

From $\L_V$, $\A$ collects accepted ballots as pairs
$$(\mathbf{C}^v,\mathbf{V})=((c_1,\ldots,c_b),(v_1,\ldots,v_b))$$
with $c_j=E_{cr}$ and $v_j=E_{vt}$.

\parhead{Verifiable mixnet.} 
Each member in $\A$ then runs a publicly verifiable mixnet 
to unlink inputs from outputs while preserving within-ballot
associations~\cite{neff2001VerifiableShuffling,
bayergroth2012EfficientShuffle}:
\begin{equation*}
    \mathbf{C}^{r}_s=\mix(\mathbf{C}^{r}),\qquad (\mathbf{C}^{v}_s,\mathbf{V}_s)=\mixk(\mathbf{C}^{v},\mathbf{V})
\end{equation*}

\parhead{Deterministic tags.}
Over the shuffled ciphertexts, 
$\A$ collectively computes deterministic tags for 
each ballot credential with publicly verifiable proofs \cite{froelicher2017UnLynx,
weber2007CRLinearDeterministicFingerprint}:
\begin{equation*}
    \mathbf{C}^{r}_t\leftarrow\tagg(\mathbf{C}^{r}_s),\qquad \mathbf{C}^{v}_t\leftarrow\tagg(\mathbf{C}^{v}_s)
\end{equation*}
Tags are equal if and only if the underlying 
plaintext (public keys) are equal.

\parhead{Linear-time join.} 
Let $J=\{\,j\mid \mathbf{C}^v_t[j]\in \mathbf{C}^r_t\,\}$. 
Keep exactly the matched vote ciphertexts:
\begin{equation*}
    \mathbf{V}_r=\{\ \mathbf{V}_s[j]\ :\ j\in J\ \}
\end{equation*}
Because each voter contributes exactly one entry in $\L_R$ and 
$\L_V$ accepts at most one ballot per credential, 
every real-credential ballot is kept once, 
and any ballot cast using a credential not in $\L_R$
is discarded.

\parhead{Decryption and proof bundle.} 
$\A$ finally verifiably decrypts $\mathbf{V}_r$ to obtain 
plaintext votes $\mathbf{V}_d$ as the final tally $X$.
The bundle $P$ consists of 
the mixnet proofs, tags and tag-computation proofs, and
the decryption proofs.

Finally, $A$ verifiably decrypts $\mathbf{V}_r$ as $\mathbf{V}^d_r$ using
distributed ElGamal decryption protocol,
which also generates zero-knowledge 
proofs that vote $\mathbf{v}_i$ is the correct decryption of $\mathbf{V}^d_{r_i}$.

\subsection{Related Work}
Koenig et al.\cite{koenig2011PreventingFloodingVotes} 
first proposed the approach of 
admitting only ballots cast with registrar-issued 
credentials to prevent ledger flooding and
enable linear-time filtering using 
the deterministic-tag techniques inspired by earlier 
works~\cite{smith2005CRLinear,
weber2007CRLinearDeterministicFingerprint}. 
Their design, however, involves participation from
the talliers during the voting phase to filter out 
duplicates and reject non-issued credentials.
In contrast, \sysname does not rely on talliers 
being online during the voting phase because each ballot 
carries $c_\pk$ in the clear and must satisfy 
$c_\pk \in \L_C$ with a valid signature under $c_\pk$, 
allowing the ledger to reject duplicates and 
inadmissible ballots.
This operational difference is important for
real deployments where at least one tallying component
is air-gapped to harden privacy 
guarantees~\cite{swisspost2024Architecture}.

\subsection{Standing Votes}
This scheme supports standing votes for voters facing 
extreme coercion --
those unable to obtain unsupervised access to any trustworthy device for voting,
or who are subject to a thorough physical search by a coercer immediately after registration.
During registration, 
a voter under extreme coercion may delegate their voting rights 
by selecting a political party on the kiosk.
The kiosk then encrypts the political party's public key
as the voter's public credential tag: 
$$c_{\pc}=\elgamal.\enc(\A^{\pk},c_{\pk}^{P})$$

The voter then leaves the booth holding only fake credentials.
At election time, 
each party $P$ casts exactly one ballot under $c_{\pk}^{P}$. 
During tallying, after mix and tag computation, 
the tag-join retains the party's ballot with 
a multiplicity equal to the number of delegations to $P$.
For auditability, 
the authority also decrypts the credential component of 
the unique party ballot and proves that the revealed key 
belongs to the political party, allowing delegating 
voters to check that the party voted as publicly stated.

\parhead{Coercion-evidence.}
Standing votes can provide an aggregate, 
privacy-preserving signal of coercive pressure:
the number of standing votes cast for each party is revealed at 
the end of the election, without identifying 
which voters delegated. 
This signal helps inform the electorate about
the prevalence of coercion and vote-buying,
and therefore could introduce policy responses
aimed at reducing such undue influence; this is in line
with the notion of coercion-evidence~\cite{Grewal2013CoercionEvidence}.

\section{Prototype implementation code size}
\label{apx:code-size}

\begin{table}[t]
\footnotesize
\centering
\begin{tabularx}{\columnwidth}{|X|r|r|r|r||r|}
\hline
\textbf{Component}	& \textbf{Go} & \textbf{Python}
				& \textbf{C} & \textbf{PHP}
					& \textbf{Total} \\
\hline
\name-Core       	& 1876  & 0       & 0   & 0    	& \textbf{1876}	\\
\name-Peripheral 	& 757   & 0       & 0   & 0    	& \textbf{757}	\\
\name-UI 	 	    & 0     & 2901    & 0   & 1408 	& \textbf{4309}	\\
\sysname-Tally      & 1816  & 0       & 424 & 0    	& \textbf{2240}	\\
\hline
\textbf{Total}		& 4449	& 2901	  & 424 & 1408	& \textbf{9182}	\\
\hline
\end{tabularx}
\vspace{0.25cm}
\caption{Lines of code across \sysname sub-systems. C code in \sysname-Tally is to integrate with a Bayer Groth Shuffle C library~\cite{anderspkd2021grothShuffleCLibrary}}
\label{tab:code-size}
\end{table}

As a rough proxy metric for implementation complexity,
we present in \cref{tab:code-size}
a breakdown of the code size of the current prototype implementation.
The components are as follows:

\begin{itemize}[leftmargin=*]
    \item \textbf{\name-Core}:
	This component contains the core logic and cryptographic operations as outlined in \cref{sec:formal-scheme}.
    \item \textbf{\name-Peripheral}:
	This component includes the logic required
	for creating and processing QR codes
	as well as operating peripheral devices
	for \cref{subsec:setup} and \cref{subsec:eval-latency}.
    \item \textbf{\name-UI}:
	This component includes the user interface logic
	for the usability study,
	featuring a web server for managing the user survey
	and the user interface definitions.
    \item \textbf{\sysname-Tally}:
	This component comprises the logic and             cryptographic operations
	detailed in \cref{apx:tallying-scheme}.
\end{itemize}

The \name-Core and \name-Peripheral components
are the ones used in this paper for evaluation of
user-observable delays in \cref{subsec:eval-latency}.
The \name-Core and \sysname-Tally components
are the ones primarily used in our evaluation
of the computational efficiency of the ``end-to-end''
registration, voting, and tallying process
described in \cref{sec:eval:compare-reg}.
The \name-UI component contains our prototype registration user interface;
it is currently used
only in the companion usability study~\cite{evoteconscience}
and is not yet integrated into the rest of the \name and \sysname prototype
that this paper focuses on.
\section{Project and paper-submission history}
\label{sec:submission-history}

We briefly summarize here
the development, writing, and submission history of this paper.
Although not customary,
we feel that disclosing such a summary
may help serve research transparency in general,
and we hope might in particular help future researchers (e.g., new PhD students)
facing the challenges and frustrations of carrying out
and publishing high-quality research,
especially research crossing multiple domains
such as systems, security/privacy, and usability.

This work commenced in fall 2020,
coincident with the first author's entry into EPFL's PhD program.
The system's basic concept, initial prototype,
and first submission-ready paper were complete by spring 2021.
The peer review and publication process, however,
involved six unsuccessful submission and revision cycles
over the next four years,
before being finally accepted only in this seventh cycle.
We outline below these submission cycles
and the main reasons for each rejection based on our reading
of the peer review feedback,
followed by a few lessons we learned in the process.

\paragraph{April 2021 submission to IEEE S\&P '22, 5 reviews:}
The reviewers found the idea interesting and promising,
but were concerned with the lack of systematic evidence of usability
or formal security proofs.
This motivated our first usability study of the system,
and our formal security analysis included in subsequent submissions.
Reviewers were also unconvinced of the system's novelty
over JCJ~\cite{juels2010CoercionResistantElections},
which we addressed with rewrites to clarify the areas of novelty.

\paragraph{October 2021 submission to USENIX Security '22, 5 reviews:}
Reviewers were glad to see usability results,
but were understandably concerned
that our first user-study population,
41 PhD students at EPFL,
was insufficiently diverse.
This motivated our subsequent 150-participant usability study.
Other reviewers were unconvinced that the work achieved coercion resistance,
due to misunderstandings of the threat model and design,
leading us to major rewrites for clarity.
One reviewer again perceived limited novelty over JCJ.

\paragraph{February 2023 submission to USENIX Security '23, 5 reviews:}
Reviewers were unconvinced
that voters could distinguish their real and fake credentials
or could use them effectively for voting,
and had concerns with the use of printed envelopes for IZKP challenges.
One reviewer perceived a lack of novelty
over Civitas~\cite{clarkson2008Civitas}
and MarkPledge~\cite{neff2006MarkPledge}.
Our conclusion was that we could not adequately cover
both the system and the usability details in the space available in one paper,
leading us to split the latter into a separate paper~\cite{evoteconscience}.

\paragraph{August 2023 submission to IEEE S\&P '24, 4 reviews:}
This submission received three ``Accept'' ratings
and only one ``Weak Reject'',
but the reviewers rejected the submission on the grounds that
the threat model, election setup processes, and novelty over JCJ
were still not sufficiently clear.
By this time we had identified a clear pattern
that security/privacy reviewers
tend to focus primarily on the system's cryptographic design and novelty,
despite our attempts to emphasize clearly and explicitly that this work's focus and novelty
is not cryptographic but in its systems-design and usability elements.
This tendency is understandable in retrospect
because most of the closely-related prior-work precedents are cryptography-focused~\cite{juels2010CoercionResistantElections,clarkson2008Civitas,neff2006MarkPledge}.\footnote{
    We have heard anecdotally from colleagues that even in venues narrowly scoped around e-voting, such as \href{https://www.e-vote-id.org}{E-Vote-ID}, it can be difficult to publish e-voting papers that are not cryptography papers.  This experience raises the question: does the field of cryptography in effect "own" the topic of e-voting?  How many other application areas across computer science are effectively ``owned'' by particular sub-fields, to the degree of making it difficult or impossible to publish research on that application outside of the sub-field traditionally most interested in that application?
}
This pattern led to our decision to retarget the work
towards systems venues,
where we hoped to find reviewers more interested in
the work's more central and novel systems contributions.

\paragraph{April 2024 submission to SOSP '24, 7 reviews:}
As we had hoped, the reviewers expressed more general appreciation for
the work's ``end-to-end'' systems-design elements
and less concern over cryptographic novelty.
Due to the extreme space pressures of attempting
to convey the big picture, explain the system's entire e-voting pipeline,
and address all the past reviewers' concerns, however,
as the SOSP reviewers pointed out,
we had pushed too many of the technical details into appendices
and left the main paper too high-level and insufficiently detailed or standalone.
This led to another major rewrite to the main text,
to summarize the additional level of technical detail
in the limited space available.
The reviewers ``also felt it would be better for the authors to wait to
submit this paper until the companion paper describing the user study had
been published.''
The usability paper had in fact been accepted for publication in March 2024 but was not yet in its final published form.

\paragraph{December 2024 submission to OSDI '25, 5 reviews:}
This submission's reviews were positive overall
and indicated that we had finally succeeded in making the main paper
sufficiently clear and standalone to convince the reviewers
of the technical correctness and general value of the work's contributions.
Some reviewers were concerned with the paper's fit for OSDI, however,
and whether
``there were significant systems innovations or ideas
that could be widely applicable to other systems.''
This was a particular concern to reviewers due to
a new ``Submission Scope and Relevance'' clause that had been added to
\href{https://www.usenix.org/conference/osdi25/call-for-papers}{OSDI's call for papers}
just that year.
As a result, the paper was rejected
on the basis of these scope concerns.
This led us to expand the discussion of potential broader systems impact
in the introduction and the new \cref{apx:broader-app}
in the next and final submission.

\paragraph{April 2024 submission to SOSP '25, 5 reviews:}
As with OSDI '25, the reviewers again generally accepted
the work's technical correctness and merit,
while again expressing concerns
over whether a systems conference was the best fit.
We are enormously grateful to SOSP's reviewers and organizers
for maintaining SOSP's long tradition of taking ``\href{https://sigops.org/s/conferences/sosp/2025/cfp.html}{a broad view of systems}''
and accepting this paper
despite these scope and fit concerns.
We hope that SOSP will continue to accept high-quality research outside of the usual systems topics.

In summary, we set out in 2020 on a systems-security research project
that incorporated some usability-motivated design elements.
We learned over subsequent years the lessons that
(a) we needed to do (multiple) user studies,
and ultimately a separately-published paper,
to convince reviewers of the system's plausible usability
and render the main systems paper publishable;
and
(b) we needed to shift from targeting security/privacy towards systems venues
to escape the apparent expectation that
all e-voting work
must have cryptographic novelty,
and to find reviewers willing to focus on the work's systems and usability contributions.
Given the still-present concerns over this work's systems fit, however,
it is unclear even to us where this work ideally ``belongs'';
we leave that open question to the community.
 }{}

\end{document}